\documentclass[a4paper,11pt]{article} 
\usepackage{jheppub2} 
\usepackage{hyperref}
\hypersetup{
	bookmarks=true,         
	unicode=false,          
	pdftoolbar=true,        
	pdfmenubar=true,        
	pdffitwindow=false,     
	pdfstartview={FitH},    
	pdftitle={My title},    
	pdfauthor={Author},     
	pdfsubject={Subject},   
	pdfcreator={Creator},   
	pdfproducer={Producer}, 
	pdfkeywords={keyword1} {key2} {key3}, 
	pdfnewwindow=true,      
	colorlinks=true,       
	linkcolor=red,          
	citecolor=purple,        
	filecolor=magenta,      
	urlcolor=blue,           
	linktocpage=true
}
\usepackage{graphics}
\usepackage{rotating}
\usepackage{subfigure}
\usepackage{physics}
\usepackage{amssymb}
\usepackage{xcolor}
\usepackage{pstricks}
\usepackage{amsfonts}
\usepackage{mathrsfs,amsmath}
\usepackage{epsfig}
\usepackage{amsmath,amssymb,amsthm,graphicx,latexsym}
\usepackage{float}
\usepackage{textcomp}
\usepackage{array}
\usepackage{booktabs}
\usepackage{dsfont}
\usepackage{float}
\usepackage{ulem}

\newcommand{\be}{\begin{equation}}
\newcommand{\ee}{\end{equation}}
\newcommand{\bal}{\begin{aligned}}
\newcommand{\eal}{\end{aligned}}
\newcommand{\bc}{\begin{center}}
	\newcommand{\ec}{\end{center}}
\newcommand{\bea}{\begin{eqnarray}}
\newcommand{\eea}{\end{eqnarray}}
\newcommand{\bml}{\begin{subequations}}
	\newcommand{\eml}{\end{subequations}}
\newcommand{\bfig}{\begin{figure}}
	\newcommand{\efig}{\end{figure}}

\newcommand{\slashed}{\hspace{-1.1ex}/}

\newcommand{\bmat}{\begin{pmatrix}}
\newcommand{\emat}{\end{pmatrix}}
\begin{document}
	
	\title{ \textcolor{red}{
	Open Quantum Entanglement: A study of two atomic system in static patch of de Sitter space
	}}
	
	
\author[a]{Samim Akhtar, }	
\affiliation[a]{Department of Physics, Indian Institute of Technology,  Madras, Chennai 600036, India.}		
\author[b]{Sayantan Choudhury,
	\footnote{\textcolor{violet}{\bf Corresponding author, Alternative
			E-mail: sayanphysicsisi@gmail.com}. ${}^{}$}
			\footnote{\textcolor{blue}{\bf NOTE: This project is the part of the non-profit virtual international research consortium ``Quantum Structures of the Space-Time \& Matter" }. ${}^{}$}	}
\affiliation[b]{Quantum Gravity and Unified Theory and Theoretical Cosmology Group, Max Planck Institute for Gravitational Physics (Albert Einstein Institute),
	Am M$\ddot{u}$hlenberg 1,
	14476 Potsdam-Golm, Germany.}
 \author[c,d]{
Satyaki Chowdhury,}	
\affiliation[c]{National Institute of Science Education and Research, Jatni, Bhubaneswar, Odisha - 752050,India.}
\affiliation[d]{Homi Bhabha National Institute, Training School Complex, Anushakti Nagar, Mumbai-400085, India.}	
\author[e]{Debopam Goswami,}	
\affiliation[e]{Department of Physics,
Indian Institute of Technology,
Kanpur- 208 016, India.}		
\author[c,d]{Sudhakar Panda,}	   
\author[f]{Abinash Swain}	
\affiliation[f]{Department of Physics, Indian Institute of Technology, Gandhinagar,
Palaj, Gandhinagar - 382355, India.}

\emailAdd{ samimphysx@gmail.com, sayantan.choudhury@aei.mpg.de, satyaki.chowdhury@niser.ac.in, debodebopam@gmail.com, panda@niser.ac.in, abinashswain2010@gmail.com     }

\abstract{In this work, our prime objective is to study
 non-locality and long range effect of two body correlation using quantum entanglement from various information theoretic measure in the static patch of de Sitter space using a two body Open Quantum System (OQS). The OQS is described by a system of two entangled atoms, surrounded by a thermal bath, which is modelled by a massless probe scalar field. Firstly, we partially trace over the bath field and construct the Gorini Kossakowski Sudarshan Lindblad (GSKL) master equation, which describes the time evolution of the reduced subsystem density matrix. This GSKL master equation is characterized by two components, these are-Spin chain interaction Hamiltonian and the Lindbladian. To fix the form of both of them, we compute the Wightman functions for probe massless scalar field. Using this result alongwith the large time equilibrium behaviour we obtain the analytical solution for reduced density matrix. Further using this solution we evaluate various entanglement measures, namely Von-Neumann entropy, R$e'$nyi entropy, logarithmic negativity, entanglement of formation, concurrence and quantum discord for the two atomic subsystem on the static patch of De-Sitter space. Finally, we have studied violation of Bell-CHSH inequality, which is the key ingredient to study non-locality in primordial cosmology.}


\maketitle
\flushbottom

\section{\textcolor{blue}{\bf \large Introduction}}

The theory of closed quantum systems is a very popular topic and has already been firmly established. But in practical situations no quantum system can be ideally treated as closed and its interactions with the surroundings cannot be neglected, hence it becomes essential to develop a theoretical framework to treat these non adiabatic interactions and develop a proper understanding of the quantum mechanical system. The knowledge of complete time evolutionary dynamics of a quantum mechanical system requires incorporation of the details of the thermal environment, which paves the way towards the study of {\it Open Quantum System }\cite{Lidar, OQS2, Kosloff:2013mea, OQSn, Zhou:2010nb} (OQS), where the physical system weakly interacts with the environment. In general these interactions with the environment significantly controls the time evolution of the quantum mechanical system and induces the phenomenon of {\it quantum dissipation} and gives rise to many out of equilibrium related phenomenon\cite{Choudhury:2020yaa, Bhagat:2020pcd}.
The dynamics of the reduced subsystem of OQS cannot be described using the unitary time evolution operators after integrating out the bath degrees of freedom from the theory. Correct time evolution of the system requires solving the effective {\it master equation}, which describes the non-unitary time evolution of the reduced density matrix of the system.


To deal with the subsystem in the context of OQS the entire combination of the system and its surroundings (thermal bath) is together treated as a closed quantum system and hence the evolution equations can be assumed to follow the unitary transformation rules. 
The prime assumption used in the present context is that the entire system environment combination forms a large closed quantum system.
 Therefore, its time evolution is governed by a unitary transformation generated by a global Hamiltonian which is made up of subsystem Hamiltonian, bath Hamiltonian and interaction Hamiltonian. Moreover the interaction between the system and the bath is assumed to depend on only the present moment, it carries no past memories at all. In short, the interaction is {\it Markovian} in nature. The interaction between the system and the surroundings is assumed to be weak which justifies the argument that the only effective change that can be seen over time occurs in the context of OQS. This assumption is generally useful in treating the evolution of the system when the OQS has sufficient time to relax to the equilibrium before being perturbed in presence of interaction between the system and thermal environment. However, in a special situation where the system has very fast or frequent perturbation in presence of system- thermal bath coupling, one needs to consider {\it Non Markovian} approximation.
 In this treatment additionally it has been assumed that the system is completely uncorrelated with the surroundings at initial time scale, provided the coupling between the subsystem and the environment is sufficiently weak in nature. The techniques developed in the context of OQS have proven very powerful in the context of quantum optics, statistical mechanics,  information theory, 
thermodynamics, cosmology and biology.

\begin{figure}[htb]
	\label{na}
\centering
{
	\includegraphics[width=8cm,height=6cm] {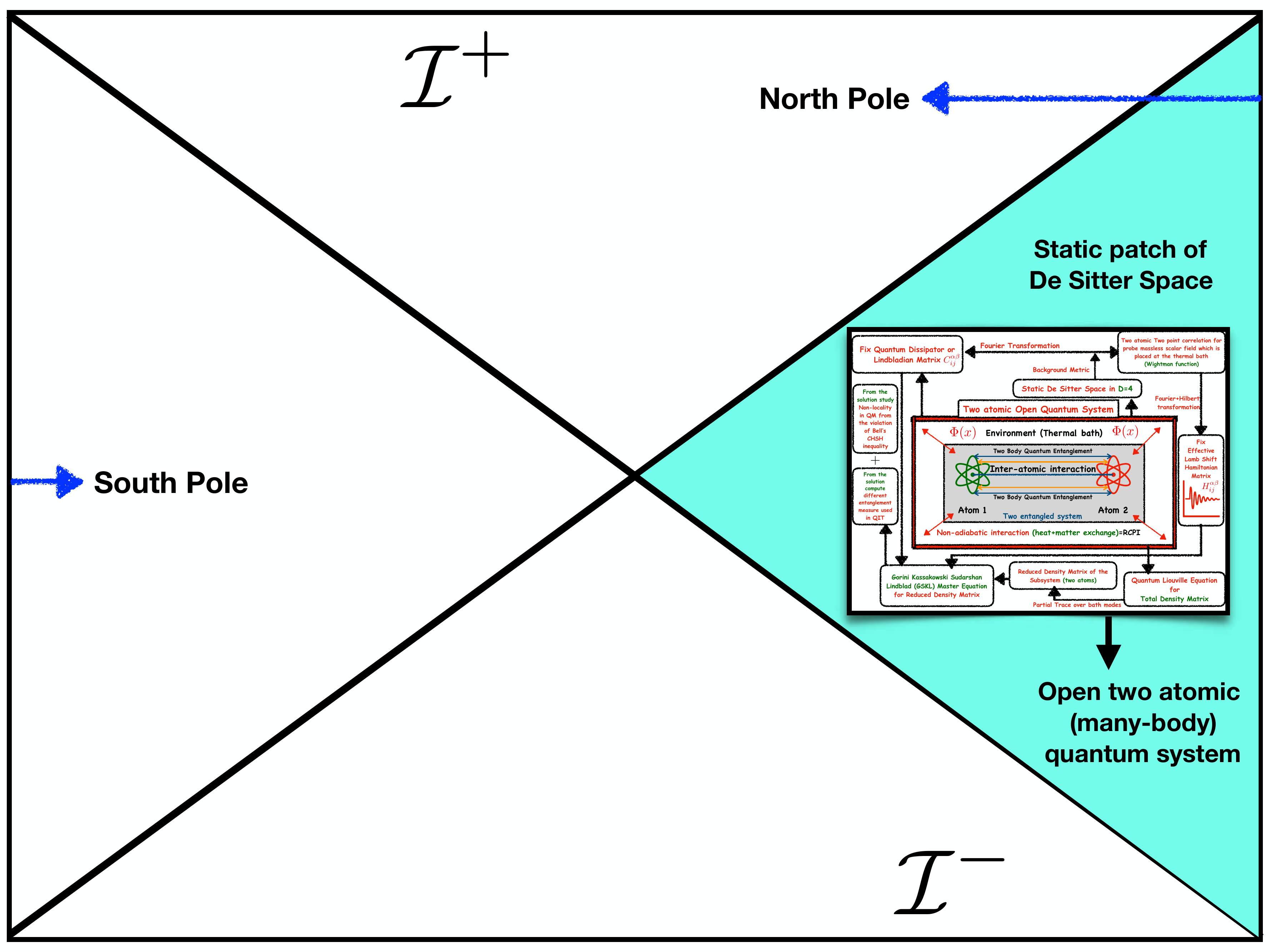}
	}
	\caption{Penrose diagram representing the static patch of the de Sitter space in which we have placed a two atomic open quantum system which is interacting with a thermal bath non-adiabatically. This diagram actually represents the causal patch of an observer sitting at the north pole, which is represented by $r=0$ is static coordinate in de Sitter space. Equivalently this can be described in global coordinate with $\theta=0$. Here the bifurcation Killing horizon for $\partial_{t}$ is represented by $r=\alpha$ where the parameter $\alpha=\sqrt{\frac{3}{\Lambda}}>0$ as the cosmological constant $\Lambda>0$ in de Sitter space. In this context the bifurcation sphere appears just at the middle of the Penrose diagram and in global coordinates it is described by $t=0$ time slice. However, the other three regions can also be filled by the static coordinate system, which is  just like Schwarzschild black holes.}
\end{figure}

On the other hand, quantum entanglement~\cite{arXiv:0901.0404} is probably the most fascinating manifestation of quantum theory in which the beauty of quantum mechanics is truly realized. It is a physical phenomenon that occurs when pairs or groups of particles are generated, interact, or share spatial proximity in ways such that the quantum state of each particle cannot be described independently of the state of the others, even when the particles are separated by a large distance. Equivalently quantum entanglement of a state which is shared by two parties is necessary but not sufficient for that state to be non-local. Quantum entanglement plays a significant role in the context of quantum computation~\cite{Horodecki:2009zz}, information and teleportation theory~\cite{Bennett:1992tv,QETr}, quantum error correction~\cite{arXiv:0905.2794,QEC},etc.There are huge applications of quantum entanglement in various contexts. In interferometry the process of entanglement is used to achieve the Heisenberg limit~\cite{Pezze:2009zz}. In multi electron atomic system the electronic shells always consists of electrons in entangled state.~\cite{ISBN 978-0-470-01187-4}. In the process of photosynthesis,entanglement is seen in the transfer of energy between light harvesting complexes and photosynthetic reaction centres.~\cite{ arXiv:0905.3787}. In living organisms like bacteria entanglement has been observed between the organism and quantized light~\cite{BIO1}. Study of various quantum information theoretic measures which quantifies the phenomenon of quantum entanglement from different types of OQS are important topics of research at present~\cite{QCQI}. Few of them, namely Von Neumann entanglement entropy, R$e'$nyi entropy, quantum discord, concurrence and entanglement of formation are calculated in this paper to study the explicit role of quantum entanglement in the reduced subsystem between the two atoms.

\begin{figure}[htb]
\centering
{
	\includegraphics[width=12cm,height=8cm] {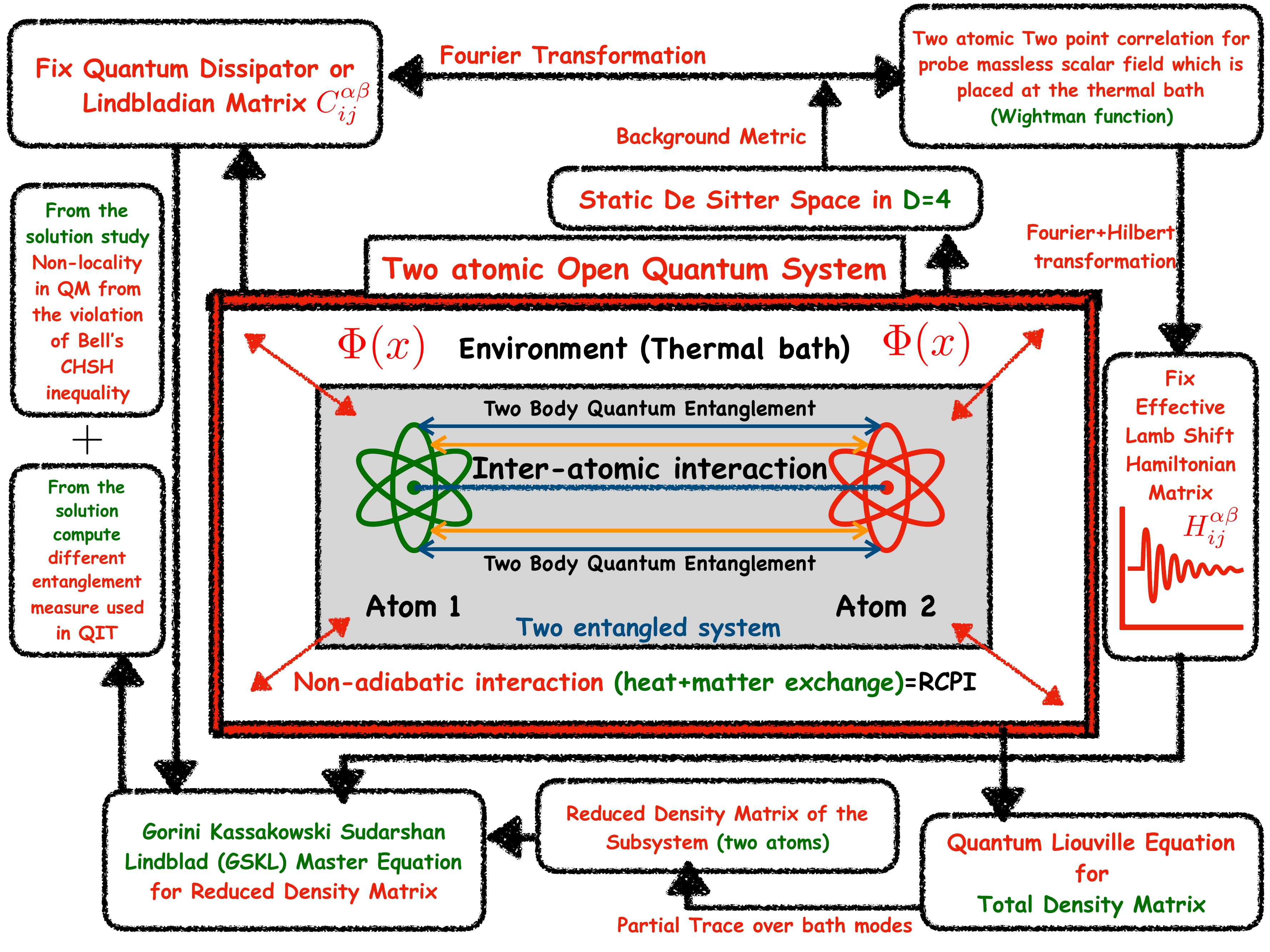}
	}
	\caption{Mnemonic chart for the two atomic (two body) entangled OQS set up.}
		\label{vvb}
\end{figure}

Many authors have studied the physics of quantum fields in curved spacetime using one and two atomic system in the context of OQS~\cite{Huang:2017yjt,Hu:2015lda,Tian:2016uwp}. A single detector system weakly interacting with a reservoir in quantized conformally coupled scalar field in De-Sitter space was investigated in the context of OQS in ~\cite{Yu:2011eq}. A similar study was done using two atoms (detectors) in ~\cite{Huang:2017yjt} where the authors have studied the evolution of the subsystem only under the effect of the {\it Lindbladian} operator for different initial states. In that paper the authors have used only concurrence as the measure to quantify entanglement.In this present work we study the entanglement generation and dynamics, not only in the early and the late time scales but also at any arbitrary time scale. For that purpose we have treated the two atomic system as an open quantum system that interacts with the massless probe scalar field acting as the bath in the de Sitter curved background.After tracing out the bath degrees of freedom, we have analytically solved the GSKL master equation, which gives the non unitary evolution of any open quantum system in constant interaction with the bath. We have taken into account the contributions from both the prime components of the {\it  master equation}, the {\it effective Lamb Shift Hamiltonian} and the {\it Lindbladian operator}, which provides a complete solution of the {\it  master equation} and gives us a proper understanding about the complete time evolution of the reduced density matrix of the two atomic subsystem. Further using this result we have computed a number of information theoretic measures to quantify the quantum entanglement and study the entanglement dynamics and its dependence on various parameters of the system and background spacetime. Additionally, we have studied quantum non-locality by establishing Bell's CHSH inequality violation in de Sitter space from the present OQS two atomic set up.

In fig.~(\ref{na}), we have explicitly shown the Penrose diagram representing the static patch of the de Sitter space in which we have placed a two atomic open quantum system which is interacting with a thermal bath non-adiabatically. This diagram represents actually the causal patch of an observer sitting at the north pole, which is represented by $r=0$ is static coordinate in de Sitter space. Equivalently this can be described in global coordinate with $\theta=0$. Here the bifurcation Killing horizon for $\partial_{t}$ is represented by $r=\alpha$ where the parameter $\alpha=\sqrt{\frac{3}{\Lambda}}>0$ as the cosmological constant $\Lambda>0$ in de Sitter space.In this context the bifurcation sphere appears just at the middle of the Penrose diagram and in global coordinates it is described by $t=0$ time slice. However, the other three regions can also be filled by the static coordinate system, which is  just like Schwarzschild black holes. In fig.~(\ref{vvb}), we have presented a mnemonic chart of the computational scheme for the present two atomic OQS set up.

The plan of this paper is as follows. In \underline{section \textcolor{red}{\ref{boqs}}}, we discuss the basics of the OQS. The objective of this section is to familiarize the reader with the prime components of open quantum systems. 
Further, in \underline{section \textcolor{red}{\ref{ehc}}} and \underline{section \textcolor{red}{\ref{lind}}}, we show the explicit construction of the two prime components of Gorini Kossakowski Sudarshan Lindblad (GSKL) master equation namely the {\it effective Lamb Shift part of the Hamiltonian} and the {\it quantum dissipator operator} or the {\it Lindbladian}.
In \underline{section \textcolor{red}{\ref{ters}}}, we explicitly calculate the analytical solution of the Gorini Kossakowski Sudarshan Lindblad (GSKL) master equation~ \cite{Lindblad:1975ef,Gorini:1975nb}  for the case of two entangled atoms which mimics the role of Unruh-De-Witt detectors, which are minimally coupled to a probe massless scalar field placed in the thermal bath. 
 In \underline{section \textcolor{red}{\ref{Von}-\ref{Quandis}}}, we  explicitly calculate various entanglement measures used in the context of quantum information theory now-a-days. To name some we calculate Von Neumann entanglement entropy~\cite{neumann}, R$e'$nyi~ entanglement entropy~\cite{ren}, Logarithmic Negativity~\cite{Plenio:2005cwa}, Concurrence~\cite{2011326.2011329}, Entanglement of formation~\cite{Wootters:1997id} and Quantum discord~\cite{discc}. Finally in \underline{section \textcolor{red}{\ref{nonlocal}}}, we study the concept of non-locality from the violation of Bell-CHSH inequality~\cite{Kukita:2017tpa} in De-Sitter space for the two atomic entangled subsystem in the context of OQS. 

\section{\textcolor{blue}{\bf \large Modelling two atomic Open Quantum System (OQS)}}
\label{boqs}

In nature we always come across open quantum systems where the system interacts with the environment. The most general form of the total Hamiltonian can be written 
\be 
H_{\bf Total}=H_{\bf System}\otimes{\cal  I}_{\bf Bath} + {\cal I}_{\bf System}\otimes H_{\bf Bath}+H_{\bf Int},
\ee
where, $H_{\bf System}$ represents the two atomic system Hamiltonian~\footnote{This simple choice of the two atomic model, representing a two qubit system is capable of capturing the effect of gravity through the well known phenomenon of \textit{Casimir Polder interaction}, which is a consequence of the fluctuations of the vacuum state of a quantum field. This Casimir interaction has proven to be an effective mean to display the non local properties of the quantum field correlations to probe the phenomenon of entanglement. The information about gravity or the background curved spacetime can be extracted from this phenomenon due to it's significant modification by the relativistic motion of the interacting systems in the curved spacetime.}, $H_{\bf Bath}$ describes the thermal bath Hamiltonian,which is described by massless probe scalar field \cite{Bhattacherjee:2019eml} which is minimally coupled to gravity in static de Sitter background and $H_{\bf Int}$ signifies the interaction between the thermal bath and the system under consideration in OQS. ${\cal I}_{\bf System}$ and ${\cal  I}_{\bf Bath}$ are basically identity operators in system and bath Hilbert space respectively.

In our two atomic OQS set up the system, bath and the interaction Hamiltonian are described by the following expressions  \cite{Bhattacherjee:2019eml}:
\bea 
\label{syshamil}
H_{\bf System}&=&\frac{\omega}{2}\sum^{2}_{\alpha=1}{\bf \hat{n}}^{\alpha}.{\bf \sigma}^{\alpha}\\
H_{\bf Bath}(\tau)&=&\int^{\infty}_{0} dr~\int ^{\pi}_{0}d\theta~\int^{2\pi}_{0}d\phi~\left[\frac{\Pi^2_{\Phi}(\tau,r,\theta,\phi)}{2}\right.\nonumber\\
&& \left.~~~~+\frac{r^2\sin^2\theta}{2}\left\{r^2~(\partial_{r}\Phi(\tau,r,\theta,\phi))^2+ \frac{\left((\partial_{\theta}\Phi(\tau,r,\theta,\phi))^2+\frac{(\partial_{\phi}\Phi(\tau,r,\theta,\phi))^2}{\sin^2\theta}\right)}{\displaystyle\left(1-\frac{r^2}{\alpha^2}\right)}\right\}\right],~~~~~~~\\
H_{\bf Int}(\tau)&=&\mu\sum^{2}_{\alpha=1}\left({\bf \hat{n}}^{\alpha}{\bf \sigma}^{\alpha}\right)\Phi(\tau,{\bf x}^{\alpha})=\mu\sum^{2}_{\alpha=1}\left({\bf \hat{n}}^{\alpha}.{\bf \sigma}^{\alpha}\right)\Phi(\tau,{r}^{\alpha},{\theta}^{\alpha},{\phi}^{\alpha})
\eea

It is important to note that, $\omega$ represents the renormalized energy level for two atoms ~\footnote{It must be noted carefully that $\omega$ appearing in the system hamiltonian is not the same as the natural frequency of the atoms, which is actually $\omega_0$. One can generally understand the significance of the background spacetime in this context. In the flat space limit, the $\omega$ appearing in the system Hamiltonian will be replaced by just $\omega_0$ along with some correction factors, whereas for curved spacetime, the $\omega$ has two major contributions, one coming from $\omega_{0}$ or the flat space limit result and the other non trivial contribution coming from the Hilbert transformation of the fourier transformed Wightman functions. This non trivial part actually captures the information about the background spacetime.}, given by:
\be
\label{ren} \begin{array}{lll}
	\displaystyle \omega=\omega_0+i\times\left\{\begin{array}{ll}
		\displaystyle [\mathcal{K}^{(11)}(-\omega_{0})-\mathcal{K}^{(11)}(\omega_{0})]~~~~~~ &
		\mbox{\small \textcolor{red}{\bf Atom~1}}  \\ 
		\displaystyle [\mathcal{K}^{(22)}(-\omega_{0})-\mathcal{K}^{(22)}(\omega_{0})]~~~~~~ & \mbox{\small \textcolor{red}{\bf Atom~2}}
	\end{array}
	\right.
\end{array}\ee	
Here $\mathcal{K}^{\alpha \alpha}(\pm \omega_{0})$ for $\alpha \in \{1,2\}$ are Hilbert transformations of Wightman function computed from the probe massless scalar field, which we have defined explicitly in later section of this paper. Also, $\omega_0$ represents the natural frequency of the two identical atoms, which we have fixed \cite{Baumann:2009ds,Choudhury:2017glj,Naskar:2017ekm,Choudhury:2015eua,Senatore:2016aui} at~\footnote{We fix the natural frequency of the two identical atoms by imposing an additional condition:
	\be \coth(\pi k \omega_0)=0~~\Longrightarrow~~\omega_0=\frac{i}{k}\left(n+\frac{1}{2}\right)~~~\forall~n\in \mathbb{Z}\ee
	We impose this condition to simplify the mathematical form of GSKL matrix which will fix the {\it Quantum dissipator} or {\it Lindbladian} operator. }:
\be \omega_0=\frac{i}{k}\left(n+\frac{1}{2}\right)~~~\forall~n\in \mathbb{Z}~~~{\rm and}~~~k=\sqrt{\alpha^2-r^2}>0, ~\alpha=\sqrt{\frac{3}{\Lambda}}>0~~{\rm as}~~\Lambda>0 \ee 
for rest of the computation performed in this paper. In this context, the atoms are characterised by the label $\alpha\in [1,2]$ and $\sigma^{\alpha}_{i}\forall i \in [1,2,3]$ are the Pauli spin matrices.

 The bath Hamiltonian for the probe massless scalar field have been expressed in the static patch of de Sitter space. The prime reason for choosing the de Sitter background can be understood from the fact that it is the simplest non trivial curved spacetime and is maximally symmetric. This unique feature of the de Sitter spacetime has led to extensive study on the quantization of the scalar fields \cite{Birrell:1982ix,Tagirov:1972vv,bunch,ford,Mottola:1984ar,Polarski:1989iu} in this spacetime, which actually acts as the bath for our model. Apart from this the significance of this spacetime can also be understood in the context of cosmology. Current cosmological observations in association with the inflation theory suggets that our universe approches the exponentially expanding de Sitter phase both in the early and the late time scales ,i.e in both the far past and the far future.

 In static patch of de Sitter space the background space time metric is described by the following infinitesimal line element:
 \be
 ds^{2}=\left(1-\frac{r^{2}}{\alpha^{2}}\right)dt^{2}-\left(1-\frac{r^{2}}{\alpha^{2}}\right)^{-1}dr^{2}-r^{2}(d\theta^{2}+\sin^{2}\theta d\phi^{2})~~~{\rm where}~~\alpha=\sqrt{\frac{3}{\Lambda}}>0
 \ee
 Instead of using the time variable as $t$ in the present context we have introduced a new rescaled time variable $\tau$, which is defined as:
 \be \tau=\sqrt{g_{00}}~t=\frac{k}{\alpha}~t=\sqrt{1-\frac{r^2}{\alpha^2}}~t~~~{\rm where}~~k=\sqrt{\alpha^2-r^2}>0
 \ee

 Under this assumption of background space time,  the massless probe scalar field is minimally coupled to the gravity in this context. Additionally, it is important to mention here that the interaction is controlled by the interaction strength or coupling parameter $\mu$ through which the bath degrees of freedom is coupled to the two atomic subsystem. For the description of the OQS we will use the traditional approach of solving the \textit{GSKL  Master Equation} which describes the non-unitary time evolution of the reduced density matrix of the subsystem.


\section{\textcolor{blue}{\bf \large Non unitary time evolution of the reduced subsystem}}
\label{nutes}

The prime objective in this section is to describe the time evolution of an OQS and to arrive at GSKL master equation which properly describes non-unitary behaviour and can be obtained by performing partial trace over the bath content i.e. the massless probe scalar field placed at the static patch of the de Sitter background space time.


The exact mathematical form of the quantum correlation part of the total density matrix can only be obtained by solving the {\it GSKL master equation}. The overall time evolution of the total density matrix is given by the following Liouville Von Neumann equation.

\be
\frac{d}{d\tau}\rho_{\bf Total}(\tau)=-i [H_{\bf Total}, \rho_{\bf Total}(\tau)]
\ee

One can perform a unitary transformation to bring the above Liouville Von Neumann equation into the following convenient form

\be
\label{von}
\frac{d}{d\tau}\rho_{\bf Total}(\tau)=-i [H_{\bf Int}(\tau), \rho_{\bf Total}(\tau)]
\ee

We can integrate the above equation to obtain 

\begin{equation*}
\rho_{\bf Total}(\tau) = \rho_{\bf Total}(0) - i \int \limits_0^t ds [H_{\bf Int}(s),\rho_{\bf Total}(s)]
\end{equation*}

By inserting this back into Equation \eqref{von} and tracing out the bath degrees of freedom, we arrive at

\be
\label{intde}
\frac{d}{dt}\rho_{\bf System}(\tau) = -\int\limits_0^t ds \text{Tr}_{\bf Bath} \{[H_{\bf Int}(t), [H_I(s), \rho_{\bf Total}(s)]] \}
\ee

where we have taken 
$$ \text{Tr}_B\{[H_{\bf Int}(t),\rho_{\bf Total}(0)]\} = 0$$
which means that initially the interaction does not create any dynamics in the bath. 

Equation \eqref{intde} is the integro-differential equation which is non-Markovian in nature as the dynamics at a particular time depends on its past evolution. This is significantly difficult to calculate. Besides on the right hand side of Equation \eqref{intde} we still have the total density operator.

Since, we are primarily concerned about the two atomic system, by doing the following approximations we bring this non-Markovian integro-differential equation into a Gorini-Kossakowski-Sudarshan-Lindblad (GKSL) \cite{Gorini:1975nb, Lindblad:1975ef} master equation which is Markovian in nature and can be solved easily.

\begin{itemize}
	\item \textbf{\textit{Born Approximation:}} Generally in the treatment of any open quantum system the density matrix is usually represented by $$\rho_{\bf Total}(t)=\rho_{\bf System}(t)\otimes \rho_{\bf Bath}(t)+\rho_{\bf Correlation}(t)$$ We assume that the interaction between the bath and the system is switched on at time t=0, which is a good enough assumption because finding such a time scale is always posiible before which there was no interaction i.e ($\rho_{\bf Correlation}(0)=0$) . Besides, we make a stronger claim that the coupling between the system and the bath is weak such that the influence of the bath is small. This is known as \textit{Born approximation}. This approximation is important since we can ignore the initial corelation between the atomic system and bath. In this weak coupling limit situation, the time evolved density matrix is reduced to the form $$\rho_{\bf Total}(t) \approx \rho_{\bf System}(t) \otimes \rho_{\bf Bath}(t) $$
	This assumption reduces the Master equation to Redfield master equation which is still non-Markovian in nature.

	\item \textbf{\textit{Markov Approximation:}} To obtain a Markovian master equation, we assume the relaxation time($\tau _{R}$) to be more compared to the bath correlation time($\tau _{B}$), i.e, {$\displaystyle \tau _{R}\gg \tau _{B}$}
	
	With this approximation we may take the bath as nearly constant $i.e.$ $\displaystyle \rho_{\bf Bath}(t) \approx \rho_{\bf Bath}(0)$
	
	As a result the total density operator can be simplified to 
	$$\rho_{\bf Total}(t) = \rho_{\bf System}(t) \otimes \rho_{\bf Bath}(0) $$
	
	This approximation grouped with the Born approximation is often regarded as the \textit{Born-Markov approximation}. However, with this approximation alone the resulting master equation does not guarantee to generate a quantum synamical semigroup. We need one more approximation to write the Master equation in Lindblad form.

	\item \textbf{\textit{Secular Approximation:}} With this approximation \textit{we can average out and discard highly oscillating terms, comapred to the system timescale of interest, in the Markovian master equation}.

\end{itemize}

With the aid of all these approximations we obtain the following markovian master equation

\be
\frac{d}{d\tau}\rho_{\bf System}(\tau)=-i[H_{\bf eff}(\tau),\rho_{\bf System}(\tau)]+{\cal L}[\rho_{\bf System}(\tau)]
\ee

where $H_{\bf eff}$ is the effective Hamiltonian and ${\cal L}[\rho_{\bf System}(\tau)]$ is the {\it quantum dissipator} or the {\it Lindbladian operator}. These are discussed in more detail in the following sections.

\section{\textcolor{blue}{\bf \large Effective Hamiltonian Construction}}
\label{ehc}
For our system, the effective Hamiltonian can be expressed as
\bea
H_{\bf eff}&=&H_{\bf System}+H_{\bf Lamb~ Shift}=\underbrace {\frac{\omega}{2}\sum_{\alpha=1}^2 {\bf n}^{\alpha}.{\bf \sigma}^{\alpha}}_{\textcolor{red}{\bf Two~Atomic~System}}-\underbrace{\frac{i}{2}\sum_{\alpha \beta=1}^{2} \sum_{ij=1}^3 H_{ij}^{\alpha \beta}({{\bf n}_i}^{\alpha}.{\sigma_i}^{\alpha})({{\bf n}_j}^{\alpha}.{\sigma_j}^{\alpha})}_{\textcolor{red}{\bf Lamb~ Shift=Heisenberg~ spin~ chain}},~~~~~~~
\eea 
where the first term in the effective Hamiltonian, physically represents the Hamiltonian of the two atomic system whereas the second term is known as the \textit{ Lamb Shift Hamiltonian}\cite{Tian:2016uwp,Zhou:2010nb}, which characterizes the {\it atomic Lamb Shift} that occurs due to the interaction between the massless free probe scalar field with the two atomic system under consideration in the background of static de Sitter space. It actually measures a shift in the  energy levels due to this interaction.

Here ${{\bf n}^\alpha}$ and ${{\bf n}^\beta}$ represent the normal unit vectors of the two atoms under consideration in the present OQS set up. The angles between the normal unit vectors and Pauli spin matrices are characterized by the three {\it Euler Angles} $\alpha$,$\beta$ and $\gamma$. However, we consider here that these {\it Euler angles} for two atoms are different to get more general result. 

Therefore the {\it Lamb Shift Hamiltonian} can be re-expressed in terms of {\it Euler angles} as:
\bea
H_{\bf Lamb~ Shift}&=&-\frac{i}{2}\sum_{\alpha \beta=1}^{2} \sum_{ij=1}^3 H_{ij}^{\alpha \beta}({{\bf n}_i}^{\alpha}.{\sigma_i}^{\alpha})({{\bf n}_j}^{\alpha}.{\sigma_j}^{\alpha})=-\frac{i}{2}\sum_{\alpha \beta=1}^{2} \sum_{ij=1}^3 H_{ij}^{\alpha \beta} \cos(\alpha_i^{\alpha}) \cos(\alpha_j^{\beta})\sigma_i^{\alpha}\sigma_j^{\beta}.~~~~~~~
\eea
In this present context, we define the following sets of Pauli  operators for the two atoms using tensor product:
\bea
\sigma_i^1&=&\sigma_i \otimes \sigma_0 ~~~~~~~~(\textcolor{red}{\bf Atom 1})
,~~~~~~~~~~
\sigma_i^2=\sigma_0 \otimes \sigma_i ~~~~~~~~(\textcolor{red}{\bf Atom 2})
\eea
Here $\sigma_0$ is the 2$\times$2 identity matrix and $\sigma_i$ ($i=1,2,3$) are usual Pauli matrices. In our case, we have changed the basis for representing the effective Hamiltonian from the  $\sigma_1$,$\sigma_2$,$\sigma_3$ to the $\sigma_+$,$\sigma_-$,$\sigma_3$ basis which along with the identity matrix forms a complete basis for the space of 2$\times$2 matrices. The change of basis basically reduces the number of differential equations that is obtained from the GSKL master equation and makes them simpler to solve.

In the transformed basis $\sigma_+$ and $\sigma_-$ are defined as: follows: 
\begin{eqnarray}
 \sigma_+&=&\frac{1}{2}(\sigma_1+i \sigma_2) 
=\begin{pmatrix}
0 & &  & & & 1\\
0 & &  & & &0
\end{pmatrix},~~~~~~~~
\sigma_-=\frac{1}{2}(\sigma_1-i \sigma_2)
=\begin{pmatrix}
0 & & & & &  0\\
1 & & & & &  0
\end{pmatrix}.\end{eqnarray}
Therefore the operators $\sigma_+^1$,and $\sigma_-^1$ for \textcolor{red}{\bf Atom 1} is defined as:
\begin{eqnarray}
 \sigma_+^1&=&(\sigma_+ \otimes \sigma_0) 
=\begin{pmatrix}
0 & & &  & & \sigma_0\\
0 & & &  & & 0
\end{pmatrix}
,~~~~~~~~~
\sigma_-^1=(\sigma_-\otimes \sigma_0)
=\begin{pmatrix}
0 & & & & &  0\\
\sigma_0 & & &  & & 0
\end{pmatrix}
\end{eqnarray}
Similarly, the operators $\sigma_+^2$,and $\sigma_-^2$ for \textcolor{red}{\bf Atom 2} is defined in a similar way as: 
\begin{eqnarray}
 \sigma_+^2&=&(\sigma_0 \otimes \sigma_+) 
=\begin{pmatrix}
\sigma_+ & 0\\
0 & \sigma_+
\end{pmatrix}
,~~~~~~~~~
\sigma_-^2=(\sigma_0\otimes \sigma_-)
=\begin{pmatrix}
\sigma_- &  & 0\\
0 &  & \sigma_-
\end{pmatrix}
\end{eqnarray}

Now we transform the Hamiltonian in the new basis which is equivalent to diagonalising the $n_i^{\alpha}\sigma_i^{\beta}$ term, and basically reduces it to the $\sigma_3$ model. Hence the terms of $H_{\bf System}$ in the diagonalized basis respectively becomes: 
\begin{equation}
\textcolor{red}{\bf Atom~1:}~~~ H_1 =\frac{\omega}{2} n_i^1\sigma_i^1=\frac{\omega}{2} n_i\left(\sigma_i \otimes \sigma_0\right) \Longrightarrow \frac{\omega'}{2} \sigma_3 \otimes \sigma_0
\end{equation}
\be
\textcolor{red}{\bf Atom~2:}~~~ H_2 =\frac{\omega}{2} n_i^2\sigma_i^2=\frac{\omega}{2} n_i\left(\sigma_0 \otimes \sigma_i \right) \Longrightarrow \frac{\omega'}{2} \sigma_0 \otimes \sigma_3
\ee
where $\omega'$ is the modified frequency. When the term $n_i\sigma_i$ is diagonalized in the new basis, a factor of $\sqrt{n_3^2+n_+n_{-}}$ arises, which can be incorporated in the frequency as the modification factor. Thus the frequency gets modified by this process and is given by: 
\be 
\omega'=\omega \sqrt{n_3^2+n_+n_-}.
\ee
Here, $n_3$, $n_+$ and $n_-$ are the normal unit vectors of the two atoms in the new basis defined as:
\bea
n_+&=&\frac{1}{2}(n_1+i n_2)=\frac{1}{2}(\cos \alpha_1+ i\cos \alpha_2)
,~~~~~
n_-=\frac{1}{2}(n_1-i n_2)=\frac{1}{2}(\cos \alpha_1-i\cos \alpha_2)
\eea

In the new diagonalized basis (already mentioned earlier) the above mentioned {\it Lamb Shift} Hamiltonian part reduces to the following simplified form:
\be
H_{\bf Lamb~ Shift}=-\frac{i}{2}\sum_{\alpha, \beta=1}^2 \sum_{i,j=\pm}^3  H_{ij}^{\alpha \beta} \sigma_i^\alpha \sigma_j^\beta
\ee
In this context, the effective Hamiltonian matrix elements, $H_{ij}^{\alpha \beta}$, is given by: 
\begin{equation}
  H_{ij}^{\alpha \beta}={\it A^{\alpha \beta}}\delta_{ij}-i {\it B^{\alpha \beta}}\epsilon_{ijk}\delta_{3j}-{{\it A^{\alpha \beta}}\delta_{3i}\delta_{3j}}
\end{equation}
where $A^{\alpha \beta}$ and  $B^{\alpha \beta}$ for the two atomic system are defined as :
    \bea
    &&A^{\alpha \beta}=\frac{\mu^2}{4}[K^{\alpha \beta}(\omega_0)+K^{\alpha \beta}(-\omega_0)],~~~~~~~~
  B^{\alpha \beta}=\frac{\mu^2}{4}[K^{\alpha \beta}(\omega_0)-K^{\alpha \beta}(-\omega_0)],
\eea
where $ K^{\alpha \beta}$ is basically the {\it Hilbert Transform} of the \textcolor{red}{\bf Wightman function} (two point correlator) computed from the probe massless scalar field placed at the bath and is given by the following expression:
\be
 K^{\alpha \beta}(\pm \omega_0)=\frac{P}{\pi i}\int_{-\infty}^{\infty}d\omega \frac{\mathcal{G^{\alpha \beta}(\pm \omega)}}{\omega \pm \omega_0}=\frac{P}{\pi i}\int_{-\infty}^{\infty}d\omega \frac{1}{\omega \pm \omega_0} \int_{-\infty}^{\infty}d\Delta \tau e^{\pm i \omega \Delta \tau} G^{\alpha \beta}(\Delta \tau)
 \ee
where P is the principal value of the integral. Here $\mathcal{G^{\alpha \beta}}$ is the {\it Fourier transform} of the of the \textcolor{red}{\bf Wightman function} in the frequency ($\omega$) space and can be expressed as:
\be
\mathcal{G^{\alpha \beta}}(\pm \omega_0)=\int_{-\infty}^{\infty}d\Delta \tau~ e^{\pm \iota \omega \Delta \tau}~G^{\alpha \beta}(\Delta \tau),
\ee
where $\omega_0$ is the energy difference between the ground and excited states of the atoms and $G^{\alpha \beta}$ is the forward two atomic Wightman Function which is defined as~\footnote{For more details on the computation of the two atiomic Wightman function of the probe massless scalar field in the static de Sitter background geometry, its Fourier transform and its Hilbert transform see the Appendix~\ref{sk}, Appendix~\ref{gskl} and Appendix~\ref{effham}.}:
\be
G^{\alpha \gamma}(\Delta \tau=\tau -\tau^{\prime})=\langle \Phi(x_\alpha,\tau) \Phi(x_\gamma,\tau^{\prime})\rangle_{\beta}=\frac{\int_{\bf SK} {\cal D}\Psi~\langle \Psi|\exp(-\beta~H_{\bf bath})~\Phi(x_\alpha,\tau) \Phi(x_\gamma,\tau^{\prime})|\Psi\rangle}{\int_{\bf SK} {\cal D}\Psi~\langle \Psi|\exp(-\beta~H_{\bf bath})|\Psi\rangle}.
\ee
In the above equations $\mu$ the coupling parameter, represents the interaction strength between the system and the external thermal bath (i.e the gravitationally coupled scalar) field degrees of freedom. The structure of the elements of the coefficient matrix $H_{ij}^{\alpha \beta}$ can be computed in terms of the Wightman function of the external free probe massless scalar field in static De-Sitter background, which finally fix the structure of the effective Hamiltonian in the present scenario.

\section{\textcolor{blue}{\bf \large Quantum Dissipator  or Lindbladian Construction}}
\label{lind}
The concept of fluctuation and dissipation in the context of OQS is introduced into the system  by the additional contribution of the Lindbladian operator in the time evolution equation of the reduced subsytem density matrix. The second term in the {\it Gorini~Kossakowski~Sudarshan~Lindblad~Master (GSKL)} equation is actually characterised as the {\it Lindbladian} or {\it Quantum Dissipator} which in our present context can be written as:
\bea
\mathcal{L[\rho_{\bf{System}}(\tau)]}&=&\frac{1}{2}\sum_{i,j=1}^3 \sum_{\alpha ,\beta=1}^2 C_{ij}^{\alpha \beta}\left[2(n_j^\beta \cdot \sigma_j^\beta)\rho_{\bf{System}(\tau)}(n_i^\alpha \cdot \sigma_i^\alpha)-\left\{(n_i^\alpha \cdot \sigma_i^\alpha)(n_j^\beta \cdot \sigma_j^\beta),\rho_{\bf System}(\tau)\right\}\right],~~~~~~
\eea
where $\rho_{\bf System}$ is the reduced subsystem density matrix of the two entangled atomic system obtained after partially tracing over the external bath scalar field degrees of freedom. The coefficient matrix $C_{ij}^{\alpha \beta}$ is known as the {\it Gorini~Kossakowski~Sudarshan~Lindblad} (GSKL) matrix, which is constructed under the weak coupling limiting approximation on the coupling parameter $\mu$ as appeared in the interaction Hamiltonian. In the context of OQS, the {\it Lindbladian} captures the effect of dissipation.

In the transformed basis, the {\it  Lindbladian} can be re-expressed as:
\be
\mathcal{L[\rho_{\bf{System}}(\tau)]}=\frac{1}{2}\sum_{i,j=\pm}^3 \sum_{\alpha, \beta=1}^2 C_{ij}^{\alpha \beta}\left[2 \sigma_j^{\beta}\rho_{\bf{System}}(\tau) \sigma_i^\alpha-\left\{\sigma_i^\alpha \sigma_j^\beta,\rho_{\bf{System}}(\tau)\right\}\right].
\ee
The matrix GSKL matrix $C_{ij}^{\alpha \beta}$ is given by the following expression:
\be
C_{ij}^{\alpha \beta}=\tilde{A}^{\alpha \beta} \delta_{ij}-i \tilde{B}^{\alpha \beta}\epsilon_{ijk}\delta_{3k}-\tilde{A}^{\alpha \beta}\delta_{3k}\delta_{3j}
\ee
where the quantities $\tilde{A}^{\alpha \beta}$ and $\tilde{B}^{\alpha \beta}$ for the two atomic system is defined as:
\be
\tilde{A}^{\alpha \beta}=\frac{\mu^2}{4}\left[\mathcal{G^{\alpha \beta}}(\omega_0)+\mathcal{G^{\alpha \beta}}(-\omega_0)\right]=\frac{\mu^2}{4}\int_{-\infty}^{\infty}G^{\alpha \beta}(\Delta \tau)\left
[e^{\iota \omega_0 \Delta \tau}+e^{-\iota \omega_0 \Delta \tau}\right]
\ee
\be
\tilde{B}^{\alpha \beta}=\frac{\mu^2}{4}\left[\mathcal{G^{\alpha \beta}}(\omega_0)-\mathcal{G^{\alpha \beta}}(-\omega_0)\right]=\frac{\mu^2}{4}\int_{-\infty}^{\infty}G^{\alpha \beta}(\Delta \tau)\left
[e^{i \omega_0 \Delta \tau}-e^{-i \omega_0 \Delta \tau}\right]
\ee
The components of $C_{ij}^{\alpha \beta}$ matrix are given in the {\bf \large  Appendix \ref{gskl}}.

\section{\textcolor{blue}{\bf \large Time evolution of the reduced subsystem density matrix}}
\label{ters}
The density matrices for \textcolor{red}{\bf Atom~1} and  \textcolor{red}{\bf Atom~2} are given by following expressions in the {\it Bloch sphere} representation:
\bea
&&\textcolor{red}{\bf Atom~1:}~~~~~~\rho_1(\tau)=\frac{1}{2}\left(I+\sum_{i=1}^3 a_i (\tau)\sigma_i \right)=\frac{1}{2}\left[I+{\bf a}(\tau) \cdot {\bf \sigma} \right].
\\
&&\textcolor{red}{\bf Atom~2:}~~~~~~\rho_2(\tau)=\frac{1}{2}\left(I+\sum_{j=1}^3 b_j(\tau) \sigma_j \right)=\frac{1}{2}\left[I+{\bf b}(\tau) \cdot {\bf \sigma} \right].
\eea
Since the two atoms are initially not entangled, the density matrix for the system can be written as the direct product of the two individual density matrices i.e.
\bea
\rho_{\bf System}(\tau)&=&\rho_1(\tau)\otimes \rho_2(\tau)=\frac{1}{2}\left(I+\sum_{i=1}^3 a_i(\tau) \sigma_i \right)\otimes\frac{1}{2}\left(I+\sum_{j=1}^3 b_j(\tau) \sigma_j \right)\nonumber \\
&=&\frac{1}{4}\left[\sigma_0\otimes \sigma_0 +\sum_{j=1}^3 b_j(\tau) \sigma_0 \otimes \sigma_j + \sum_{i=1}^3 a_i(\tau) \sigma_i \otimes \sigma_0+\sum_{i,j=1}^3 a_i(\tau) b_j(\tau) \sigma_i \otimes \sigma_j \right].~~~~~~~~
\eea
Note that in the above equation the identity matrix is denoted by $\sigma_0$.
Now we define:
 \bea  && a_i(\tau)\equiv a_{i0}(\tau)~~~~~~~~~~~\forall~~i=1,2,3, \\
 && b_j(\tau)\equiv a_{0j}(\tau)~~~~~~~~~~~\forall~~j=1,2,3, \\
 &&a_i(\tau) b_j(\tau)\equiv a_{ij}(\tau)~~~~~\forall~~i,j=1,2,3.\eea 
 Using these definitions the density matrix $\rho_{\bf System}(\tau)$ for the reduced subsystem can be re-expressed as:
\be
\rho_{\bf System}(\tau)=\frac{1}{4}\left[\sigma_0 \otimes \sigma_0 +\sum_{i=1}^3 a_{0i}(\tau)(\sigma_0 \otimes \sigma_i)+\sum_{i=1}^3 a_{i0}(\tau)(\sigma_i \otimes \sigma_0)+\sum_{i,j=1}^3 a_{0i}(\tau)(\sigma_i \otimes \sigma_j)\right] 
\ee
In the new transformed basis i.e in terms of $\sigma_+$, $\sigma_-$ and $\sigma_3$ basis $\rho_{\bf System}(\tau)$ can be written as:
\be
\rho_{\bf System}(\tau)=\frac{1}{4}\left[\sigma_0 \otimes \sigma_0 +\sum_{m=+,-}^3 a_{0m}(\tau)(\sigma_0 \otimes \sigma_m)+\sum_{m=+,-}^3 a_{m0}(\tau)(\sigma_m \otimes \sigma_0)+\sum_{m,n=+,-}^3 a_{mn}(\tau)(\sigma_i \otimes \sigma_j)\right].~~~~~~~~~
\ee
In this new basis the reduced subsystem density matrix, in terms of the {\it Bloch vectors} can be explicitly written as:
\begin{eqnarray}
 \rho_{\bf System}(\tau) 
=\frac{1}{4}\begin{pmatrix}
1+a_{03}+a_{30}+a_{33} &  0 &  0 &  a_{++}\\
0 &  1-a_{03}+a_{30}-a_{33} &  a_{+-} &  0\\
0 &  a_{-+} &  1+a_{03}-a_{30}-a_{33} &  0\\
a_{--} & 0 &  0 &  1-a_{03}-a_{30}+a_{33}
\end{pmatrix},~~~~~
\end{eqnarray}
where the Hermiticity and ${\rm Tr}(\rho_{\bf System}(\tau) )=1$ 
 property of the density matrix has been used to find the matrix elements explicitly.

\subsection{\textcolor{blue}{ Large Scale time dependent solution}}
Solving the master equation in the new basis i.e. $\sigma_+$, $\sigma_-$ and $\sigma_3$ basis in the large time limit ($\tau=\infty$) we get the following solution for the components of the density matrix as given by:
\be
\begin{aligned}
\label{boun}
 a_{03}(\infty)=a_{30}(\infty)&=&-{\rm tanh(\pi k \omega)} \\
 a_{33}(\infty)&=&{\rm tanh^2(\pi k \omega)} \\
 a_{++}(\infty)=a_{--}(\infty)&=&0 \\
 a_{+-}(\infty)=a_{-+}(\infty)&=&0
\end{aligned}
\ee
where all other {\it Bloch vector} components are zero.

We can get the large time reduced density matrix from the above solution, which can be expressed as
\bea
\rho_{\bf System}(\infty)&&=\frac{1}{4}\left[\sigma_0 \otimes \sigma_0 + a_{03}(\infty)(\sigma_0 \otimes \sigma_3)+a_{30}(\infty)(\sigma_3 \otimes \sigma_0)+ a_{33}(\infty)(\sigma_3 \otimes \sigma_3)\right]\nonumber\\
&&=\frac{1}{4}\begin{pmatrix}
1+2a_{03}(\infty)+a_{33}(\infty) & & 0 & & 0 & & 0\\
0 & & 1-a_{33}(\infty) & & 0 & & 0 \\
0 & & 0 & & 1-a_{33}(\infty) & & 0 \\
0 & & 0 & & 0 & & 1-2a_{03}(\infty)+a_{33}(\infty)
\end{pmatrix} ~~~~~~~~
\eea
 The large time behaviour basically demonstrates the  equilibrium behaviour of the system. Hence the solution of the Bloch vectors obtained in the large time scale is applicable in any basis. These solutions can therefore be used as the boundary conditions for obtaining the finite time solution of the density matrix. Writing the density matrix in terms of the solutions of the Bloch vectors as,
\begin{eqnarray}\label{fk1}
 \rho_{\bf System}(\infty) 
=\frac{1}{4}\begin{pmatrix}
(1-{\rm tanh(\pi k \omega)})^2 & & 0 & & 0 & & 0\\
0 & & 1-{\rm tanh^2(\pi k \omega)} & & 0 & & 0 \\
0 & & 0 & & {\rm 1-tanh^2(\pi k \omega)} & & 0 \\
0 & & 0 & & 0 & & {\rm (1+tanh(\pi k \omega))^2}
\end{pmatrix} ~~~~~~~~
\end{eqnarray}
On the other hand, from quantum statistical mechanics one can compute the expression for the density matrix at finite temperature and large time limit, which is given by the following expression:
\bea\label{fk2}
\rho_{\bf System}(\infty)&&=\frac{e^{-\beta H_{\bf System}}}{{\rm Tr}(e^{-\beta H_{\bf System}})}\nonumber\\
&=&\frac{1}{4}\begin{pmatrix}
\left(1-{\rm tanh\left(\frac{\beta \omega}{2}\right)}\right)^2 & & 0 & & 0 & & 0\\
0 & & 1-{\rm tanh^2\left(\frac{\beta \omega}{2}\right)} & & 0 & & 0 \\
0 & & 0 & & {\rm 1-tanh^2\left(\frac{\beta \omega}{2}\right)} & & 0 \\
0 & & 0 & & 0 & & {\rm \left(1+tanh\left(\frac{\beta \omega}{2}\right)\right)^2}
\end{pmatrix}.~~~~~~~~
\eea
Comparing Eq~(\ref{fk1}) and  Eq~(\ref{fk2}), we obtain the following result:
\be
T=\frac{1}{\beta}=\frac{1}{2\pi k}=\frac{1}{2 \pi \sqrt{\alpha^2-r^2}}, ~~~{\rm where}~~\alpha=\sqrt{\frac{3}{\Lambda}}>0.
\ee
which physically represents the equilibrium temperature of the  thermal bath at large time scale. 

In ref.~\cite{Bhattacherjee:2019eml} we have explicitly shown that the temperature of the thermal bath can be computed in terms of the {\it Gibbons Hawking} temperature and {\it Unruh} temperature, which can be expressed as \cite{Choudhury:2017bou}: 
\bea T=\sqrt{T^2_{\bf GH}+T^2_{\bf Unruh}}=\frac{1}{2\pi \alpha}\sqrt{1+\frac{r^2}{\alpha^2-r^2}}, ~~~{\rm where}~~\alpha=\sqrt{\frac{3}{\Lambda}}>0.\eea
Here {\it Gibbons Hawking} and {\it Unruh} temperature is defined in the present context as:
\be T_{\bf GH}=\frac{1}{2\pi \alpha},~~
T_{\bf Unruh}=\frac{a}{2\pi}=T_{\bf GH}\frac{r}{\sqrt{\alpha^2-r^2}}, ~~{\rm where}~~~a=2\pi T_{\bf GH}\frac{r}{\sqrt{\alpha^2-r^2}}=\frac{1}{\alpha}\frac{r}{\sqrt{\alpha^2-r^2}}.\ee
Now, we know that in static patch of the de Sitter space\cite{Akhmedov:2013vka,Spradlin:2001pw} the curvature is determined by the following expression for the Ricci scalar:
\be R=\frac{12}{\alpha^2}>0,~~~{\rm where}~~\alpha=\sqrt{\frac{3}{\Lambda}}>0.\ee
Consequently, the equilibrium temperature of the thermal bath can be re-expressed in terms of the curvature of the static patch of the de Sitter space as:
\be
T=\frac{1}{\beta}=\frac{1}{2\pi k}=\frac{1}{2 \pi \sqrt{\left(\frac{12}{R}\right)-r^2}}=\frac{\sqrt{R}}{2\pi}\frac{1}{\sqrt{12-Rr^2}}.
\ee
Similarly, the {\it Gibbons Hawking} and {\it Unruh} temperature can be re-expressed in terms of the curvature of the static patch of the de Sitter space as:
\bea T_{\bf GH}&=&\frac{\sqrt{R}}{4\sqrt{3}\pi},~~~~~
T_{\bf Unruh}=\frac{a}{2\pi}=T_{\bf GH}\frac{\sqrt{R}r}{\sqrt{12-Rr^2}}.\eea
In this context, one can consider the following limiting situations:
\begin{enumerate}
\item \textcolor{red}{\bf \underline{Flat space limit:}}\\
Flat space limit is characterised by the following condition:
\be R\rightarrow0,\Longrightarrow \alpha\rightarrow\infty,\Longrightarrow T_{\bf GH},T_{\bf Unruh}\rightarrow 0,\Longrightarrow T\rightarrow 0,\Longrightarrow k\rightarrow\infty\Longrightarrow k>>L.\ee
Here $L$ represents the Euclidean distance between the two atoms. In this case, we will get back the result obtained in the Minkowski flat space inertial case where $k>>L$.
\item  \textcolor{red}{\bf \underline{Zero acceleration limit:}}\\
The zero acceleration limit is characterised by the following condition:
\be a\rightarrow0,\Longrightarrow r\rightarrow 0,\Longrightarrow T_{\bf GH}\neq 0,T_{\bf Unruh}\rightarrow 0,\Longrightarrow T\rightarrow T_{\bf GH},\Longrightarrow  k<<L.\ee
 In this case, we will get back the result obtained in the limiting case where $k<<L$.
\end{enumerate}

\subsection{\textcolor{blue}{ Arbitrary time dependent general solution}}

To obtain the finite time solution of the Bloch vector components, we use the $\sigma_+$, $\sigma_-$ and $\sigma_3$ basis. Substituting the components of the density matrix in the GSKL master equation, in this new basis, we obtain the evolution equations for the Bloch vector components.

\bea
\dot{a}_{03}(\tau)&=&\frac{1}{4}(A^{12}+A^{21})(a_{++}-a_{--})+\frac{1}{2}(\tilde{A}^{21}-\tilde{A}^{12})(a_{++}-a_{--})+\frac{i}{2}(\tilde{B}^{12}+\tilde{B}^21)(a_{+-}+a_{-+})+4i \tilde{B}^{22}\nonumber
\\
\dot{a}_{03}(\tau)&=&\frac{1}{4}(A^{12}+A^{21})(a_{++}-a_{--})+\frac{1}{2}(\tilde{A}^{21}+\tilde{A}^{12})(a_{++}-a_{--})+\frac{i}{2}(\tilde{B}^{12}+\tilde{B}^{21})(a_{+-}+a_{-+})+4i \tilde{B}^{11}\nonumber
\\
\dot{a}_{++}(\tau)&=&(A^{12}+A^{21})(a_{03}+a_{30})+i a_{++}(B^{11}+B^{22})+2\omega a_{++}+2 \tilde{A}^{22}a_{+-}+2 \tilde{A}^{11}a_{-+}\nonumber\\
&&~~~~~~~~~~~~~~~~~~~~~~~~~~+2\tilde{A}^{21}(a_{03}-a_{30}-2a_{33})+2\tilde{A}^{12}(-a_{03}+a_{30}-2a_{33})
\nonumber\\
\dot{a}_{+-}(\tau)&=&i (-B^{12}+B^{21})(a_{30}-a_{03})+i a_{12}(B^{11}-B^{22})+2i \tilde{B}^{21}(a_{03}-a_{30}+2a_{33})\nonumber\\
&&~~~~~~~~~~~~~~~~~~~~~~~~~~-2i \tilde{B}^{12}(a_{03}-a_{30}+2a_{33})+2\tilde{A}^{11}a_{--}+2\tilde{A}^{22}a_{++}
\nonumber\\
\dot{a}_{-+}(\tau)&=&i (B^{12}-B^{21})(a_{03}-a_{30})+i a_{21}(-B^{11}+B^{22})-2i \tilde{B}^{21}(a_{03}+a_{30}+2a_{33})\nonumber\\
&&~~~~~~~~~~~~~~~~~~~~~~~~~~-2i \tilde{B}^{12}(-a_{03}-a_{30}+2a_{33})+2\tilde{A}^{11}a_{++}+2\tilde{A}^{22}a_{--}
\nonumber\\
\dot{a}_{--}(\tau)&=&(A^{12}+A^{21})(-a_{03}-a_{30})+i a_{--}(B^{11}+B^{22})+2\omega a_{--}+2 \tilde{A}^{22}a_{-+}+2 \tilde{A}^{11}a_{+-}\nonumber\\
&&~~~~~~~~~~~~~~~~~~~~~~~~~~+2\tilde{A}^{21}(a_{03}-a_{30}-2a_{33})+2\tilde{A}^{12}(-a_{03}+a_{30}-2a_{33})
\nonumber\\
\dot{a}_{33}(\tau)&=&-(\tilde{A}^{12}+\tilde{A}^{21})a_{++}-(\tilde{A}^{21}+\tilde{A}^{12})a_{--}+4i \tilde{B}^{11}a_{03}+4i \tilde{B}^{22}a_{30}
\eea

We try to solve the evolution equations analytically in the limit $2 \pi k \omega\gg 1$,along with an additional condition on the natural frequency of the two identical atoms i.e restricting $\omega_o$ to take values such that $\coth(\pi k \omega_o)$=0. The first condition ensures that the factor $\left(1-e^{-2 \pi k \omega}\right)^{-1}$ appearing in $\tilde{A}^{\alpha \beta}$ and $\tilde{B}^{\alpha \beta}$ term of the GSKL matrix $C_{ij}^{\alpha \beta}$ reduces to unity. Substituting the components $H_{ij}^{\alpha\beta}$ and $C_{ij}^{\alpha\beta}$ (calculated in {\bf Appendix}~\ref{gskl} and \ref{effham} ) in the evolution equations we obtain the following simplified equations .

\be
\label{evo}
\bal
\dot{a}_{03}(\tau)&= 4B_{1}+A_{1}a_{++}+B_2a_{+-}+B_2a_{-+}-A_1a_{--} \\
\dot{a}_{30}(\tau)&= 4B_1+A_1a_{++}+B_2a_{+-}+B_2a_{-+}-A_1a_{--} \\
\dot{a}_{++}(\tau)&= 4A_1a_{03}+4A_1a_{30}+2\omega a_{++} \\
\dot{a}_{+-}(\tau)&= D_2a_{03}-D_2a_{30}-4B_2a_{30} \\
\dot{a}_{-+}(\tau)&= D_2a_{03}-D_2a_{30}-4B_2a_{03}-4B_2a_{30} \\
\dot{a}_{--}(\tau)&= -4A_1a_{03}-4A_1a_{30}+2\omega a_{--} \\
\dot{a}_{33}(\tau)&= 4B_1 a_{03}+ 4B_2 a_{30}
\eal
\ee
where in the above sets of equations the following notations have been used, which are also calculated in {\bf Appendix}~\ref{effham} :

\be
\begin{aligned}
A_1(\omega) =& \frac{1}{4}(A^{12}+A^{21}) = -\frac{i \mu^2}{4L\sqrt{1+(\frac{L}{2k})^2}}~ \cosh\left((2n+1) ~\rm \sinh^{-1}\left(\frac{L}{2k}\right)\right) \\
B_1(\omega) =& i \tilde{B}^{11}=i \tilde{B^{22}} = -\frac{\mu^2}{4\pi k}\left(n+\frac{1}{2}\right) \\
B_2(\omega) =& i \tilde{B}^{12}=i \tilde{B^{21}} = - \frac{ \mu^2}{4L\sqrt{1+\left(\frac{L}{2k}\right)^2}}~ \sinh\left((2n+1) ~\rm \sinh^{-1}\left(\frac{L}{2k}\right)\right) \\
D_2(\omega) =& i(B^{12}-B^{21}) = 0
\end{aligned}
\ee

Solving the equations \eqref{evo} we obtain the finite time dependent solution of the Bloch vector components:
\bea
\label{ist}
a_{03}(\tau)&=&C_{1}e^{f_1(\omega) \tau} \frac{f_1(\omega)}{4(B_1+B_2)}+C_2e^{f_2(\omega) \tau} \frac{f_2(\omega)}{4(B_1+B_2)}+C_{3}e^{f_3(\omega) \tau} \frac{f_3(\omega)}{4(B_1+B_2)}
\\
\label{2nd}
a_{30}(\tau)&=&C_{1}e^{f_1(\omega) \tau} \frac{f_1(\omega)}{4(B_1+B_2)}+C_2e^{f_2(\omega) \tau} \frac{f_2(\omega)}{4(B_1+B_2)}+C_{3}e^{f_3(\omega) \tau} \frac{f_3(\omega)}{4(B_1+B_2)}
\\
a_{++}(\tau)&=&C_{4}e^{2\omega \tau}+C_{1}e^{f_1(\omega) \tau}\left(\frac{-2 f_1(\omega) A_1} {(-2\omega +f_1)(B_1+B_2)} +\frac{(f_1^2(\omega)+12 B_2^2}{4A_1(B_1+B_2)}\right)\nonumber\\
&&~~~~~~~~~~~~~~~+e^{f_2(\omega) \tau}\left(\frac{-2 f_2(\omega) A_1} {(-2\omega +f_2)(B_1+B_2)} +\frac{(f_2^2(\omega)+12 B_2^2}{4A_1(B_1+B_2)}\right)\nonumber\\
&&~~~~~~~~~~~~~~~~~~~~+e^{f_3(\omega) \tau}\left(\frac{-2 f_3(\omega) A_1} {(-2\omega +f_3)(B_1+B_2)} +\frac{(f_3^2(\omega)+12 B_2^2)}{4A_1(B_1+B_2)}\right)
\\
a_{+-}(\tau)&=& -C_5-\left(\frac{B_2}{B_1+B_2}\right)(C_1 e^{f_1(\omega)\tau}+C_2 e^{f_2(\omega)\tau}+C_3 e^{f_3(\omega)\tau})
\\
a_{-+}(\tau)&=& C_5-\left(\frac{2B_2}{B_1+B_2}\right)(C_1 e^{f_1(\omega)\tau}+C_2 e^{f_2(\omega)\tau}+C_3 e^{f_3(\omega)\tau})
\\
a_{--}(\tau)&=&C_4e^{2 \omega \tau}-\frac{2A_1}{B_1+B_2}\left(\frac{C_1 f_1(\omega)e^{f_1(\omega)\tau}}{-2\omega+f_1}+\frac{C_2 f_2(\omega)e^{f_2(\omega)\tau}}{-2\omega+f_2}+\frac{C_3 f_3(\omega)e^{f_3(\omega)\tau}}{-2\omega+f_3}\right)
\\
a_{33}(\tau)&=&\left(C_6+C_1e^{f_1(\omega)\tau}+C_2e^{f_2(\omega)\tau}+C_3e^{f_3(\omega)\tau}\right)
\eea
where $\ C_i~~ \forall i=1,6$ are arbitrary constants.

In the above sets of equations $f_1(\omega)$, $f_2(\omega)$, $f_3(\omega)$ can explicitly be written as: 
\be
\bal
f_1(\omega)&= -2\omega (\Delta_1(\omega))^2+(\Delta_1(\omega))^3-24\omega B_2^2+\Delta_1(\omega)(-16A_1^2(\omega)+12B_2^2(\omega)) \\
f_2(\omega)&= -2\omega (\Delta_2(\omega))^2+(\Delta_2(\omega))^3-24\omega B_2^2+\Delta_2(\omega)(-16A_1^2(\omega)+12B_2^2(\omega)) \\
f_3(\omega)&= -2\omega (\Delta_3(\omega))^2+(\Delta_1(\omega))^3-24\omega B_2^2+\Delta_3(\omega)(-16A_1^2(\omega)+12B_2^2(\omega))
\eal
\ee

The functions $\Delta_1(\omega)$,$\Delta_2(\omega)$ and $\Delta_3(\omega)$ appearing in the $f_1(\omega)$,$f_2(\omega)$ and $f_3(\omega)$ are explicitly written in the following equations.
\be
\bal
\Delta_1(\omega)&= \frac{2\omega}{3}-2^{1/3} \frac{b_1}{3} {\cal Z}(b_1,b_2)+\frac{1}{3 \times 2^{1/3}} {\cal Z}(b_1,b_2) \\
\Delta_2(\omega)&= \frac{2\omega}{3}+\frac{((1+i \sqrt{3}b_1))}{3\times 2^{2/3} {\cal Z}(b_1,b_2)}-\frac{1}{6\times 2^{1/3}}(1-i \sqrt{3}) {\cal Z}(b_1,b_2) \\
\Delta_3(\omega)&= \frac{2\omega}{3}+\frac{((1-i \sqrt{3}b_1))}{3\times 2^{2/3} {\cal Z}(b_1,b_2)}-\frac{1}{6\times 2^{1/3}}(1+i \sqrt{3}) {\cal Z}(b_1,b_2) 
\eal
\ee
where we have denoted the following symbols :~~~~~
$$b_1=-48 A_1^2+36B_2^2-4\omega^2
~~~~~~~~~~ \ \ \ \ \ \ \
b_2=288A_1^2\omega +432 B_2^2 \omega +16\omega^3$$
and we define again a new function, ${\cal Z} (b_1,b_2) = \left(b_2+\sqrt{4b_1^3+b_2^2}\right)^{1/3}$. The above equation shows that the function $\Delta_1(\omega)$ is real,  whereas  $\Delta_2(\omega)$ and  $\Delta_3(\omega)$ are complex. The arbitrary constants are determined using the equilibrium behaviour as the boundary conditions which are already mentioned in equation \ref{boun}. From the appearance it might seem that the function $e^{f_l(\omega)}$ diverges as $\tau$ tends to infinity but it can be explicitly shown that the function $f_{l}(\omega)(l=1,2,3) <0$, taking into account the leading order term in $\Delta$, which is a decaying function. Hence the function, $e^{f_l(\omega) \tau}$ tends to a finite value, which we denote by $f_l(\omega)\tau'$.

 The physically acceptable solutions to the simplified evolution equations obeying the boundary conditions are given by the following expressions:
\bea
a_{03}(\tau)&=&-\left[ g_1(\omega)e^{-|f_1(\omega)|(\tau-\tau')} \frac{|f_1(\omega)|}{4(B_1+B_2)}+g_2(\omega)e^{-|f_2(\omega)|(\tau-\tau')} \frac{|f_2(\omega)|}{4(B_1+B_2)}\right.\nonumber\\ &&\left.~~~~~~~~~~~~~~~~~~~~~~~~~~~~~~~~~~~~~~~~+g_{3}(\omega)e^{-|f_3(\omega)|(\tau-\tau')} \frac{|f_3(\omega)|}{4(B_1+B_2)}\right]
\\
a_{30}(\tau)&=& -\left[g_{1}(\omega)e^{-|f_1(\omega)| (\tau-\tau')} \frac{|f_1(\omega)|}{4(B_1+B_2)}+g_2(\omega)e^{-|f_2(\omega)| (\tau-\tau')} \frac{|f_2(\omega)|}{4(B_1+B_2)}\right.\nonumber\\
&&\left.~~~~~~~~~~~~~~~~~~~~~~~~~~~~~~~~~~~~~~~+g_{3}(\omega)e^{-|f_3(\omega)|(\tau-\tau')} \frac{|f_3(\omega)|}{4(B_1+B_2)}\right]
\\
a_{++}(\tau)&=&g_{1}(\omega)e^{-|f_1(\omega)|(\tau-\tau')}\left(\frac{2|f_1(\omega)| A_1} {-(2\omega +|f_1|)(B_1+B_2)} +\frac{f_1^2(\omega)+12 B_2^2}{4A_1(B_1+B_2)}\right)\nonumber\\
&&~~~~+g_2(\omega) e^{-|f_2(\omega)| (\tau-\tau')}\left(\frac{2|f_2(\omega)| A_1} {-(2\omega +|f_2|)(B_1+B_2)} +\frac{f_2^2(\omega)+12 B_2^2}{4A_1(B_1+B_2)}\right)\nonumber
\\
&&~~~~~~~+g_3(\omega) e^{-|f_3(\omega)| (\tau-\tau')}\left(\frac{2 |f_3(\omega)| A_1} {-(2\omega +|f_3|)(B_1+B_2)} +\frac{f_3^2(\omega)+12 B_2^2}{4A_1(B_1+B_2)}\right)
\eea
\bea
a_{+-}(\tau)&=& -g_5(\omega)-\left(\frac{B_2}{B_1+B_2}\right)\left(g_1(\omega) e^{-|f_1(\omega)|(\tau-\tau')}+g_2(\omega) e^{-|f_2(\omega)|(\tau-\tau')}\right.\nonumber\\
&&\left.~~~~~~~~~~~~~~~~~~~~~~~~~~~~~~~~~~~~~~~~~~~~~~~~~+g_3(\omega) e^{-|f_3(\omega)|(\tau-\tau')}|\right)
\\ 
a_{-+}(\tau)&=& g_5(\omega)-\left(\frac{2B_2}{B_1+B_2}\right)\left(g_1(\omega) e^{-|f_1(\omega)|(\tau-\tau')}+g_2(\omega) e^{-|f_2(\omega)|(\tau-\tau')}\right.\nonumber\\
&&\left.~~~~~~~~~~~~~~~~~~~~~~~~~~~~~~~~~~~~~~~~~~~~~~~~~~~~~+g_3(\omega) e^{-|f_3(\omega)|(\tau-\tau')}\right)
\\
a_{--}(\tau)&=&-\frac{2A_1}{B_1+B_2}\left[\frac{g_1(\omega) |f_1(\omega)|e^{-|f_1(\omega)|(\tau-\tau')}}{2\omega+|f_1|}+\frac{g_2(\omega) |f_2(\omega)|e^{-|f_2(\omega)|(\tau-\tau')}}{2\omega+|f_2|}\right.\nonumber\\
&&\left.~~~~~~~~~~~~~~~~~~~~~~~~~~~~~~~~~~~~~~~~~~~~~~~~+\frac{g_3(\omega) |f_3(\omega)|e^{-|f_3(\omega)|(\tau-\tau')}}{2\omega+|f_3|}\right] \\
a_{33}(\tau)&=&g_6(\omega)+g_1(\omega)e^{-|f_1(\omega)|(\tau-\tau')}+g_2(\omega)e^{-|f_2(\omega)|(\tau-\tau')}+g_3(\omega)e^{-|f_3(\omega)|(\tau-\tau')}
\eea
where the explicit functional forms of  $g_i(\omega)~\forall i=1,\cdots, 6$ are given below and the constant $C_4$ is zero which is consistent with the given boundary condition.

\begin{figure}[htb]
\centering
\subfigure[Bloch vector component $a_{03}$ vs Time profile.]{
	\includegraphics[width=7.8cm,height=4cm] {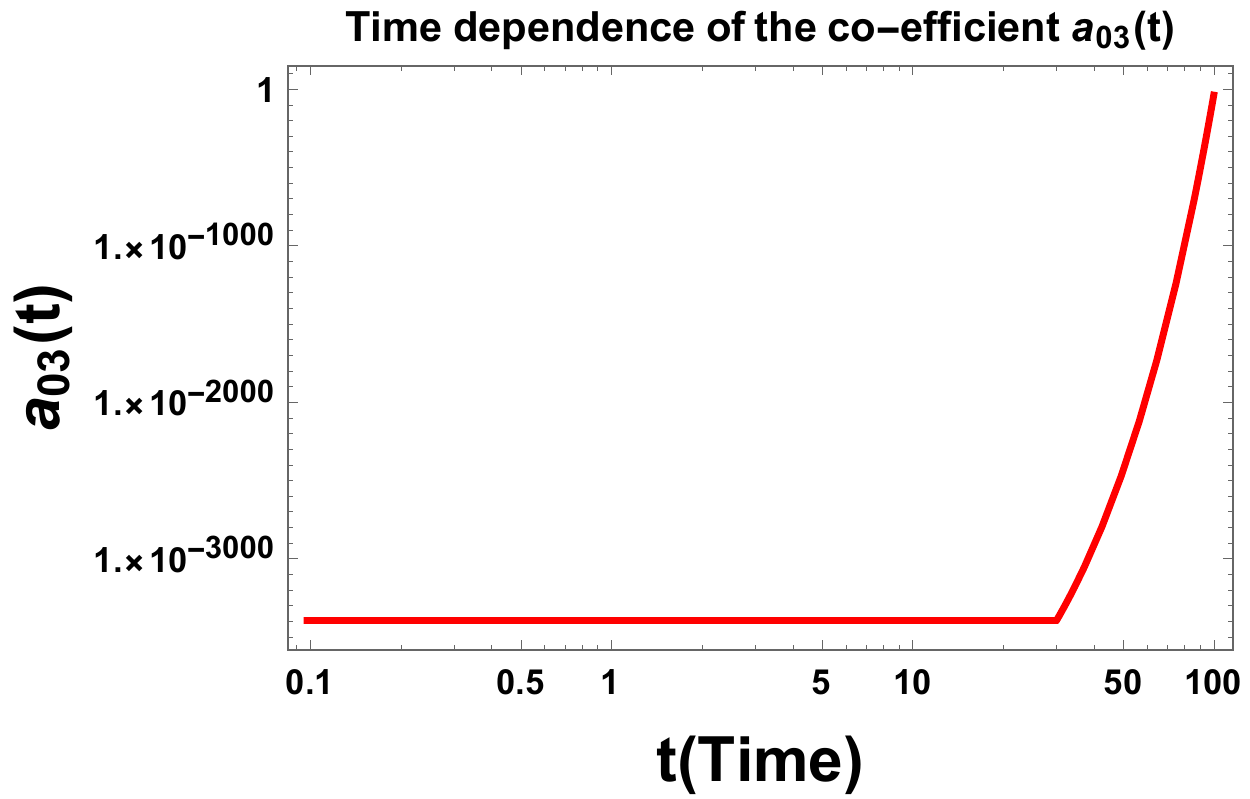}
	\label{1a}
	}
\subfigure[Bloch vector component $a_{33}$ vs Time profile.]{
	\includegraphics[width=7.8cm,height=4cm] {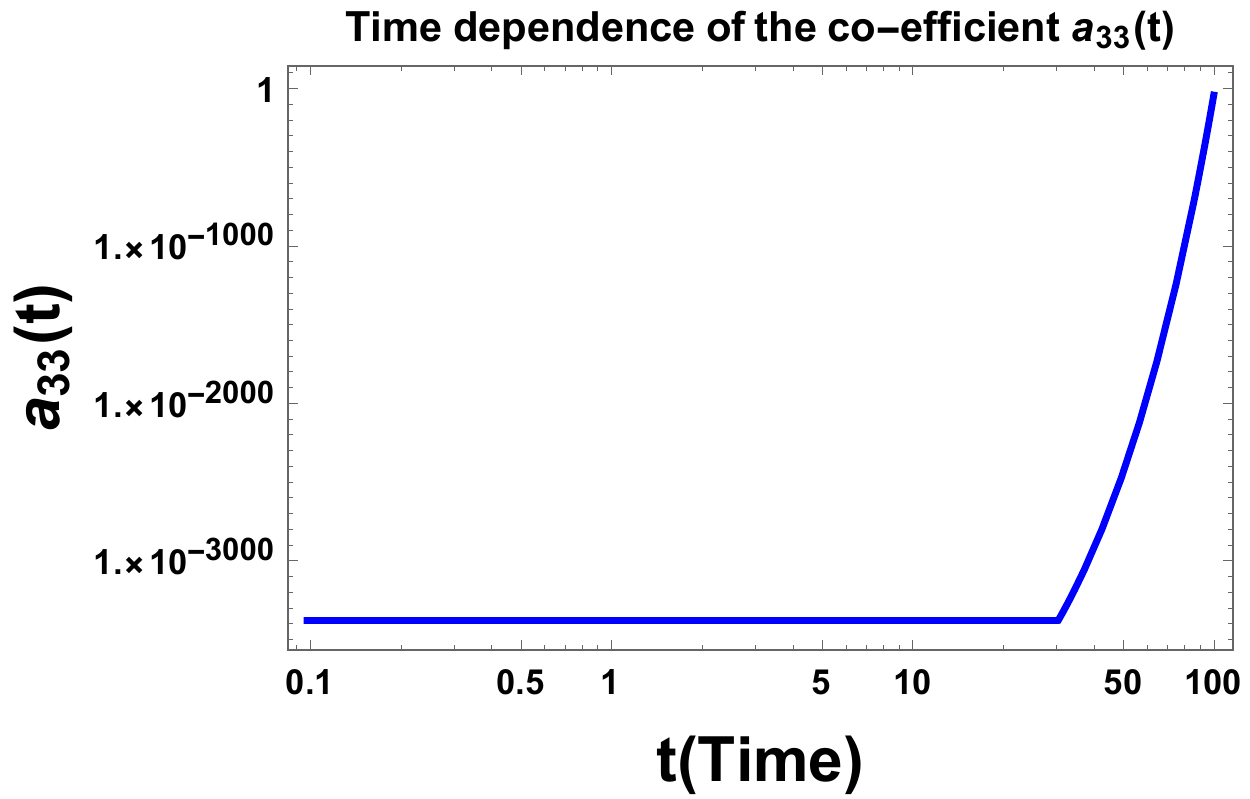}
	\label{1b}
	}
	\subfigure[Bloch vector component $a_{++}$ vs Time profile.]{
	\includegraphics[width=7.8cm,height=4cm] {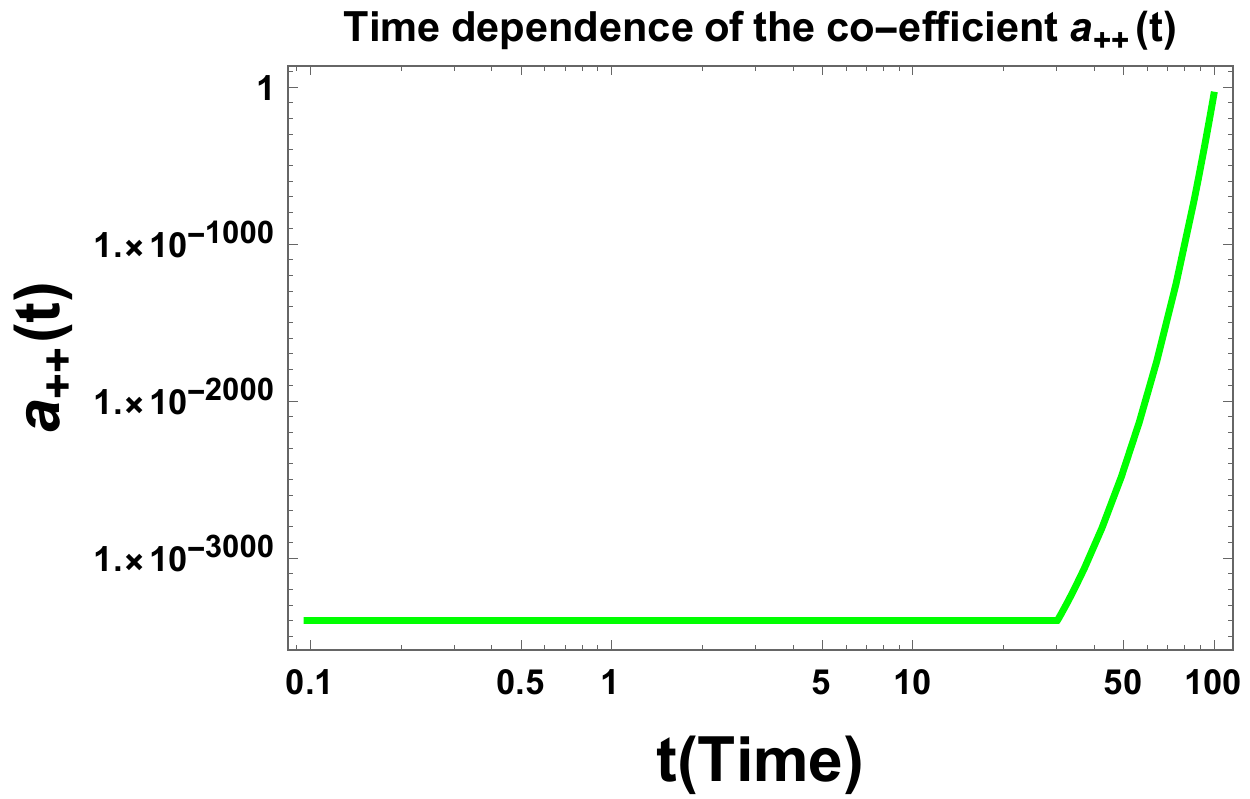}
	\label{1c}
	}
	\subfigure[Bloch vector component $a_{--}$ vs Time profile.]{
	\includegraphics[width=7.8cm,height=4cm] {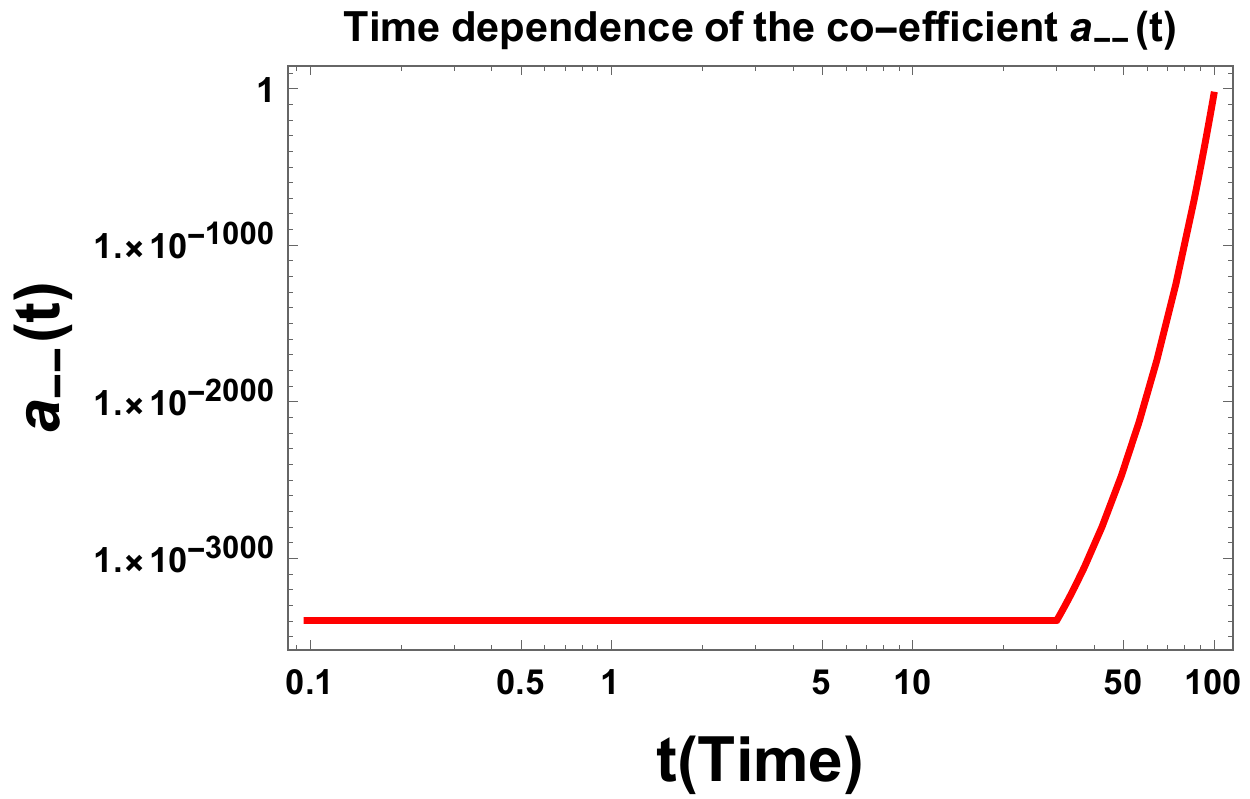}
	\label{1d}
	}
	\subfigure[Bloch vector component $a_{+-}$ vs Time profile.]{
	\includegraphics[width=7.8cm,height=4cm] {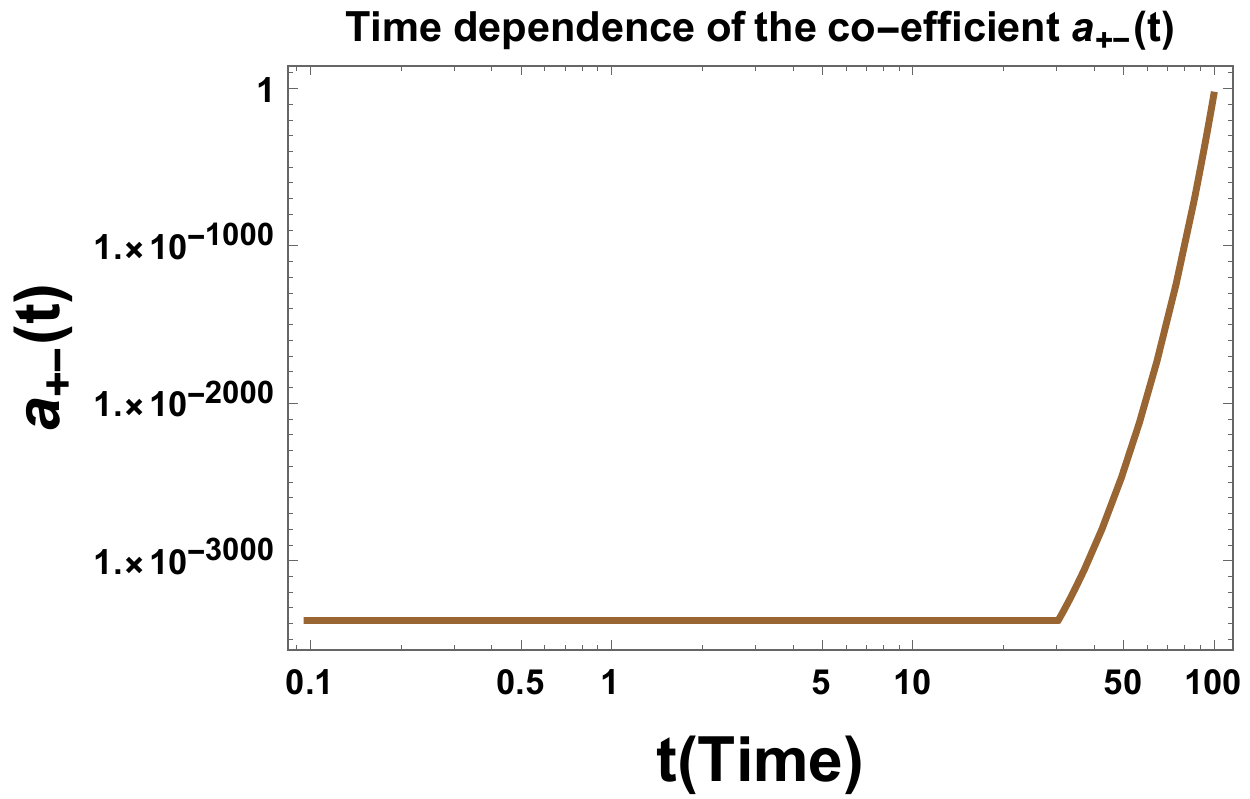}
	\label{1e}
	}
	\subfigure[Bloch vector component $a_{-+}$ vs Time profile.]{
	\includegraphics[width=7.8cm,height=4cm] {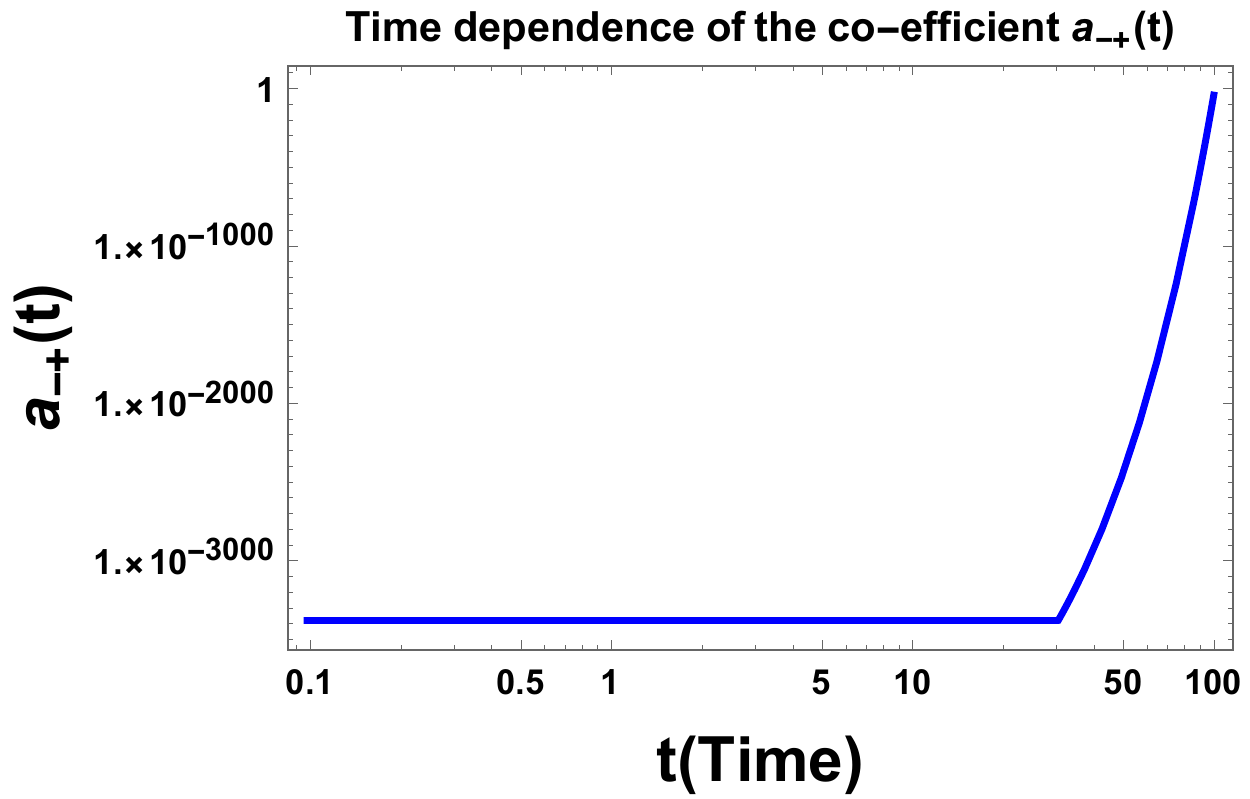}
	\label{1f}
	}
	\caption{Dependence of the Bloch vector components on Time is shown here }
	\label{gd1}
\end{figure}

\begin{figure}[htb]
\centering
\subfigure[Bloch vector component $a_{03}$ vs Frequency profile.]{
	\includegraphics[width=7.8cm,height=4cm] {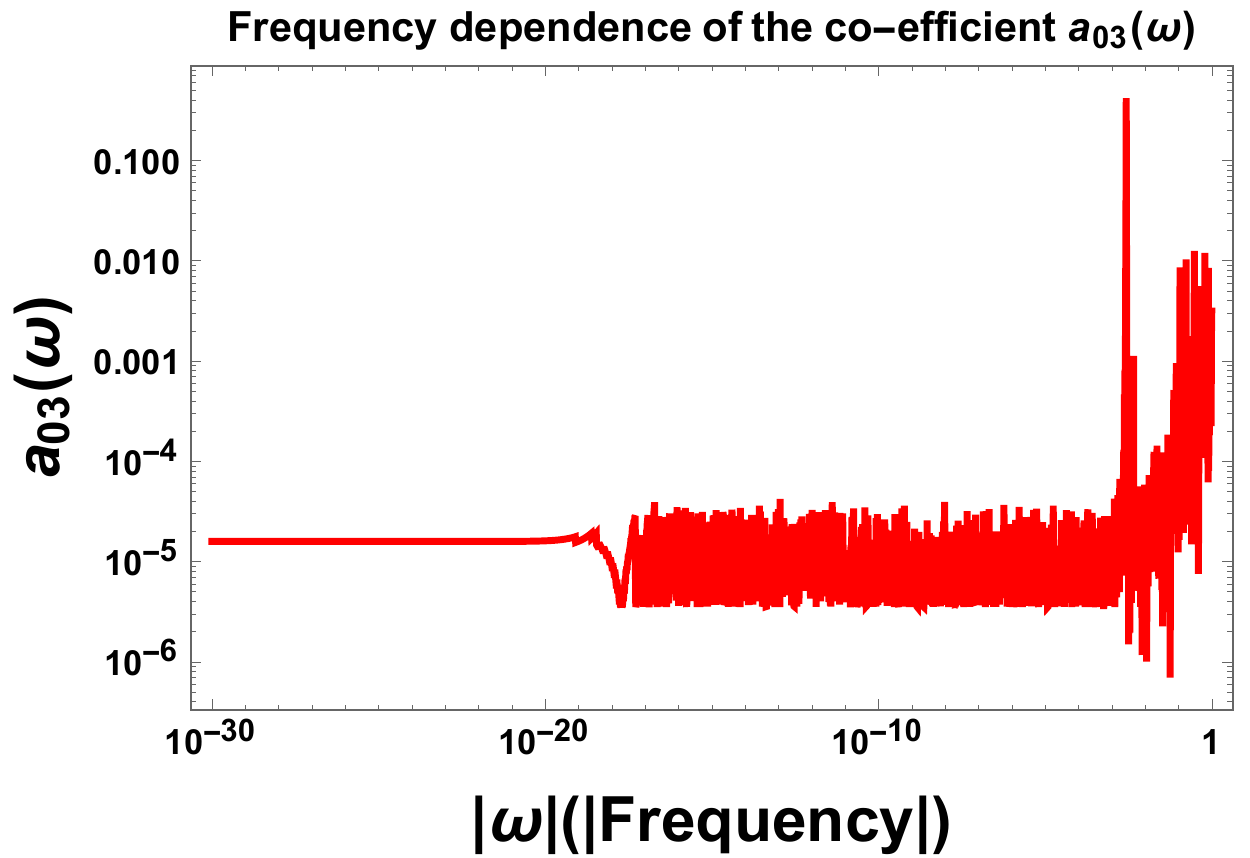}
	\label{2a}
	}
\subfigure[Bloch vector component $a_{33}$ vs Frequency profile.]{
	\includegraphics[width=7.8cm,height=4cm] {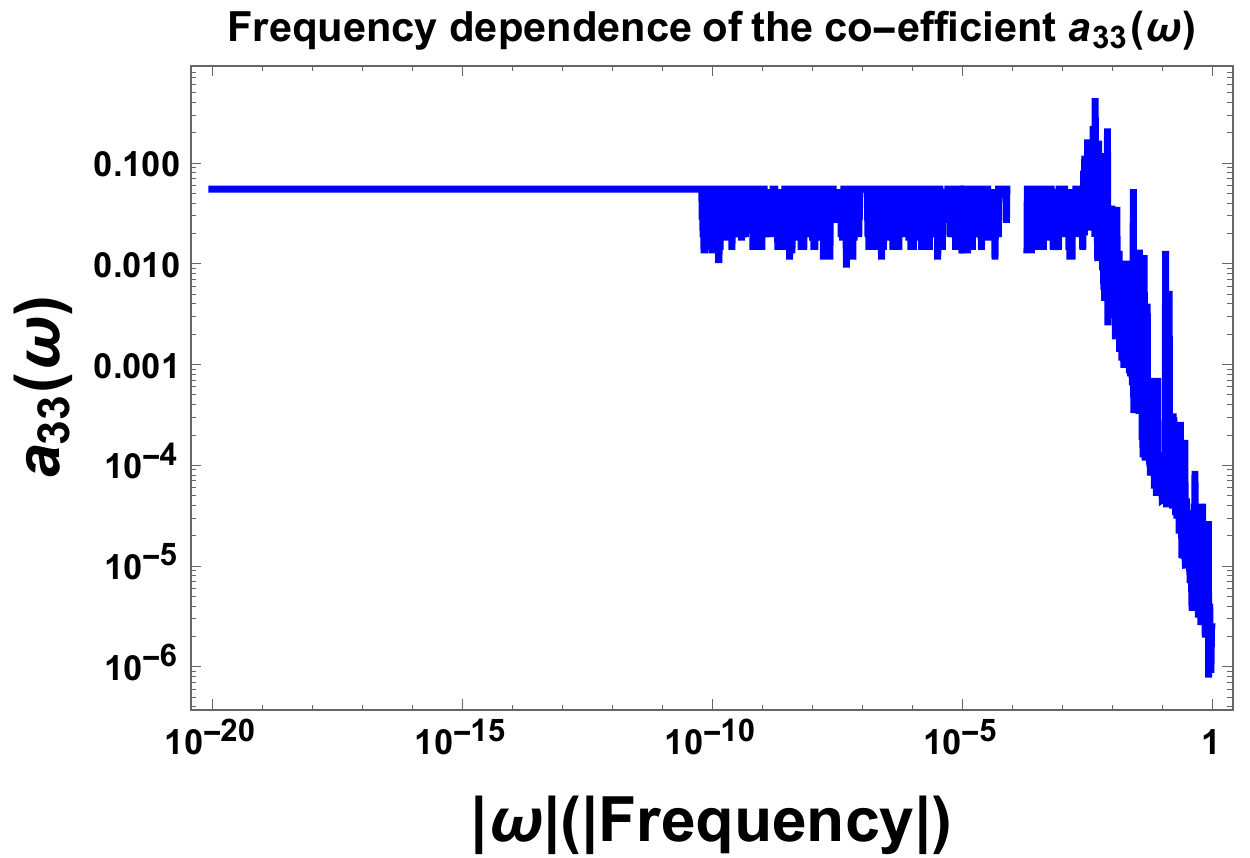}
	\label{2b}
	}
	\subfigure[Bloch vector component $a_{++}$ vs Frequency profile.]{
	\includegraphics[width=7.8cm,height=4cm] {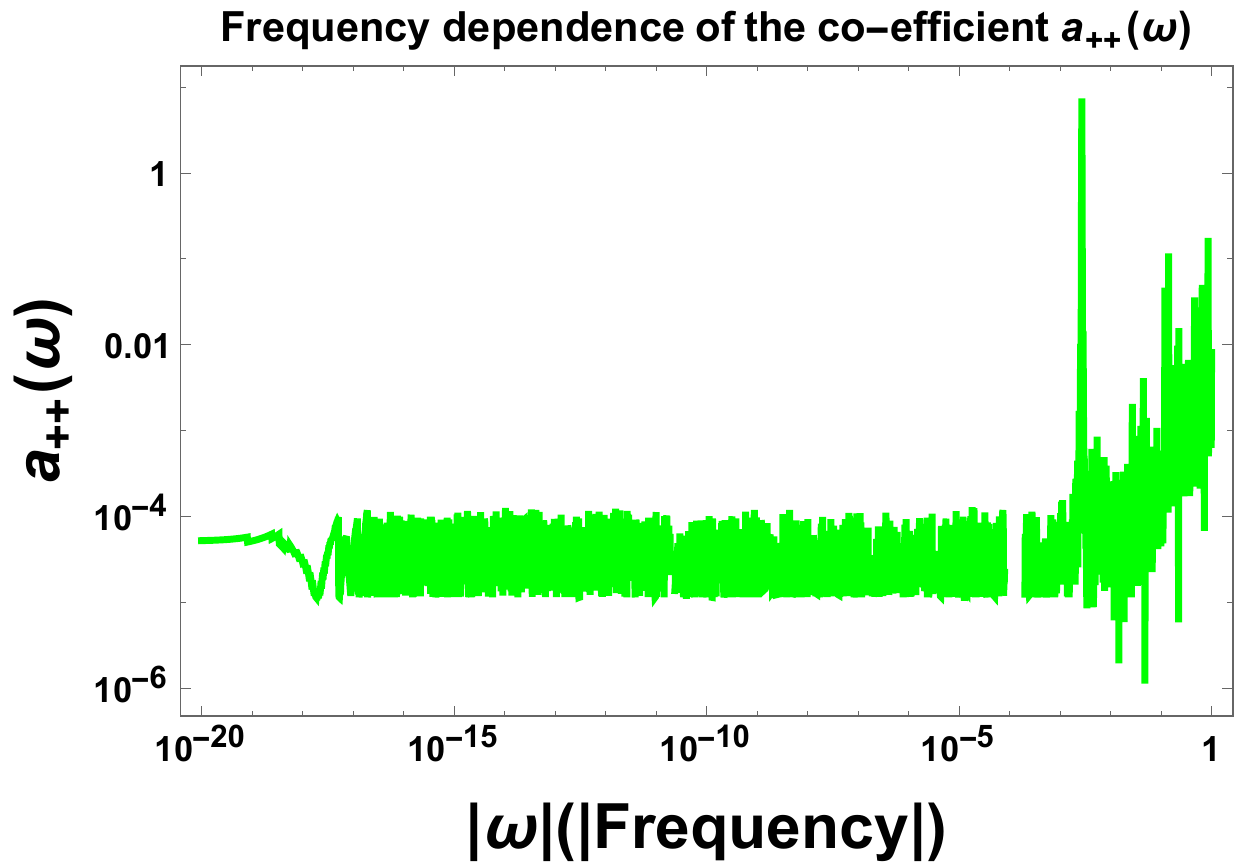}
	\label{2c}
	}
	\subfigure[Bloch vector component $a_{--}$ vs Frequency profile.]{
	\includegraphics[width=7.8cm,height=4cm] {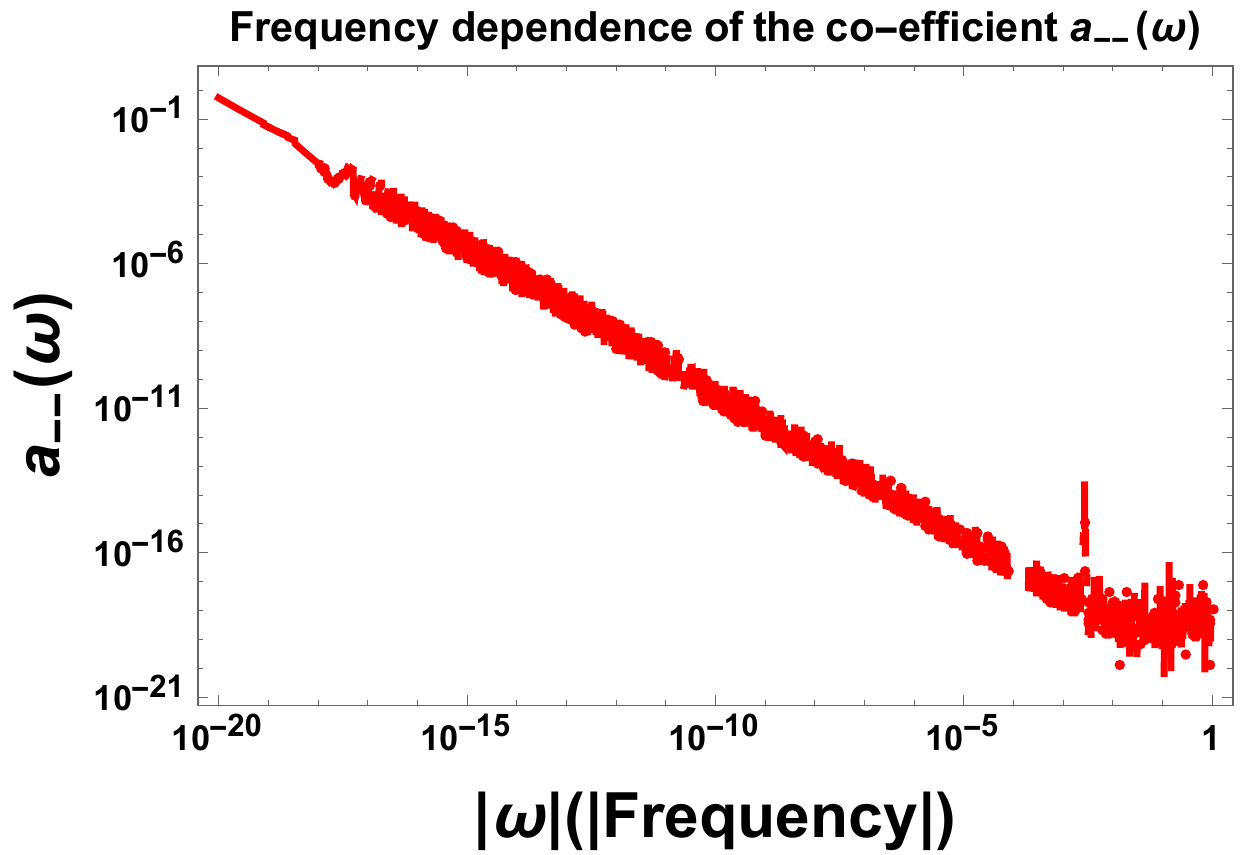}
	\label{2d}
	}
	\subfigure[Bloch vector component $a_{+-}$ vs Frequency profile.]{
	\includegraphics[width=7.8cm,height=4cm]{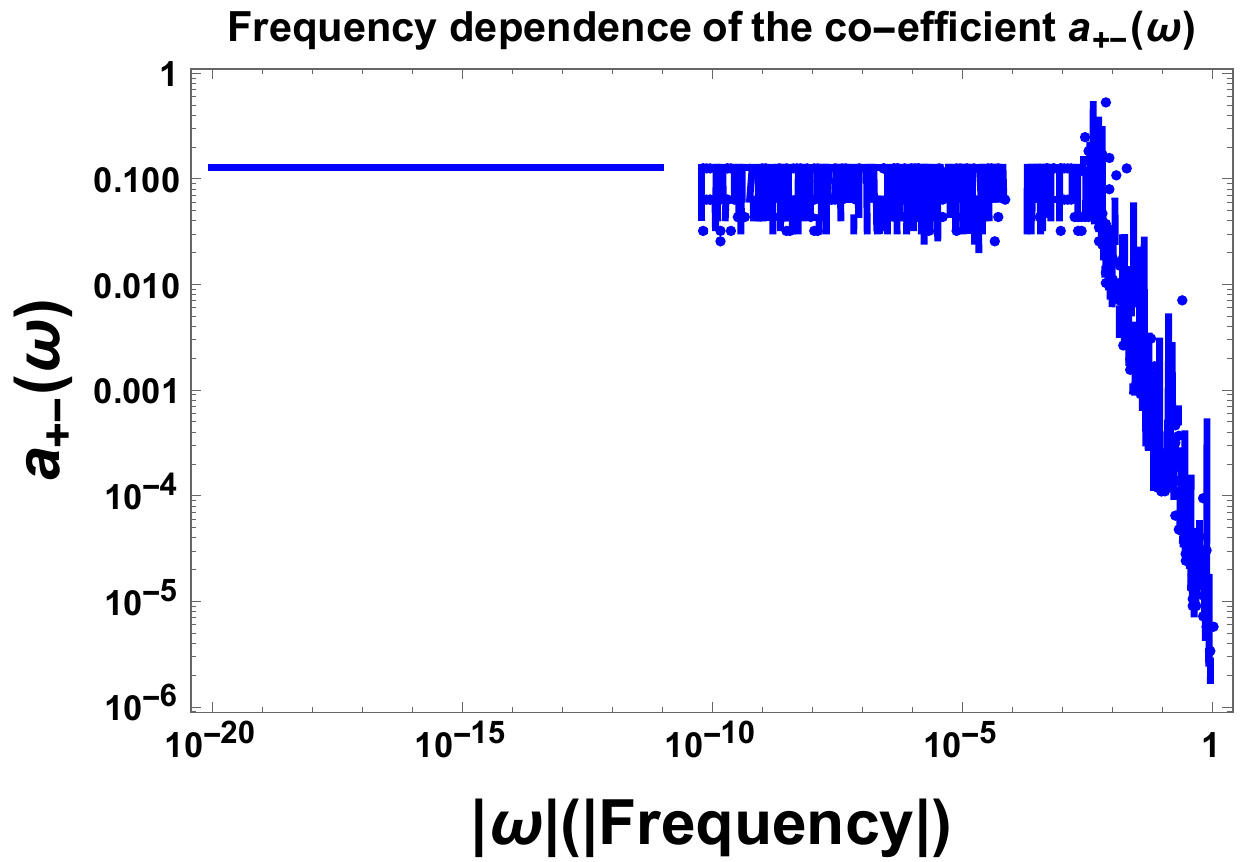}
	\label{2e}
	}
	\subfigure[Bloch vector component $a_{-+}$ vs Frequency profile.]{
	\includegraphics[width=7.8cm,height=4cm] {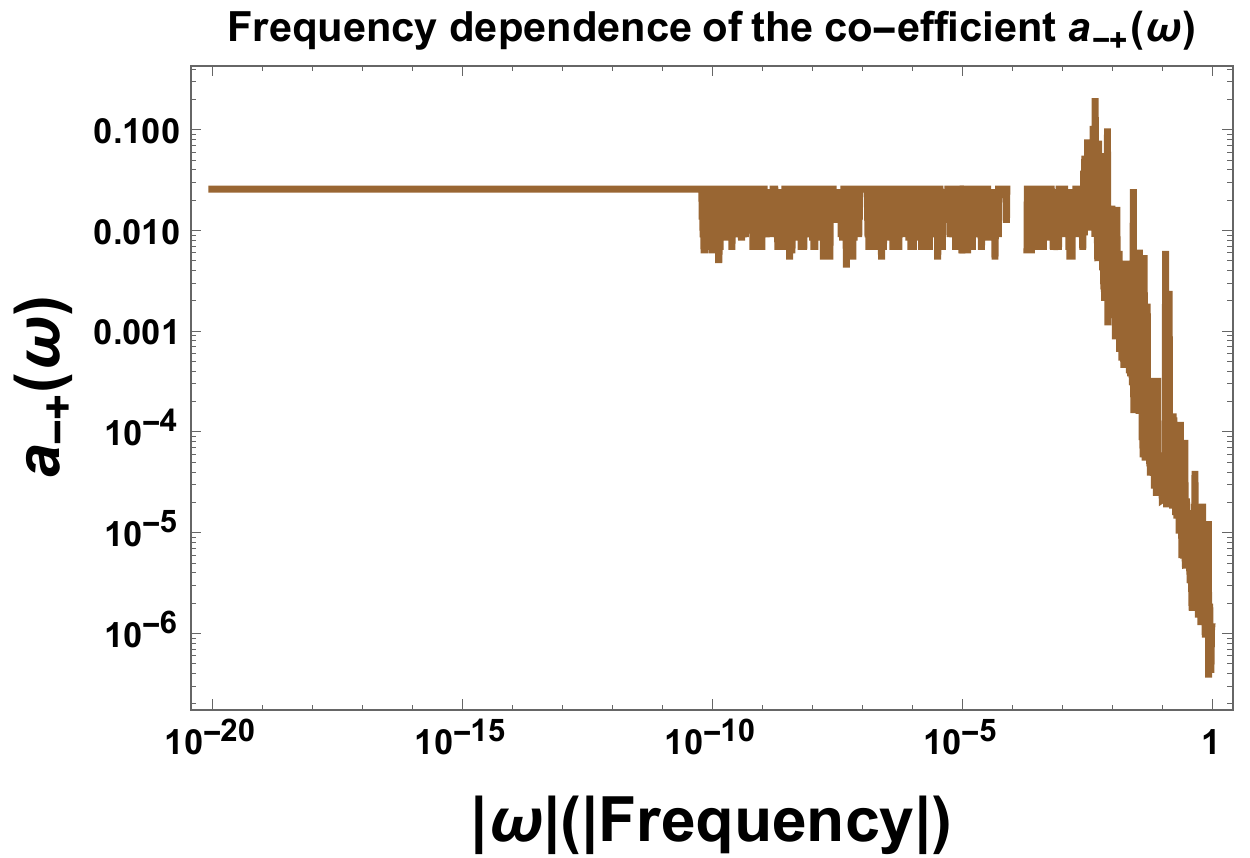}
	\label{2f}
	}
	\caption{Dependence of the Bloch vector components on frequency is shown here.}
	\label{gd2}
\end{figure}

\begin{figure}[htb]
\centering
\subfigure[Bloch vector component $a_{03}$ vs Euclidean Distance profile.]{
	\includegraphics[width=7.8cm,height=4cm] {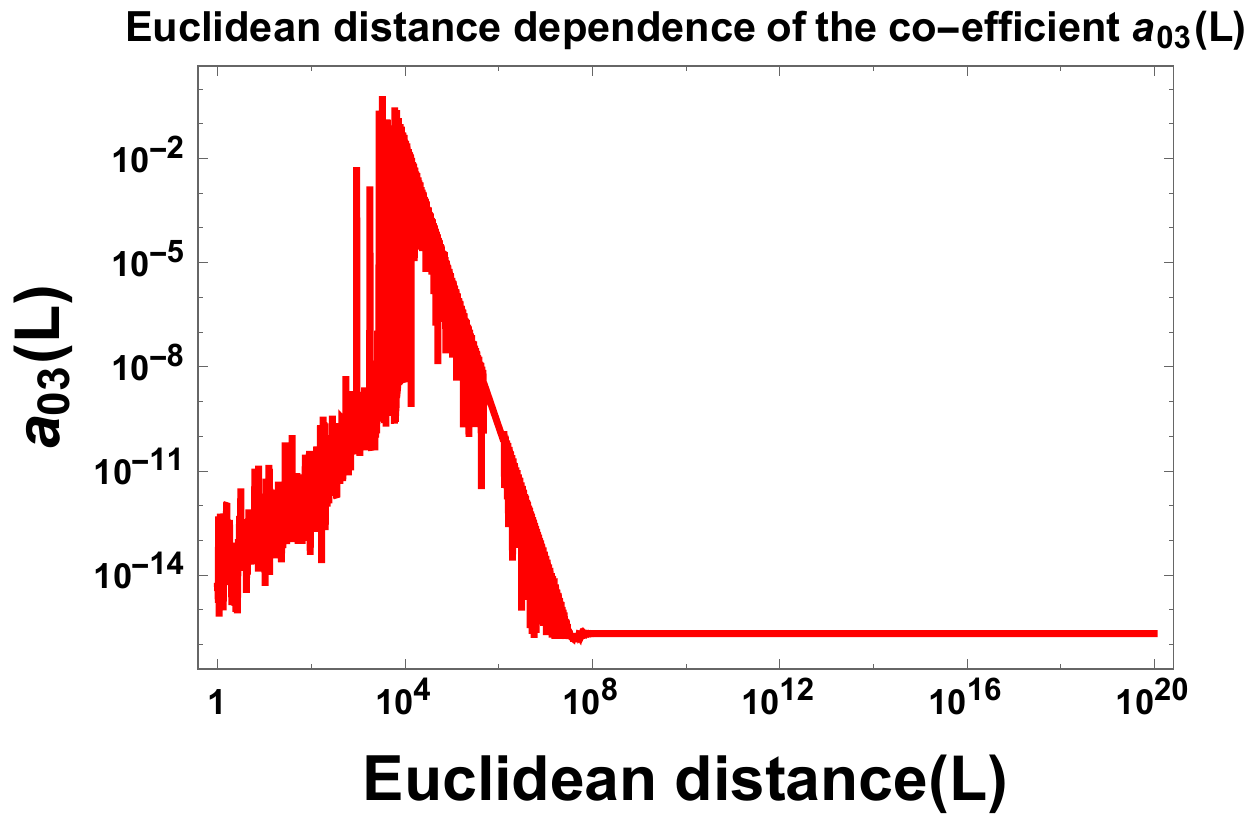}
	\label{3a}
	}
\subfigure[Bloch vector component $a_{33}$ vs Euclidean Distance profile.]{
	\includegraphics[width=7.8cm,height=4cm] {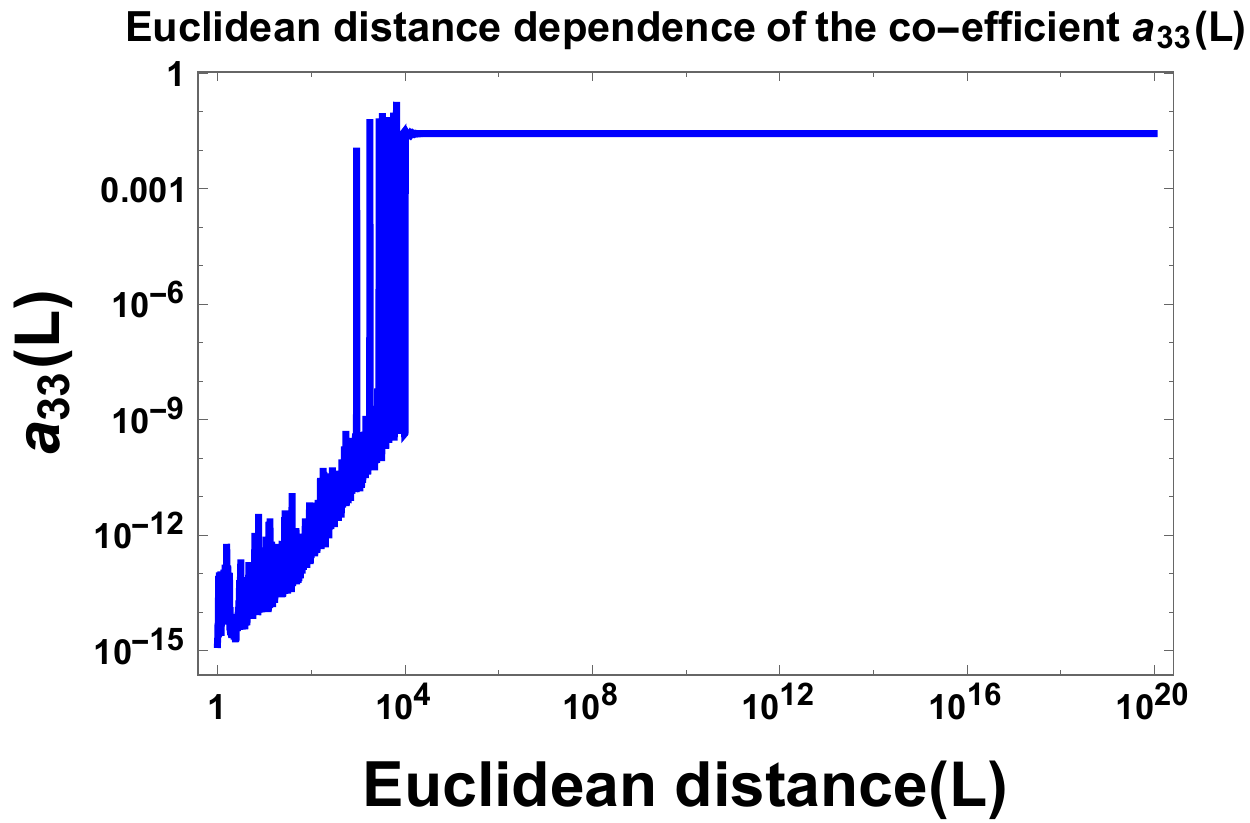}
	\label{3b}
	}
	\subfigure[Bloch vector component $a_{++}$ vs Euclidean Distance profile.]{
	\includegraphics[width=7.8cm,height=4cm] {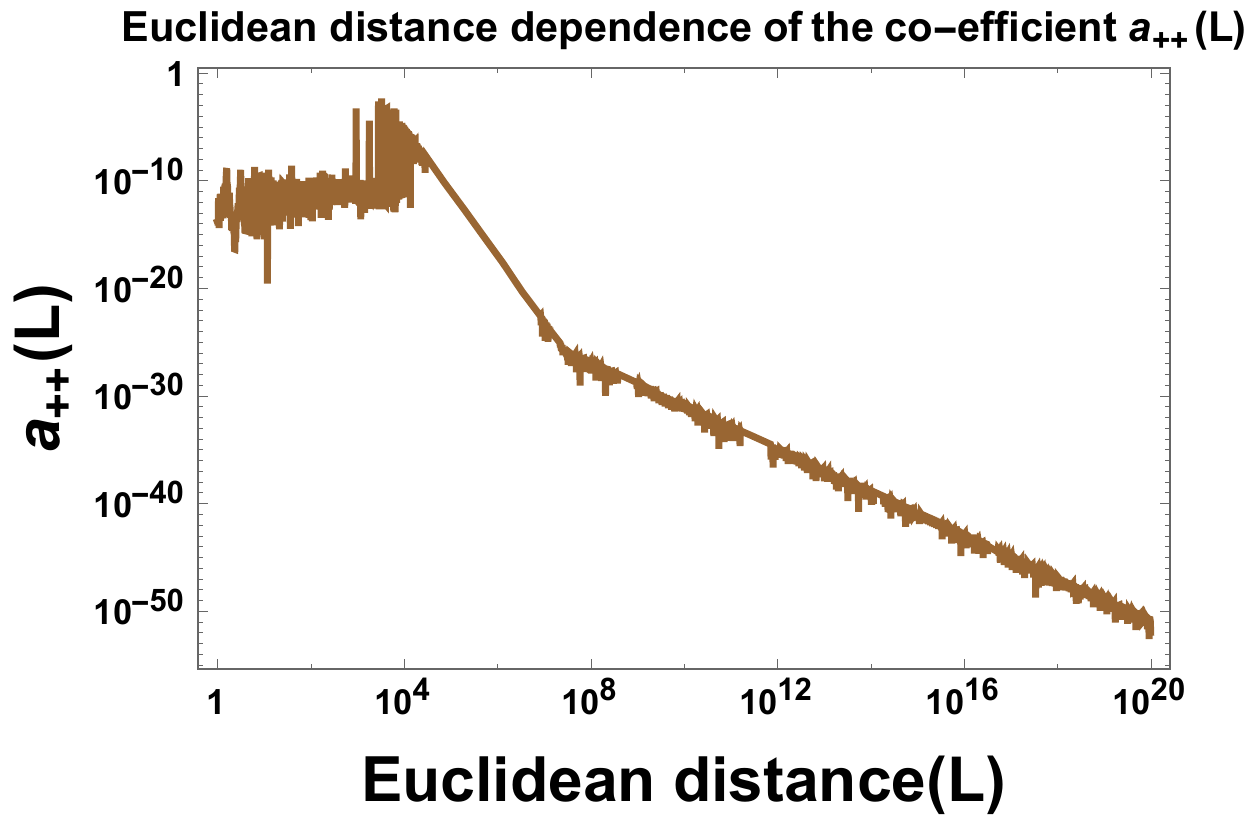}
	\label{3c}
	}
	\subfigure[Bloch vector component $a_{--}$ vs Euclidean Distance profile.]{
	\includegraphics[width=7.8cm,height=4cm] {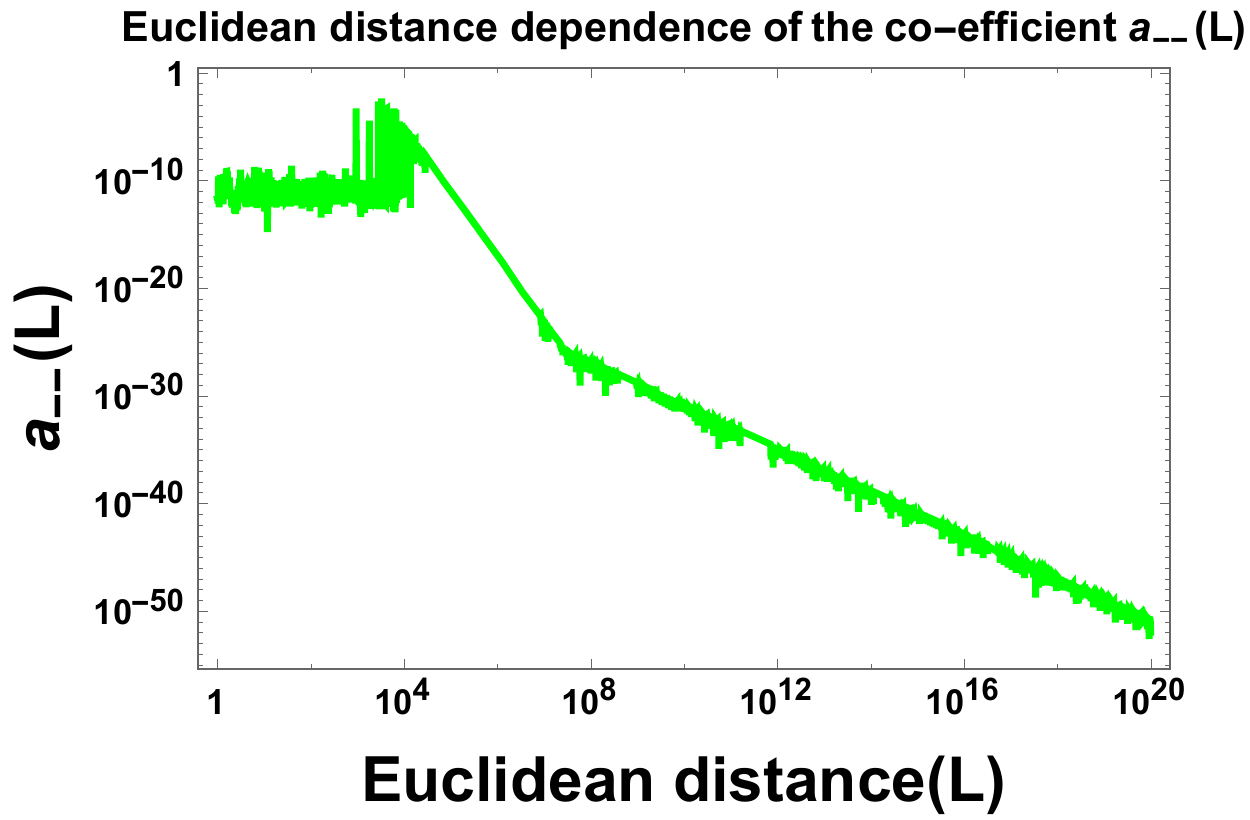}
	\label{3d}
	}
	\subfigure[Bloch vector component $a_{+-}$ vs Euclidean Distance profile.]{
	\includegraphics[width=7.8cm,height=4cm] {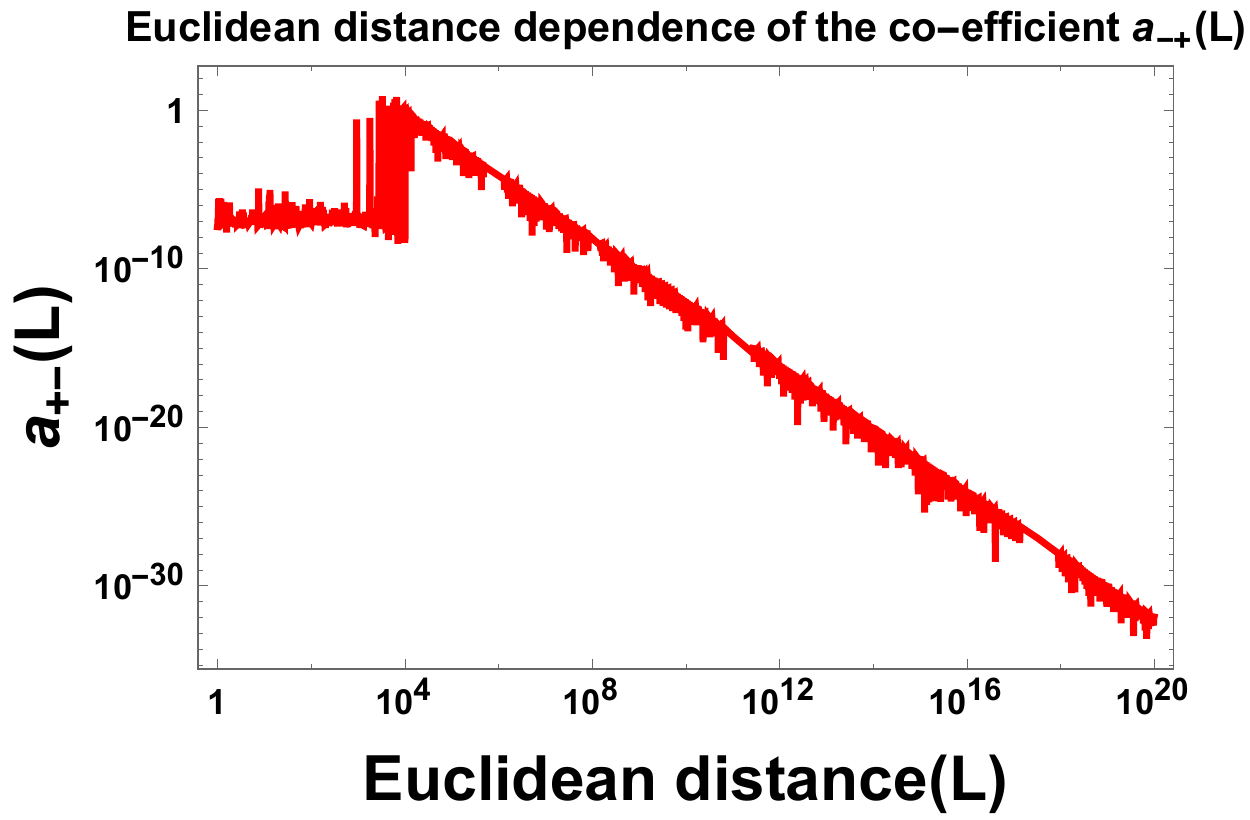}
	\label{3e}
	}
	\subfigure[Bloch vector component $a_{-+}$ vs Euclidean Distance profile.]{
	\includegraphics[width=7.8cm,height=4cm] {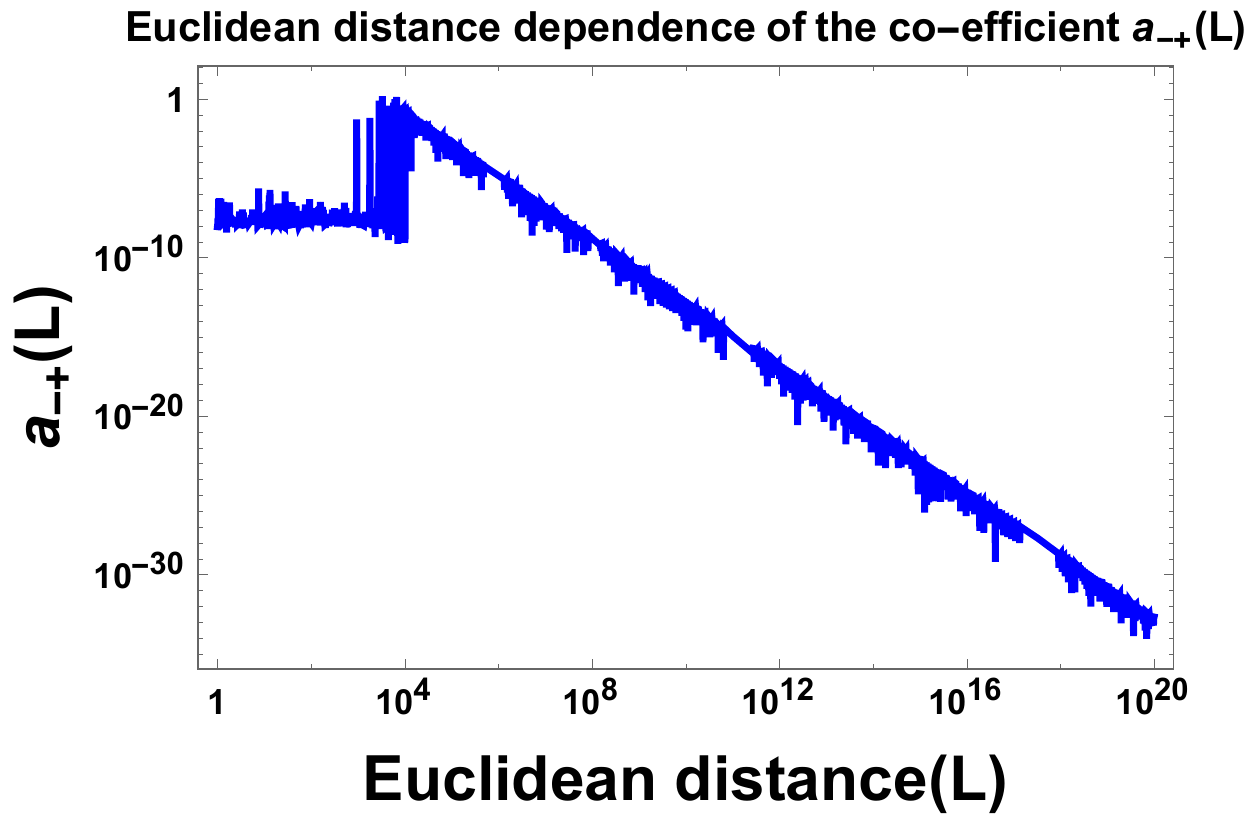}
	\label{3d}
	}
	\caption{ Dependence of the various Bloch vector components on the Euclidean distance is shown here.}
	\label{gd3}
\end{figure}

\begin{figure}[htb]
\centering
\subfigure[Bloch vector component $a_{03}$ vs k profile.]{
	\includegraphics[width=7.8cm,height=4cm] {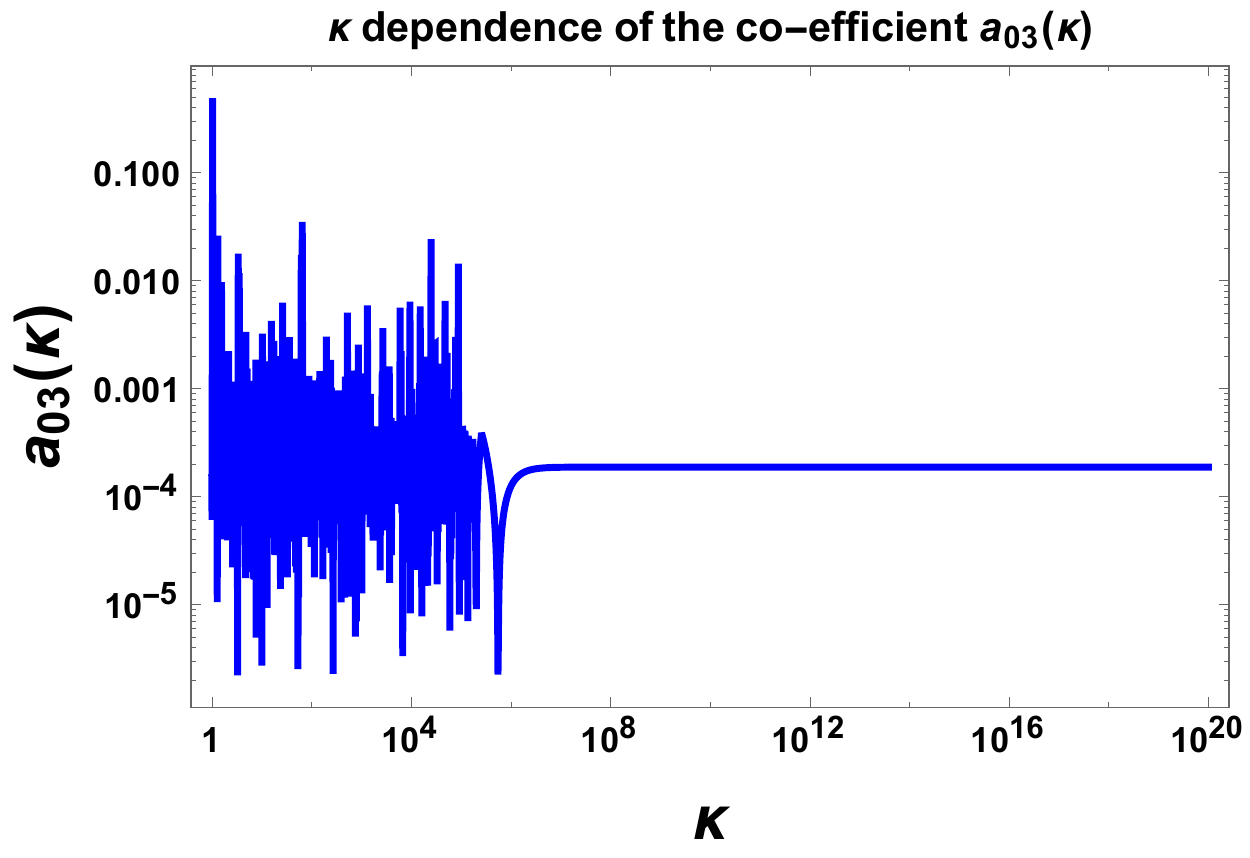}
	\label{4a}
	}
\subfigure[Bloch vector component $a_{33}$ vs k profile.]{
	\includegraphics[width=7.8cm,height=4cm] {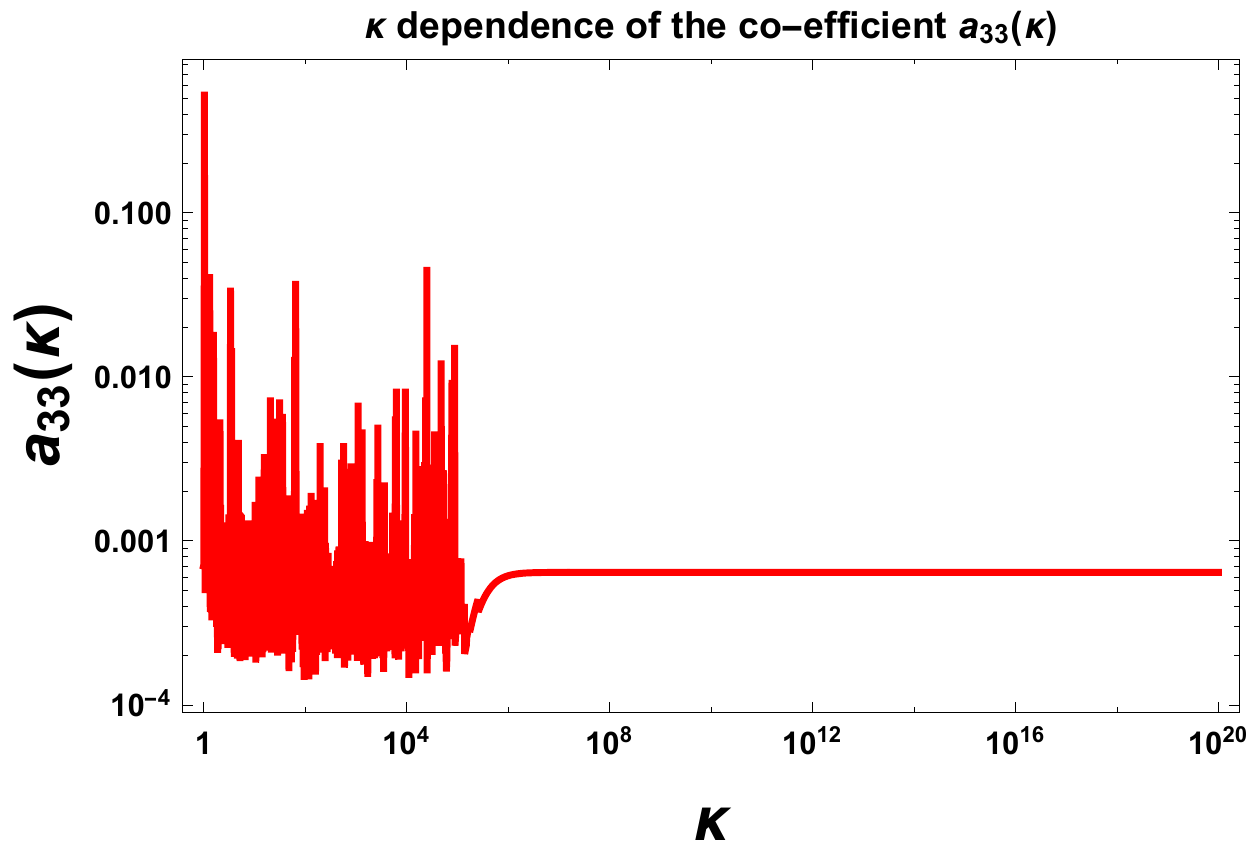}
	\label{4b}
	}
	\subfigure[Bloch vector component $a_{++}$ vs k profile.]{
	\includegraphics[width=7.8cm,height=4cm] {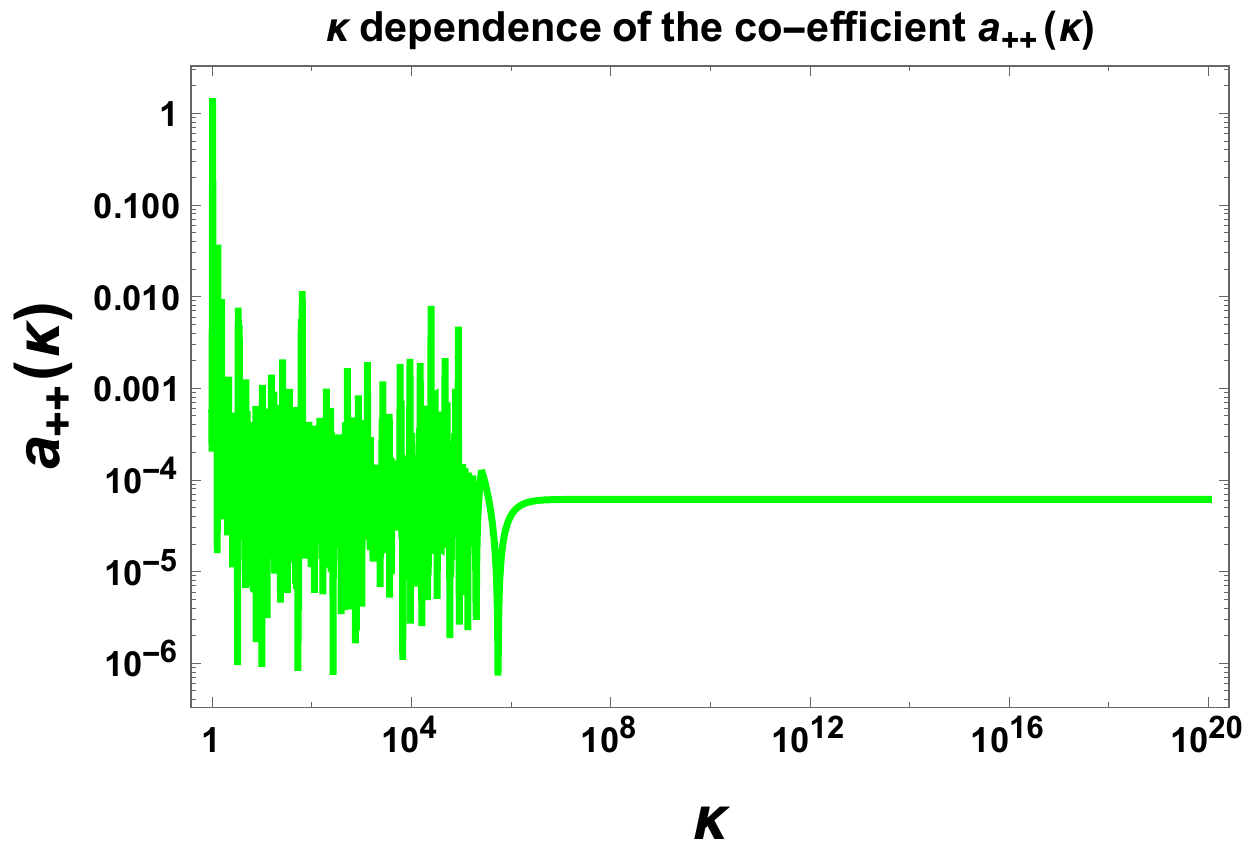}
	\label{4c}
	}
	\subfigure[Bloch vector component $a_{--}$ vs k profile.]{
	\includegraphics[width=7.8cm,height=4cm] {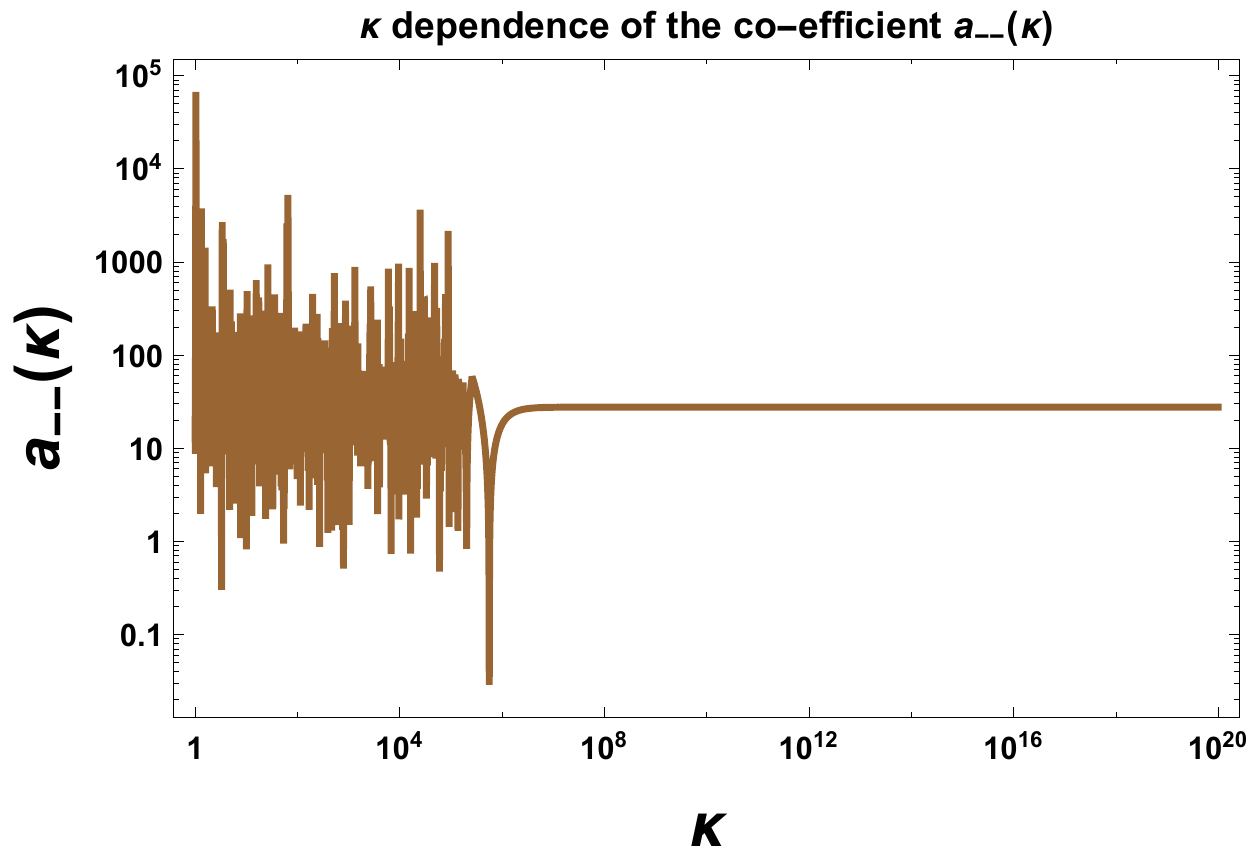}
	\label{4d}
	}
	\subfigure[Bloch vector component $a_{+-}$ vs k profile.]{
	\includegraphics[width=7.8cm,height=4cm] {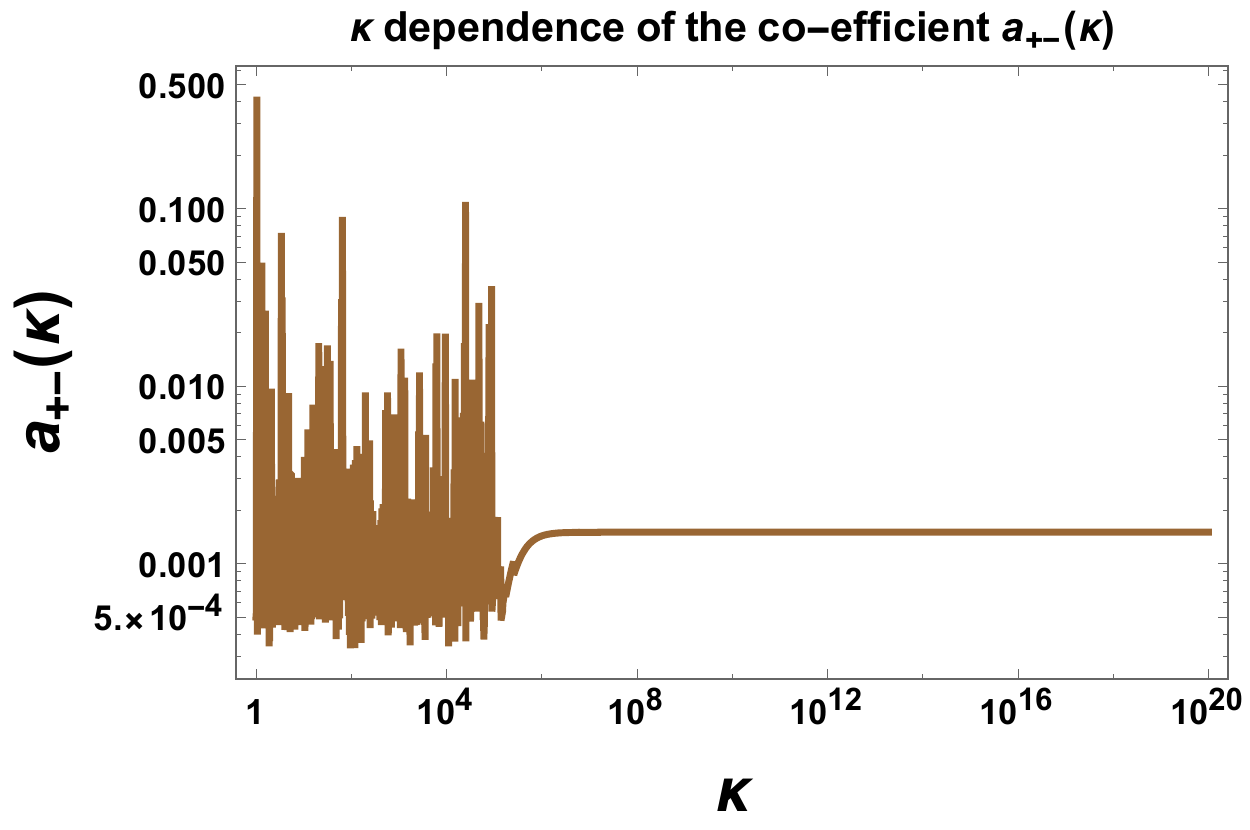}
	\label{4e}
	}
	\subfigure[Bloch vector component $a_{-+}$ vs k profile.]{
	\includegraphics[width=7.8cm,height=4cm] {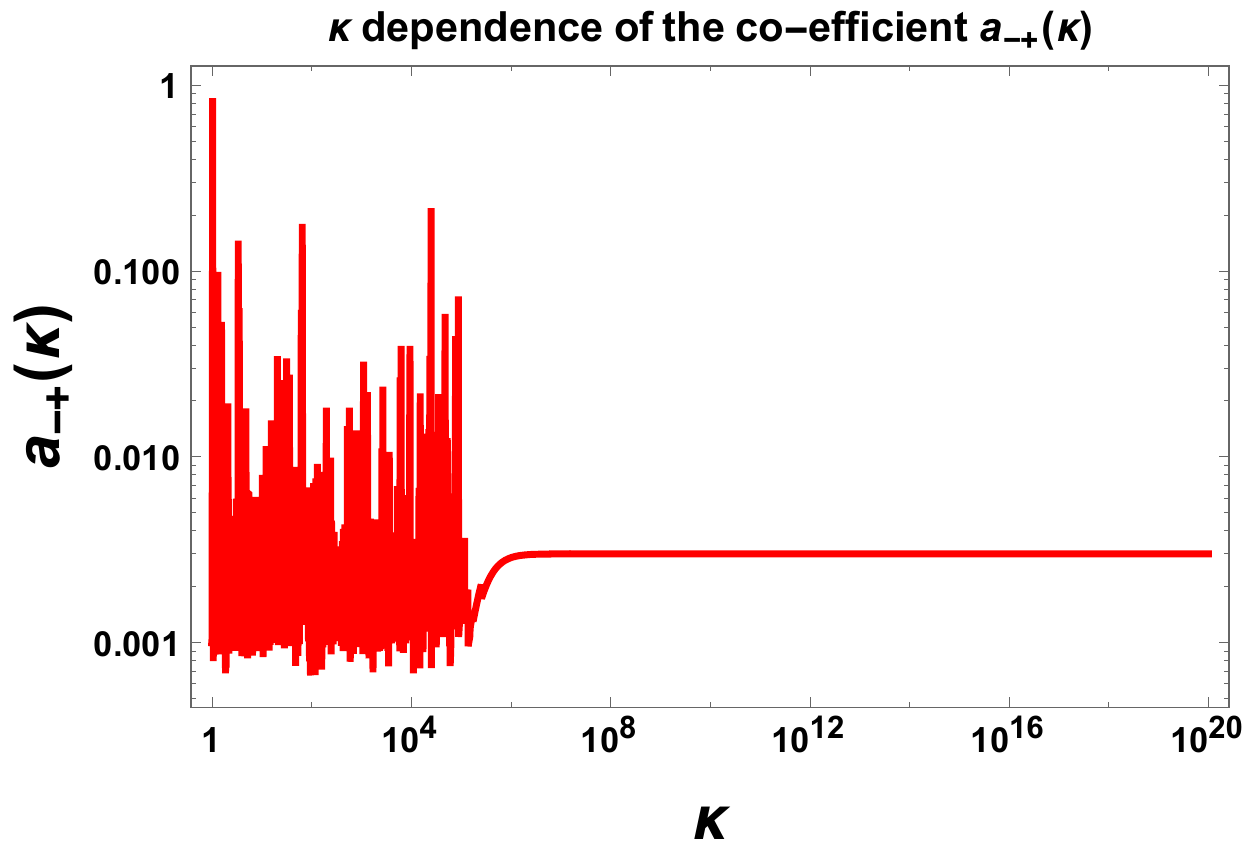}
	\label{4f}
	}
	\caption{Dependence of the Bloch vector components on k is shown here }
	\label{gd4}
\end{figure}

The arbitrary constants obtained after using the late time behaviour as the initial conditions are explicitly defined by the following expressions:~\footnote{Identifying the terms in the obtained solutions coming from the system-system interaction and the system-bath interaction is extremely essential in the study of entanglement dynamics in the context of open quantum systems. It can be seen from the effective Hamiltonian construction that it is the Lamb Shift Hamiltonian which mostly tells about the system-system interaction and the Lindbladian term mostly tells us about the system-bath interaction. The physically relevant solution satisfying the boundary condition contains three prime non zero components given as $A_1$, $B_1$ and $B_2$. The term $A_1$ appears from the lamb shift hamiltonian coefficient matrix and hence this factor gives the contribution coming from the system system interaction. The terms $B_1$ and $B_2$ on the other hand appears in this context from the GSKL master equation, which determines the degree of system bath interaction. This source of origin of various terms appearing in the solutions will prove to be essential to develop an approximate idea about the source of entanglement when various entanglement measures are computed from these obtained solutions. However due to the complicated structure of the solutions it is not possible to quantify or exactly determine the part of the computed entropies coming from the interaction between the atoms and the one coming from interaction with the environment.}
\bea
g_1(\omega)&=&\frac{1}{\left(-\frac{\mathcal{Y}_1 f_2(\omega)}{4(B_1+B_2)}+\frac{\mathcal{Y}_1 f_3(\omega)}{4(B_1+B_2)}\right)\left(-\frac{\mathcal{Y}_2 f_3(\omega)}{\mathcal{Y}_4}+\frac{\mathcal{Y}_3 f_2(\omega)}{\mathcal{Y}_4}\right)+\left(\frac{\mathcal{Y}_1 f_2(\omega)}{4(B_1+B_2)^2}+\frac{3B_2 f_3(\omega)}{4(B_1+B_2)^2}\right)\left(-\frac{\mathcal{Y}_2 f_3(\omega)}{\mathcal{Y}_4}+\frac{\mathcal{Y}_3 f_2(\omega)}{\mathcal{Y}_4}\right)}\nonumber\\
&&\times\left[\frac{-\frac{\mathcal{Y}_3}{4A_1(B_1+B_2)}\left(-\frac{\mathcal{Y}_1 f_2(\omega)}{4(B_1+B_2)}+\frac{\mathcal{Y}_1 f_3(\omega)}{4(B_1+B_2)}\right)\rm~tanh(\pi k \omega)-\mathcal{Y}_1(-\frac{\mathcal{Y}_2 f_3}{\mathcal{Y}_4}+\frac{\mathcal{Y}_3 f_2}{\mathcal{Y}_4})\rm~tanh(\pi k \omega)}{\left(-\frac{\mathcal{Y}_1f_2(\omega)}{4(B_1+B_2)}+\frac{\mathcal{Y}_1f_2(\omega)f_3(\omega)}{4(B_1+B_2)}\right)}\right],~~~~~~~~~
\\
g_2(\omega)&=&\frac{4(B_1+B_2)12B_1B_2^2+36B_2^2+B_1f_1^2+3B_2f_3^2 \rm~tanh(\pi k \omega)}{(f_2-f_3)(12 B_1 B_2^2-36B_2^3-B_1f_1^2+B_1f_1f_2+B_1f_1f_3+3B_2f_2f_3)},
\\
g_3(\omega)&=&-\frac{4}{(f_2-f_3)(12 B_1 B_2^2-36B_2^3-B_1f_1^2+B_1f_1f_2+B_1f_1f_3+3B_2f_2f_3)}\nonumber\\
&&~~~\times[12B_1^2B_2^2 {\rm ~tanh(\pi k \omega)}+48 B_1B_2^3 {\rm ~tanh(\pi k \omega)} +36B_2^4{\rm ~tanh(\pi k \omega)}+B_1^2f_1^2 {\rm ~tanh(\pi k \omega)}\nonumber\\
&&~~~~~~~~~~~~~~~~+B_1B_2f_1^2{\rm ~tanh(\pi k \omega)}+3B_1B_2f_2^2{\rm ~tanh(\pi k \omega)}+3B_2^2f_2{\rm ~tanh(\pi k \omega)}],
\\
g_5(\omega)&=&-\frac{1}{12B_1B_2^2+36B_2^3B_1f_2^2-B_1f_1f_2-B_1f_1f_3-3B_2f_2f_3}\nonumber\\
&&~~~\times[4B_1B_2f_2~{\rm tanh(\pi k\omega)}+3B_2^2f_2~{\rm tanh(\pi k\omega)}
+B_1B_2f_3~{\rm tanh(\pi k \omega)}+3B_2^2~ f_3{\rm tanh(\pi k \omega)}],
\eea
\bea
g_6(\omega)&=&\frac{1}{12B_1B_2^2+36B_2^3+B_1f_1^2-B_1f_1f_2-B_1f_1f_3-3B_2f_2f_3}\nonumber
\\
&&~~~\times[4B_1^2f_2~{\rm tanh(\pi k\omega)}{\rm tanh(\pi k\omega)}+16B_1B_2(f_2~{\rm tanh(\pi k\omega)}+12B_2^2f_3~{\rm tanh(\pi k\omega)}\nonumber\\&&~~~~~~~~~~~~+4B_1^2f_5~{\rm tanh(\pi k\omega)}+16B_1B_2~{\rm tanh(\pi k\omega)}+12B_2^2f_3~{\rm tanh(\pi k\omega)}\nonumber\\&&~~~~~~~~~~~~~~~~~~~~~-12B_1B_2^2~{\rm tanh^2(\pi k\omega)}-36B_2^3~{\rm tanh^2(\pi k\omega)}-B_1f_1^2~{\rm tanh^2(\pi k\omega)}\nonumber\\&&~~~~~~~~~~~~~~~~~~~~~~~~~~~~~~+B_1f_1f_2~{\rm tanh^2(\pi k\omega)}+B_1f_1f_3~{\rm tanh^2(\pi k\omega)}\nonumber\\&&~~~~~~~~~~~~~~~~~~~~~~~~~~~~~~~~~~~~~~~~~
~~~~~~~~~~~~~~~~~~~~~+3B_2f_2f_3~{\rm tanh^2(\pi k\omega)}].
\eea
where in writing the function $g_1(\omega)$ we have introduced the following symbols:
\bea
\mathcal{Y}_1=\left(1-\frac{B_2}{B_1+B_2}\right),~~
\mathcal{Y}_2=-12B_2^2-f_2^2,~~
\mathcal{Y}_3=-12B_2^2-f_3^2,~~
\mathcal{Y}_4=16A_1(B_1+B_2)^2.
\eea
Using these solution the variation of the Bloch vector components with respect to various parameters are plotted in fig.~\ref{gd1}, fig.~\ref{gd2}, fig.~\ref{gd3} and fig.~\ref{gd4}.

\section{\textcolor{blue}{\bf \large Entanglement Measures}}

Using these solution our next objective is to compute various entanglement measures from the present OQS set up. 

\subsection{\textcolor{blue}{ Von Neumann Entanglement Entropy}}
\label{Von}
In the context of quantum statistical mechanics and quantum information theory, {\it Von Neumann entropy} plays the role of extended version of classical {\it Gibbs entropy}. It actually measures the amount of quantum entanglement for a subsystem or reduced system of a bipartite quantum system.

In the present context, for the two atomic subsystem, which we have obtained after partially tracing over the bath degrees of freedom the Von Neumann entanglement entropy is defined in terms of the reduced density matrix ($\rho_{\bf System}$) as:
\be
S(\rho_{\bf System})=-{\rm Tr}\left[\rho_{\bf System}~\ln\left(\rho_{\bf System}\right)\right]
\ee
\begin{enumerate}
	\item It is expected from our present OQS set up that the measure of {\it Von Neumann entanglement entropy} is non zero because our subsystem, which we have obtained by partially tracing over bath degrees of freedom, is described by mixed states.
	Conversely, if a subsystem is to be characterised by pure quantum mechanical states , the entanglement measure is zero. Since in our set up we are  dealing with pure quantum mechanical states initially due to the initial uncorrelation between the system and the environment, we get zero entanglement measure from our computation. But during time evolution of the subsystem it becomes more correlated and consequently pure state transforms to a mixed state, for which we get a saturation but non-zero value of the Von Neumann entanglement entropy.
	\item 
	The Von Neumann entanglement measure for the reduced subsystem can be expressed as a sum of the contributions from the two independent and identical atoms.
	\be
	S(\rho_{\bf System}) = S(\rho_1 \otimes \rho_2) = S(\rho_1)+S(\rho_2)
	\ee
	\item Additionally, it is important to note that, the Von Neumann entanglement measure 
	is not the only reasonable measure of quantum entanglement. In the later sections we will consider other quantum entanglement measures, which are commonly used in the context of quantum information theory.
	
\end{enumerate}


In the context of two entangled atoms relevant to our OQS it can be seen from the solution of the Bloch vector components that $\rm{a_{03}}(\tau)$ and $\rm{a_{30}}(\tau)$ has the same form i.e. $a_{03}(\tau)=a_{30}(\tau)$. So in that case the Von Neumann entropy can be written as:

\be
S = \frac{1}{4} [ 8~\ln~2-(\alpha-\widetilde{\beta}) \ln(\alpha-\widetilde{\beta}) - (\alpha+\widetilde{\beta}) \ln(\alpha+\widetilde{\beta})
- (\eta-\widetilde{\gamma}) \ln(\eta-\widetilde{\gamma}) 
- (\eta+\widetilde{\gamma}) \ln(\eta+\widetilde{\gamma}) ]
\ee

where we have introduced two new functions, $\widetilde{\beta}$ and $\widetilde{\gamma}$, which are defined as:
\bea \widetilde{\beta}&=&\beta_{\bf Two~atom}=\sqrt{a_{-+}(\tau)a_{+-}(\tau)}, \ \ \ ~~~~ \ \ 
\widetilde{\gamma}=\gamma_{\bf Two~atom}=\sqrt{4a^2_{03}(\tau)+a_{--}(\tau)a_{++}(\tau)}~~~~~\eea
Here all of these time dependent coefficients $a_{03}(\tau)$, $a_{-+}(\tau)$ , $a_{+-}(\tau)$, $a_{--}(\tau)$, $a_{++}(\tau)$ and $a_{33}(\tau)$ have explicitly been computed for the two atomic OQS in the previous section.

\begin{figure}[htb]
\centering
\subfigure[Normalized Von Neumann Entanglement entropy vs Time profile.]{
	\includegraphics[width=7.8cm,height=4cm] {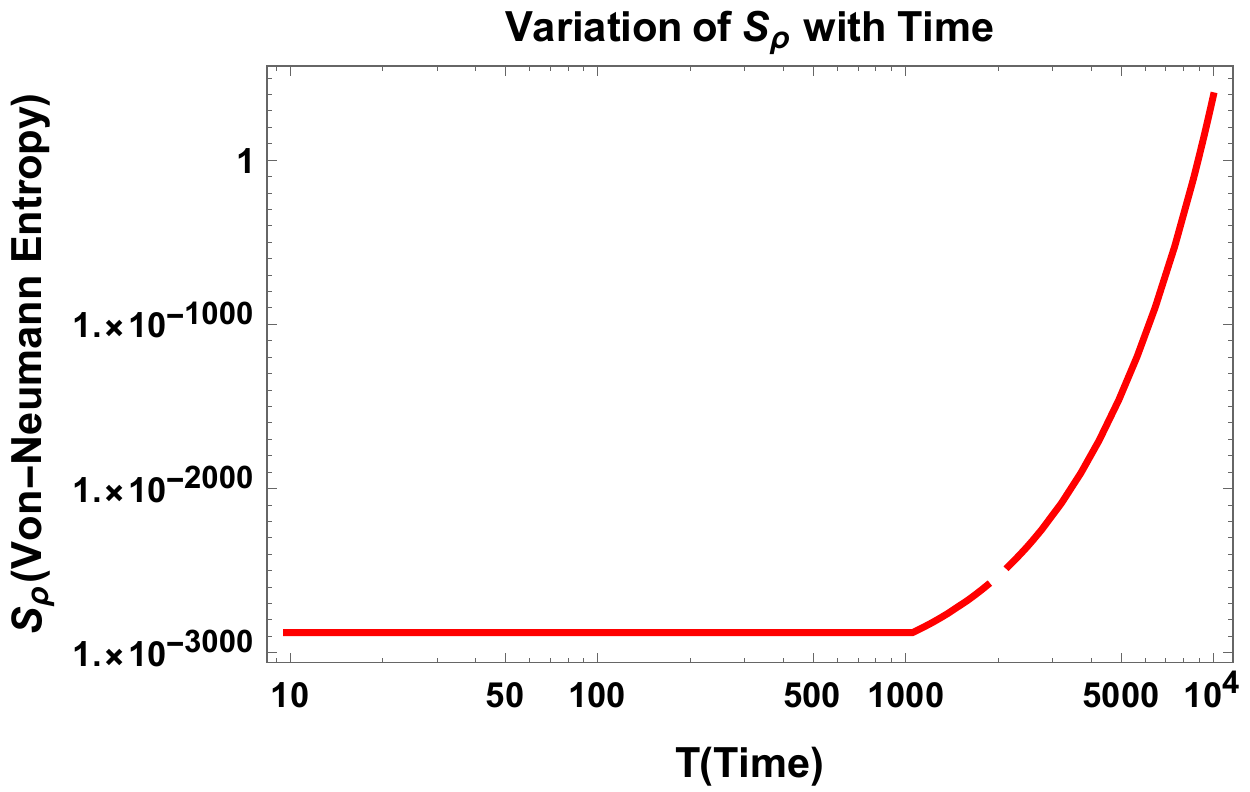}
	\label{vont}
	}
	\subfigure[Normalized Von Neumann Entanglement entropy vs $|Frequency|$ profile.]{
	\includegraphics[width=7.8cm,height=4cm] {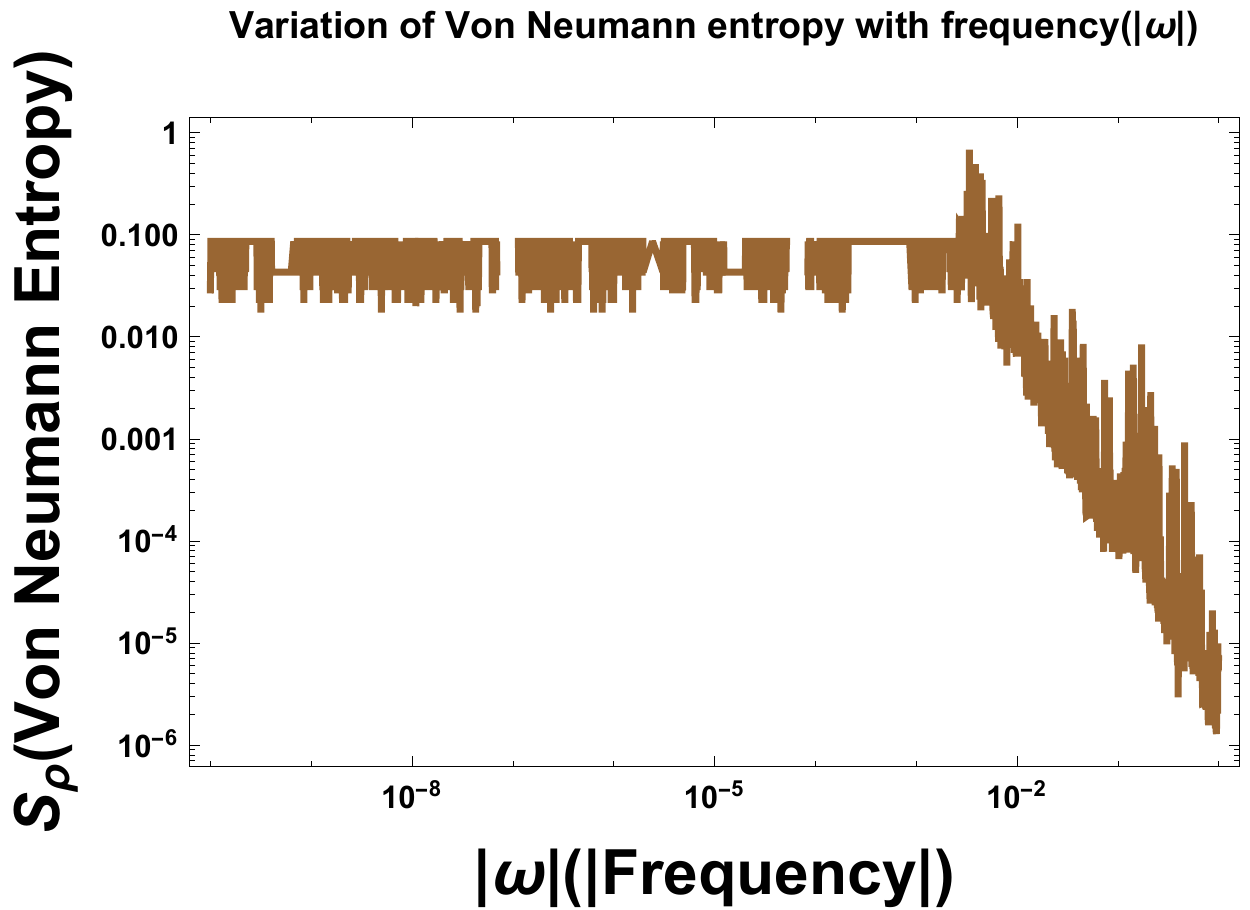}
	\label{vonw}
	}
	\subfigure[Normalized Von Neumann Entanglement entropy vs Euclidean Distance profile.]{
	\includegraphics[width=7.8cm,height=4cm] {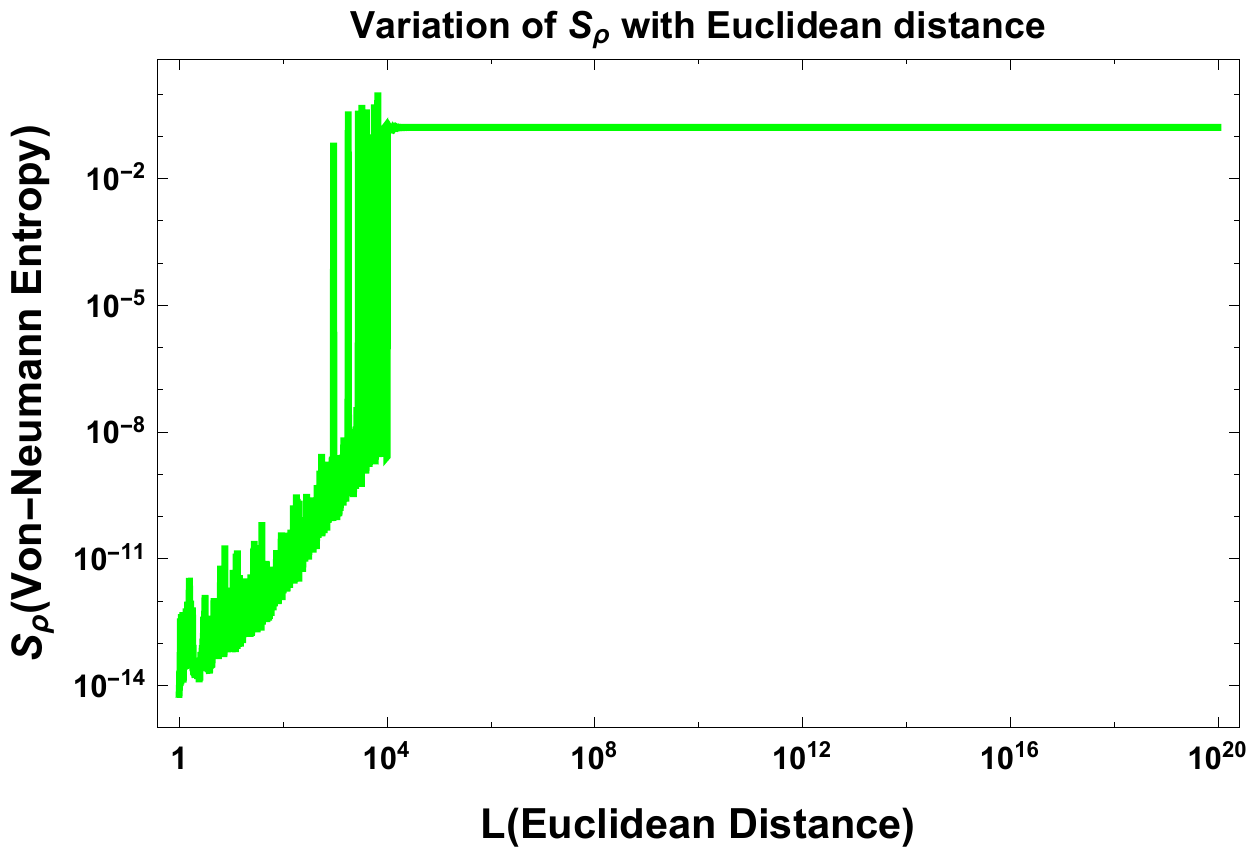}
	\label{vonl}
	}
	\subfigure[Normalized Von Neumann Entanglement entropy vs k profile.]{
	\includegraphics[width=7.8cm,height=4cm] {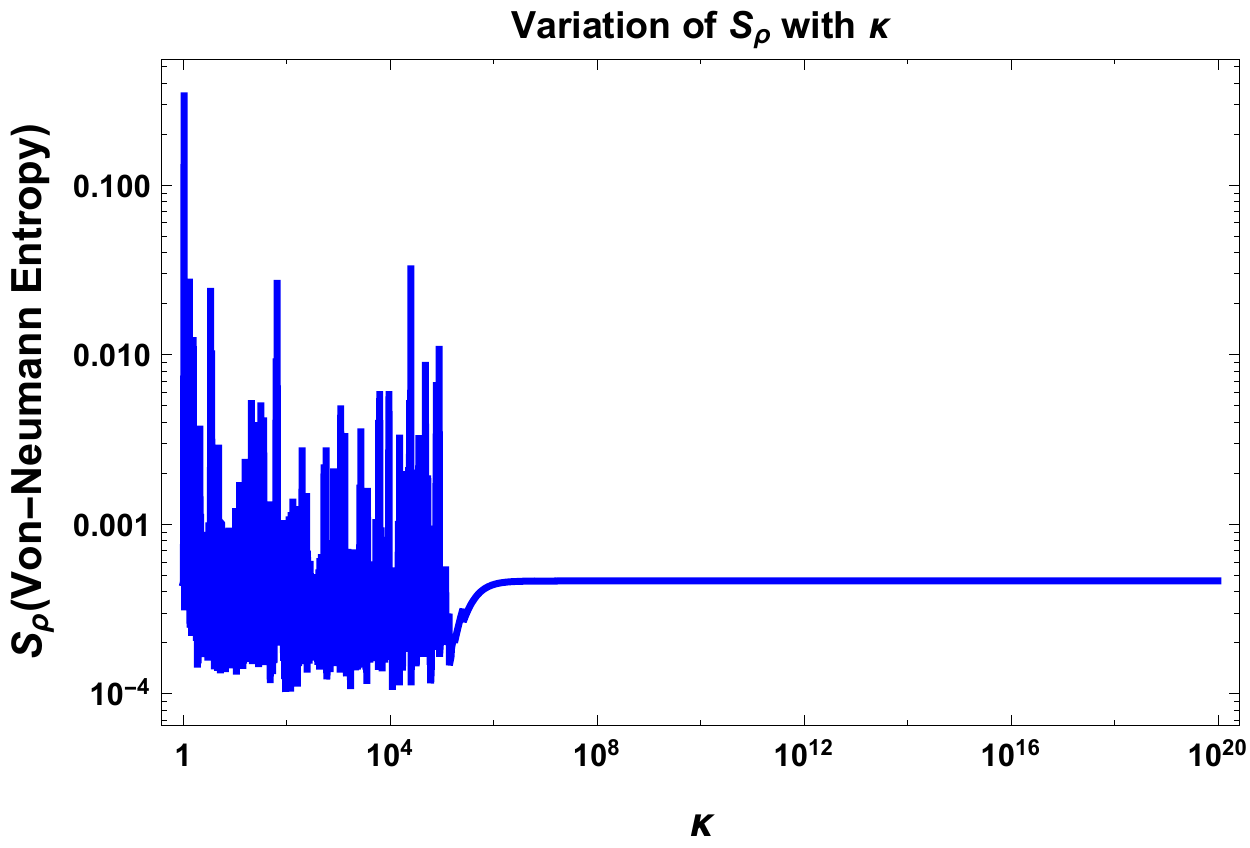}
	\label{vonk}
	}
	\caption{Normalized Von Neumann Entanglement entropy variation with various parameters is shown here. }
\end{figure}

To physically analyse this result we have plotted the behaviour of the Von Neumann entropy from the present two atomic OQS set up with respect to different useful parameters present in the theory. However, as we will see later that almost all the netanglement measures give plots of similar nature. Hence, the discussions related to all the plots are given afterwards in a separate section \ref{disc}.

\subsection{\textcolor{blue}{ R$e^{'}$nyi Entropy}}
\label{}
R$e^{'}$nyi entropy is a generalisation of various information theoretic measure which quantify randomness of a quantum mechanical system. The R$e^{'}$nyi entropy can be expressed in terms of Hartley function as:
\be
S_q(\rho_{\bf System})=\frac{1}{1-q}\ln{\rm Tr}[(\rho_{\bf System})^q]\equiv {\cal H}_{q}(\rho_{\bf System}),~~~~~q\geq 0,q\ne1
\ee
where $q$ is known as the R$e'$nyi Index.

In the case of our two atomic OQS set up, we found from the structure of the solution that $a_{03}$=$a_{30}$. Hence the expression for the R$e'$nyi entropy reduces to the following simplified form:

\be
S_q(\rho_{\bf System}) = \frac{1}{1-q} \ln\left[ 4^{-q} \left\{ (\alpha-\widetilde{\beta})^q + (\alpha+\widetilde{\beta})^q
 + (\eta-\widetilde{\gamma})^q + (\eta+\widetilde{\gamma})^q \right\} \right]
\ee

where the symbols used has already been defined in the previous section. It can be very easily verified that in this context relevant to our OQS the R$e'$nyi entropy reduces to the Von Neumann entropy in the limit q $\rightarrow 1$.

\begin{figure}[htb]
\centering
\subfigure[Renyi Entropy (q $\rightarrow$1) vs Time profile.]{
	\includegraphics[width=7.8cm,height=4cm] {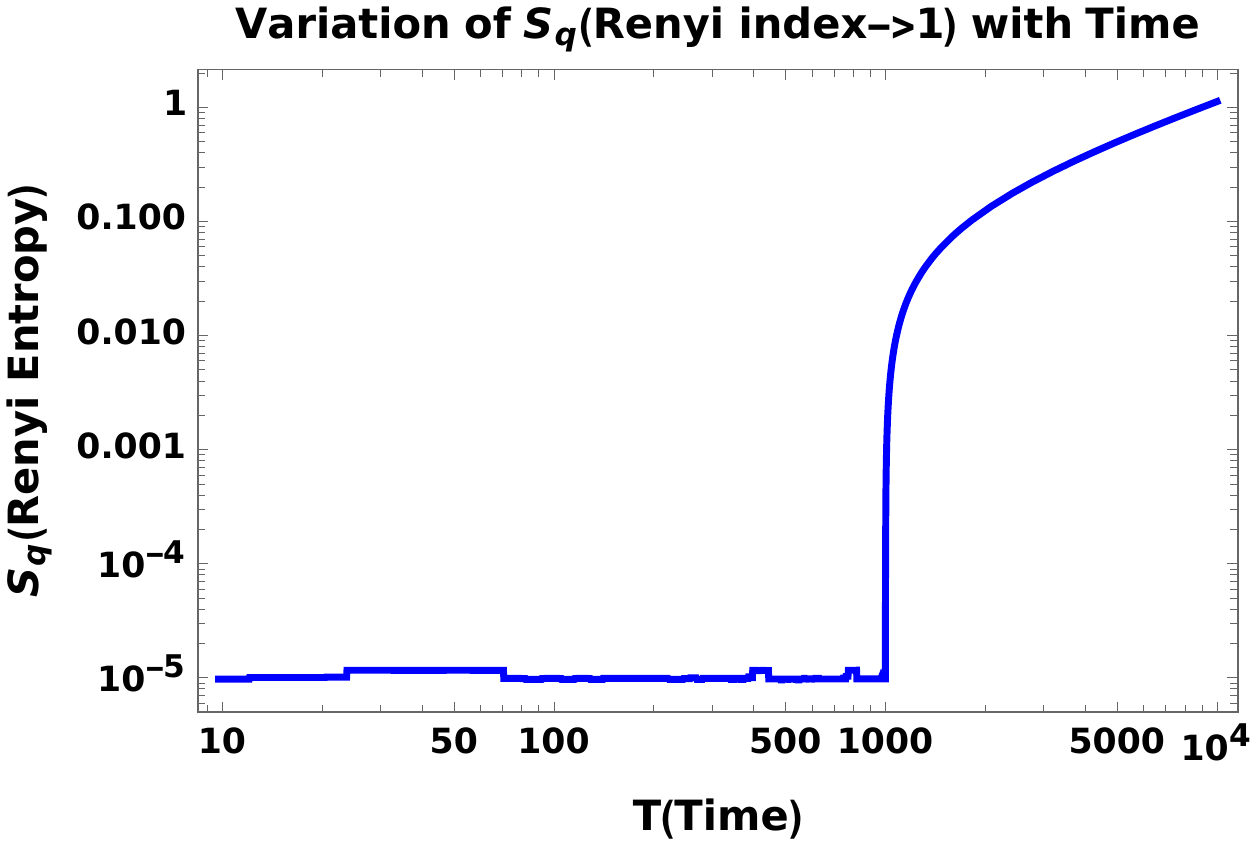}
	\label{renq1t}
	}
\subfigure[Renyi Entropy (q $\rightarrow$1) vs $|Frequency|$ profile.]{
	\includegraphics[width=7.8cm,height=4cm] {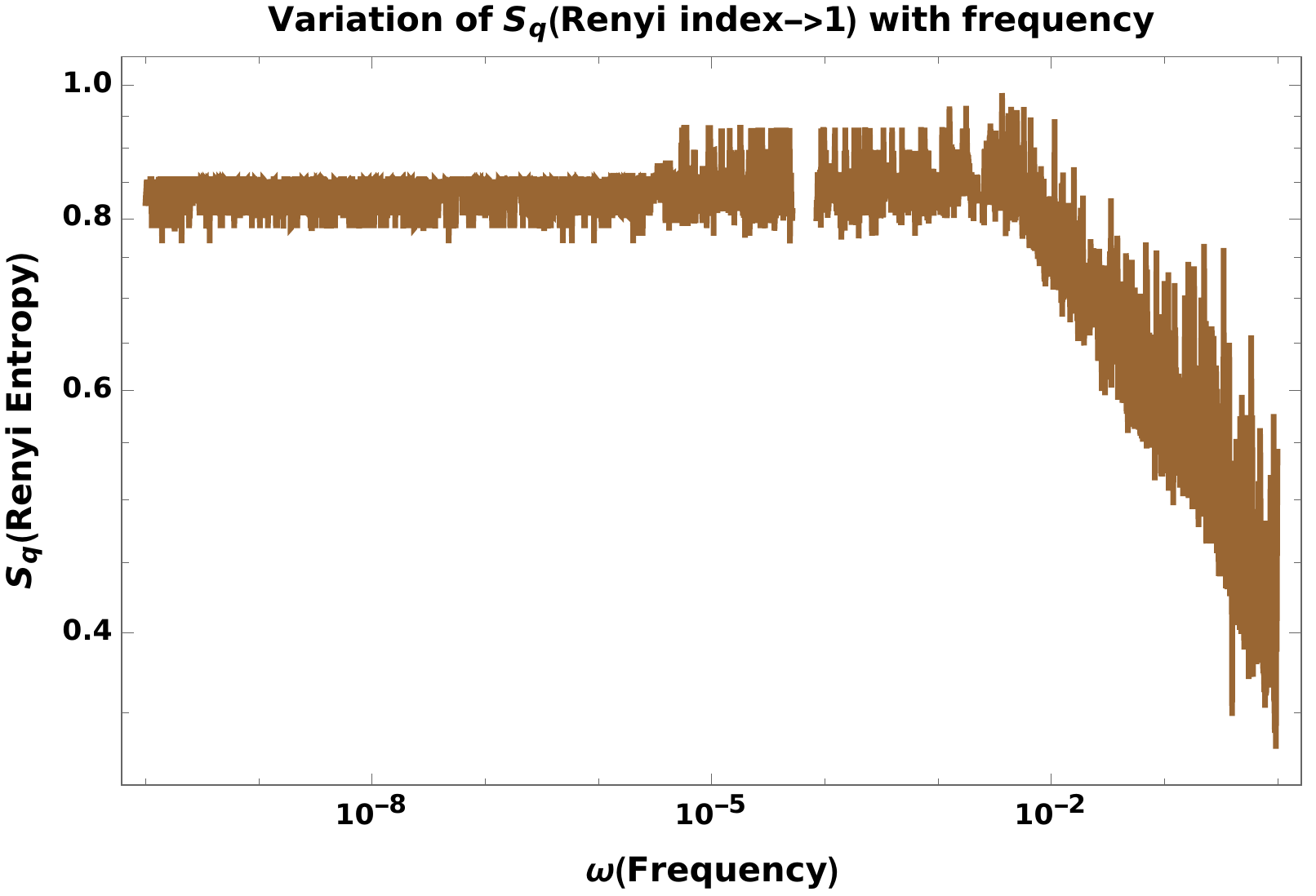}
	\label{renq1w}
	}
	\subfigure[Renyi Entropy (q $\rightarrow$1) vs Euclidean distance profile.]{
	\includegraphics[width=7.8cm,height=4cm] {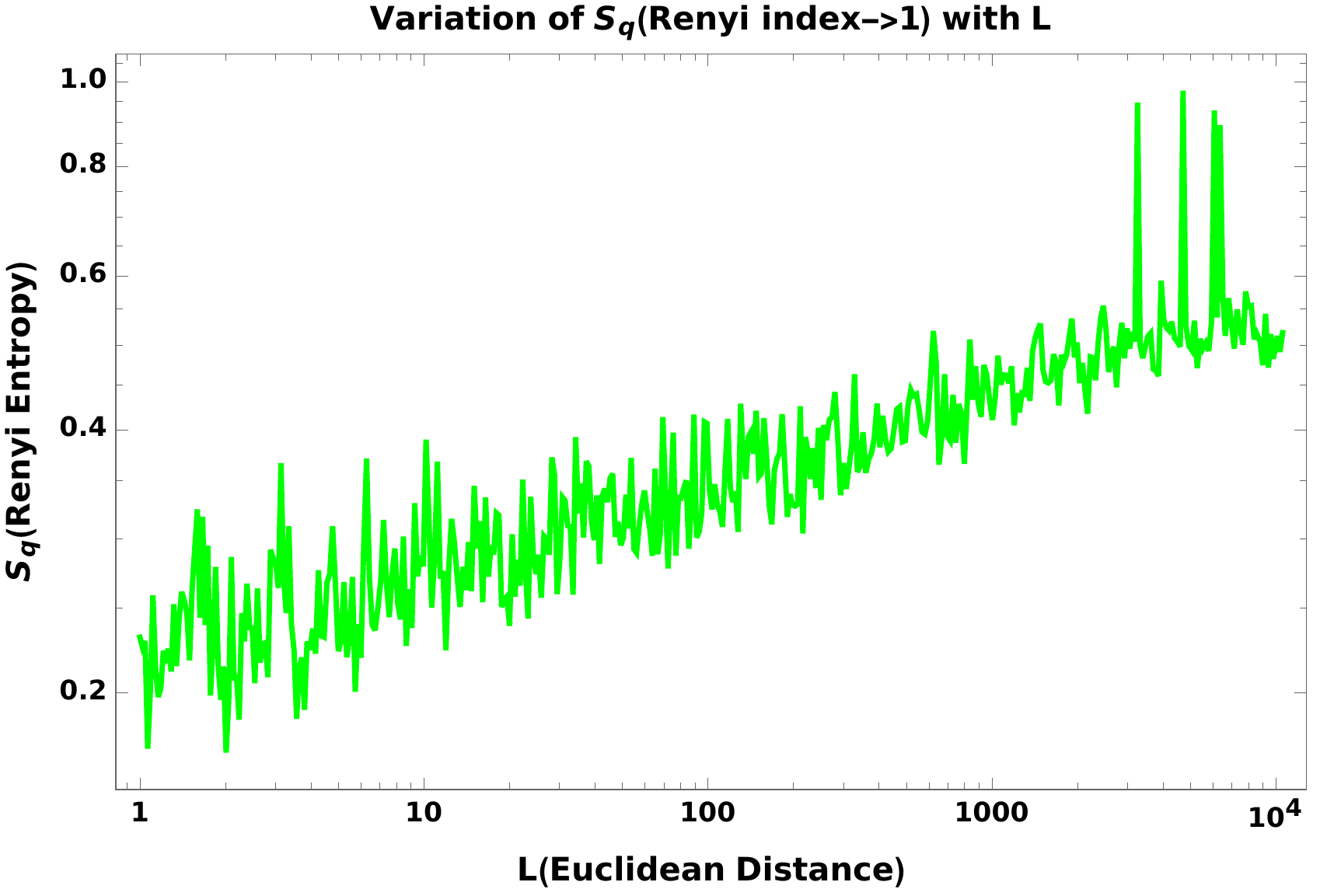}
	\label{renq1l}
	}
	\subfigure[Renyi Entropy (q $\rightarrow$1) vs k profile.]{
	\includegraphics[width=7.8cm,height=4cm] {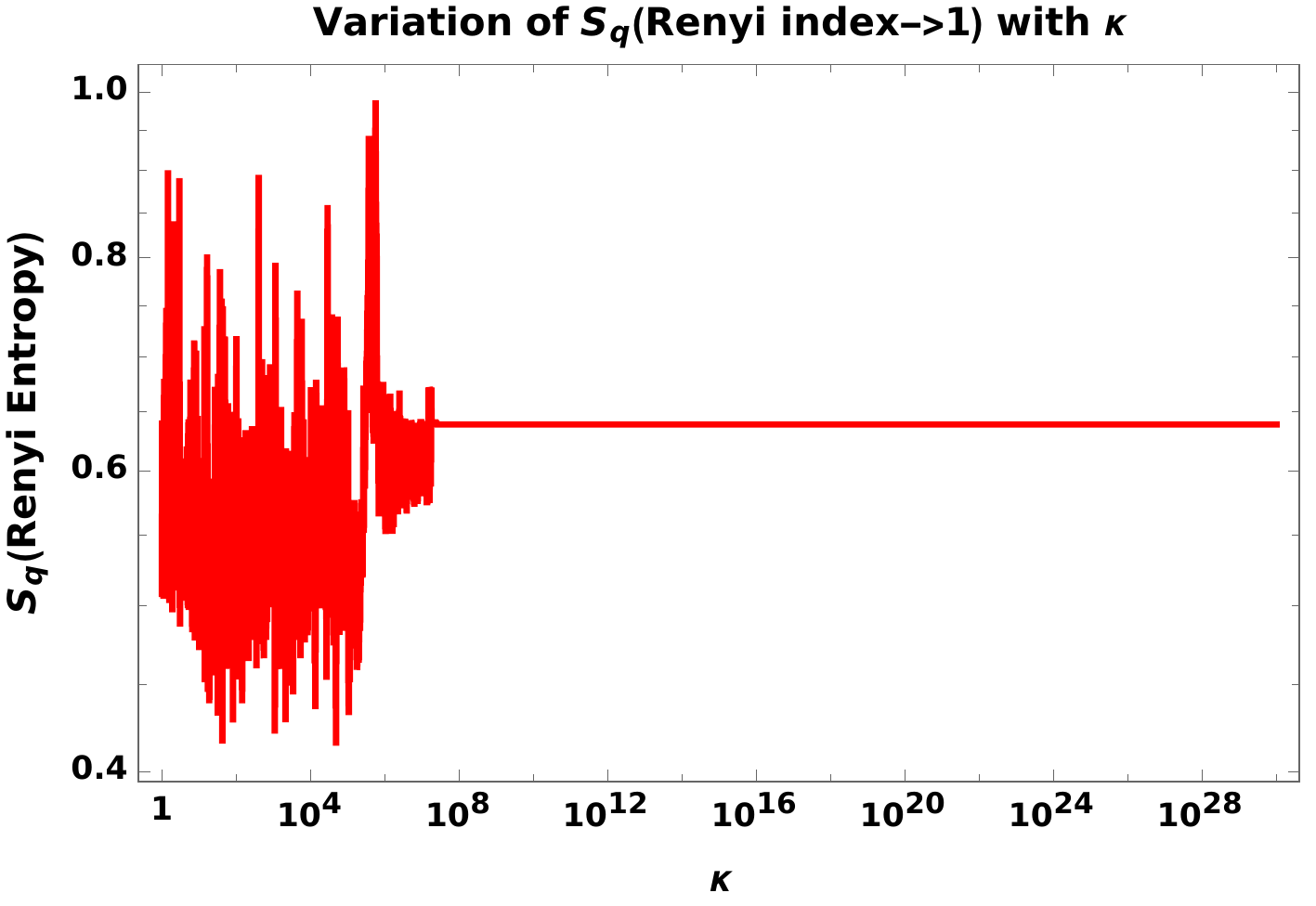}
	\label{renq1k}
	}
	\caption{ Renyi Entropy (q $\rightarrow$1) variation with various parameters are shown here.}
\end{figure}

\begin{figure}[htb]
\centering
\subfigure[Collision entropy vs Time profile.]{
	\includegraphics[width=7.8cm,height=4cm] {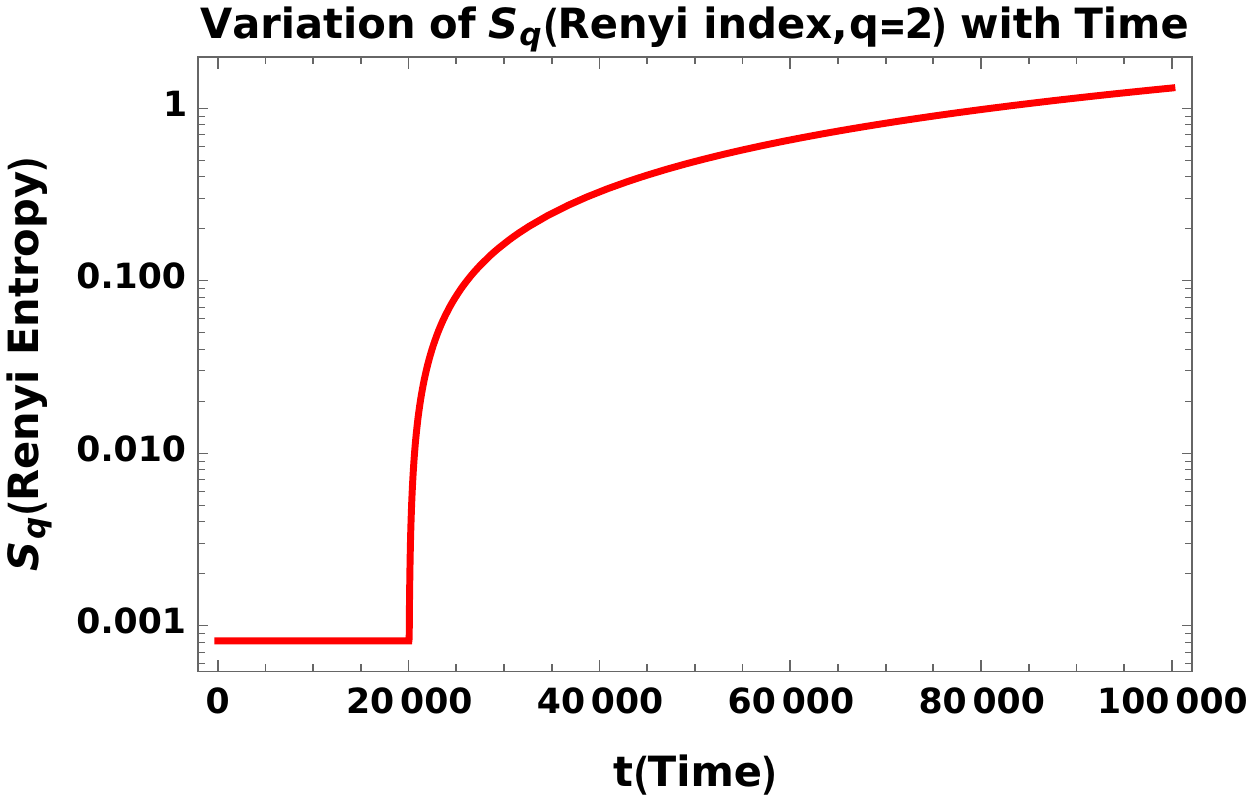}
	\label{colt}
	}
\subfigure[Collision entropy entropy vs $|Frequency|$ profile.]{
	\includegraphics[width=7.8cm,height=4cm] {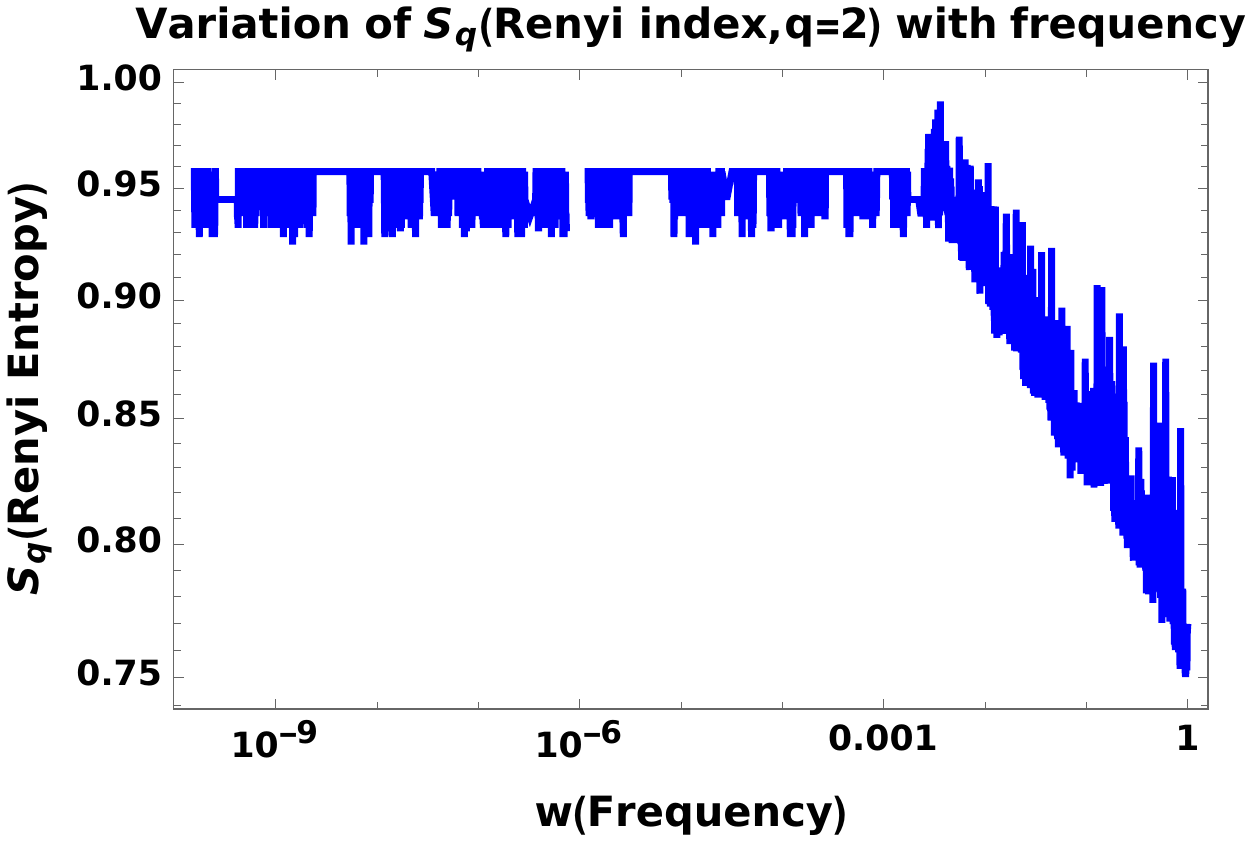}
	\label{colw}
	}
	\subfigure[Collision entropy vs Euclidean distance profile.]{
	\includegraphics[width=7.8cm,height=4cm] {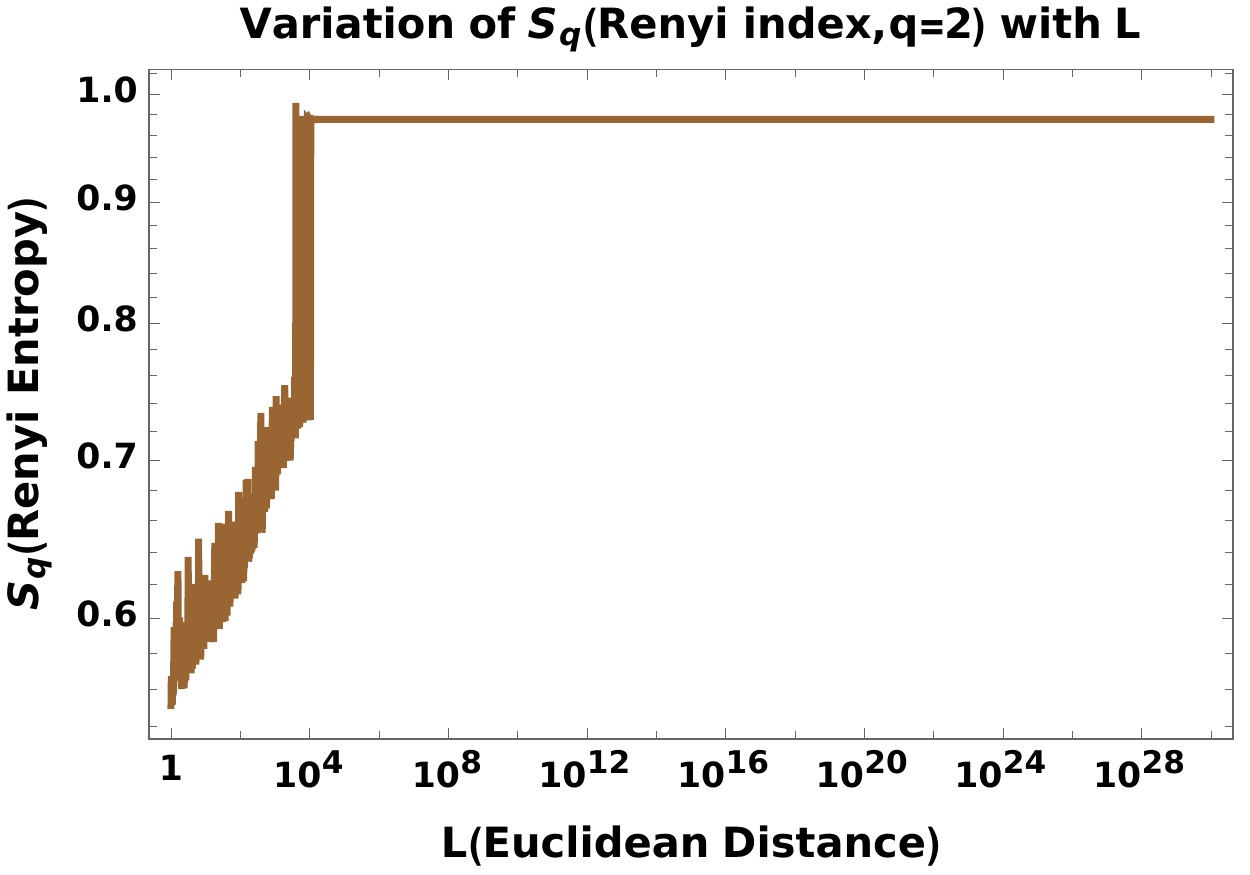}
	\label{coll}
	}
	\subfigure[Collision entropy vs k profile.]{
	\includegraphics[width=7.8cm,height=4cm] {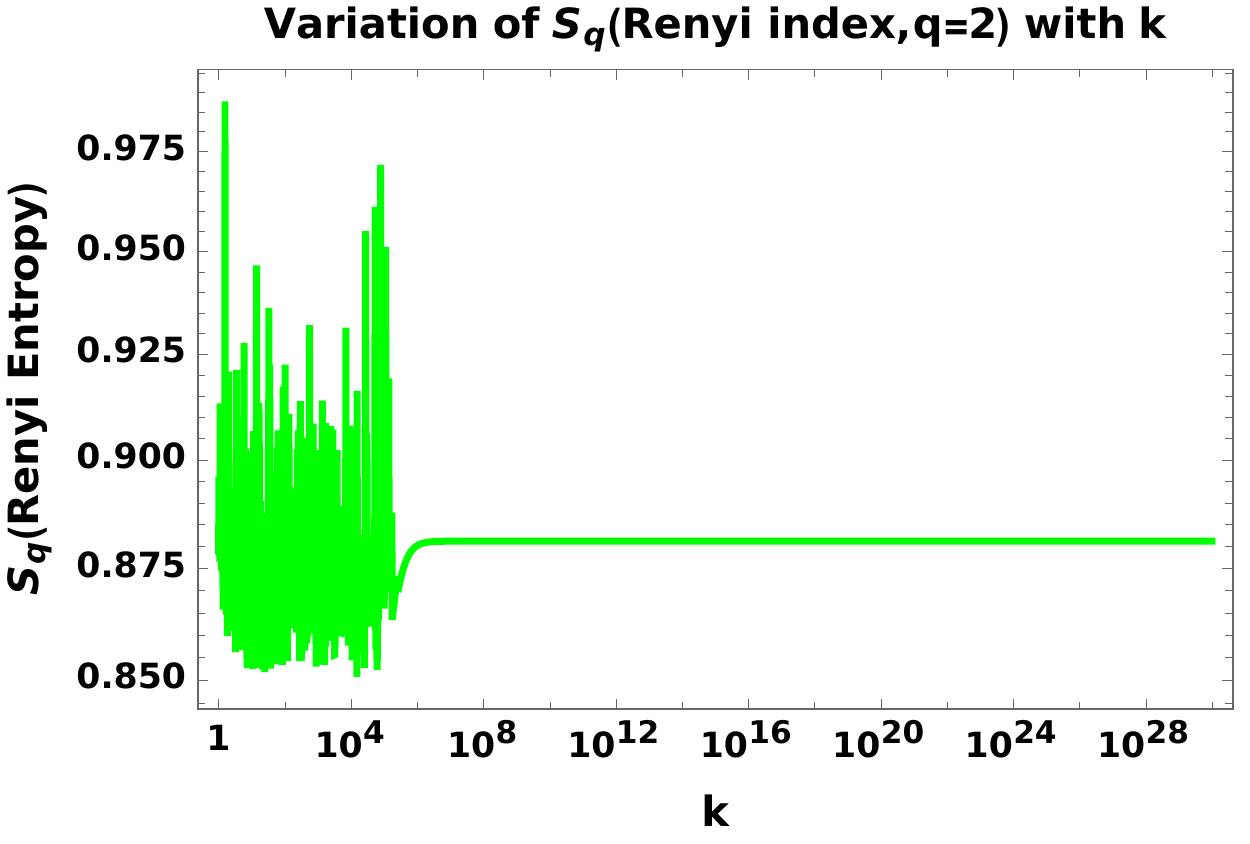}
	\label{colk}
	}
	\caption{Collision entropy variation with various parameters are shown here.}
\end{figure}

\begin{figure}[htb]
\centering
\subfigure[Min entropy vs Time profile.]{
	\includegraphics[width=7.8cm,height=4cm] {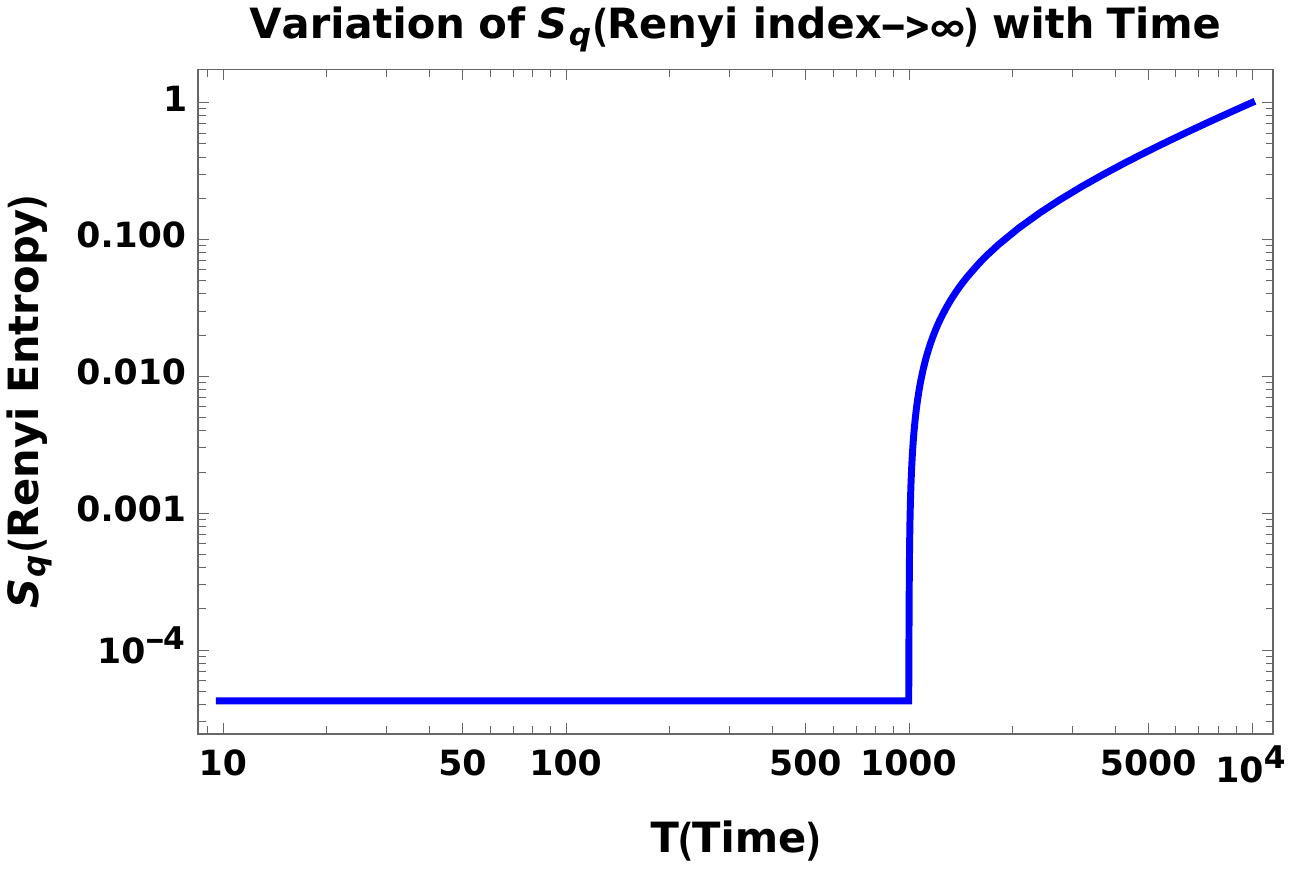}
	\label{mint}
	}
\subfigure[Min entropy vs $|Frequency|$ profile.]{
	\includegraphics[width=7.8cm,height=4cm] {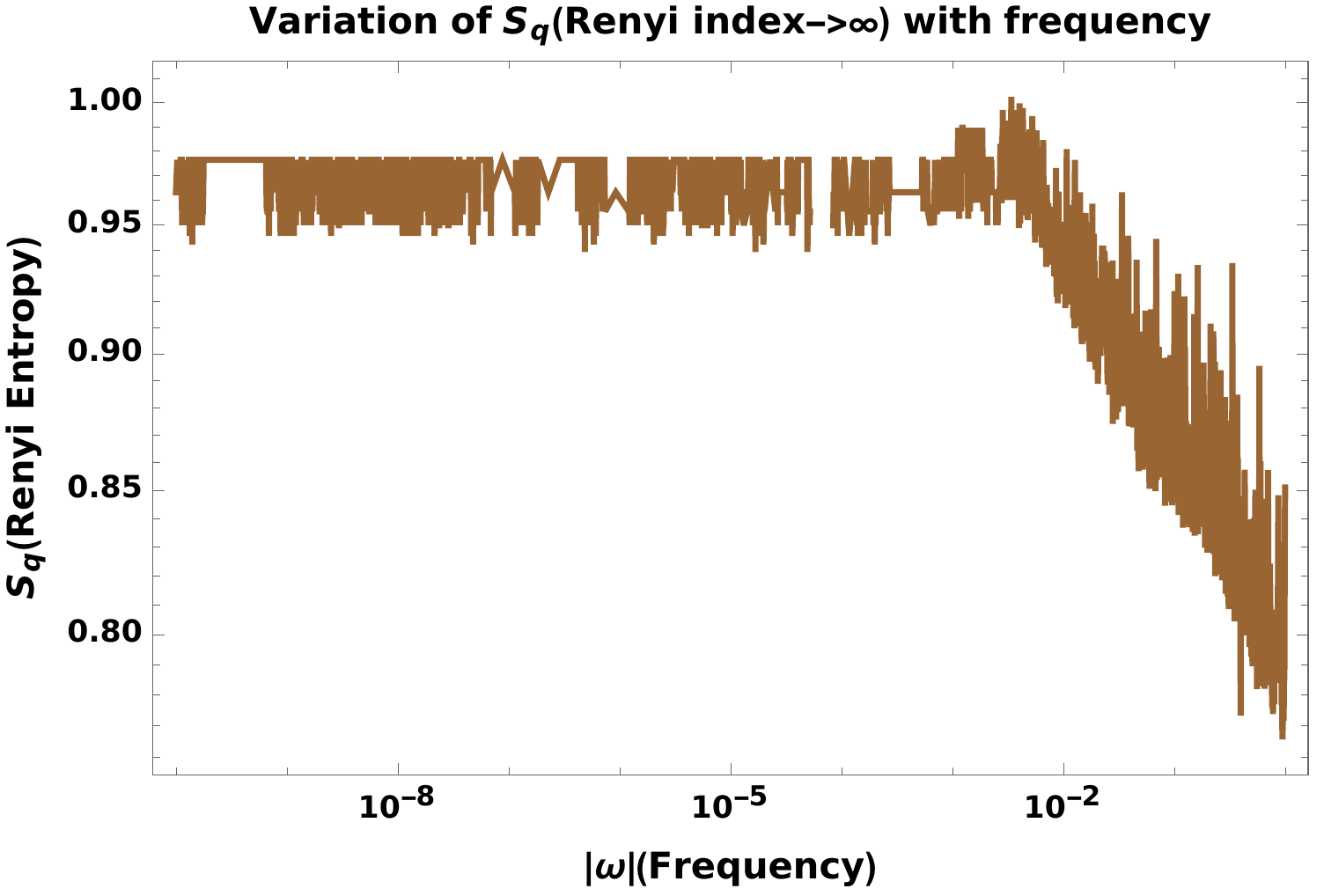}
	\label{minw}
	}
	\subfigure[Min Entanglement entropy vs Euclidean distance profile.]{
	\includegraphics[width=7.8cm,height=4cm] {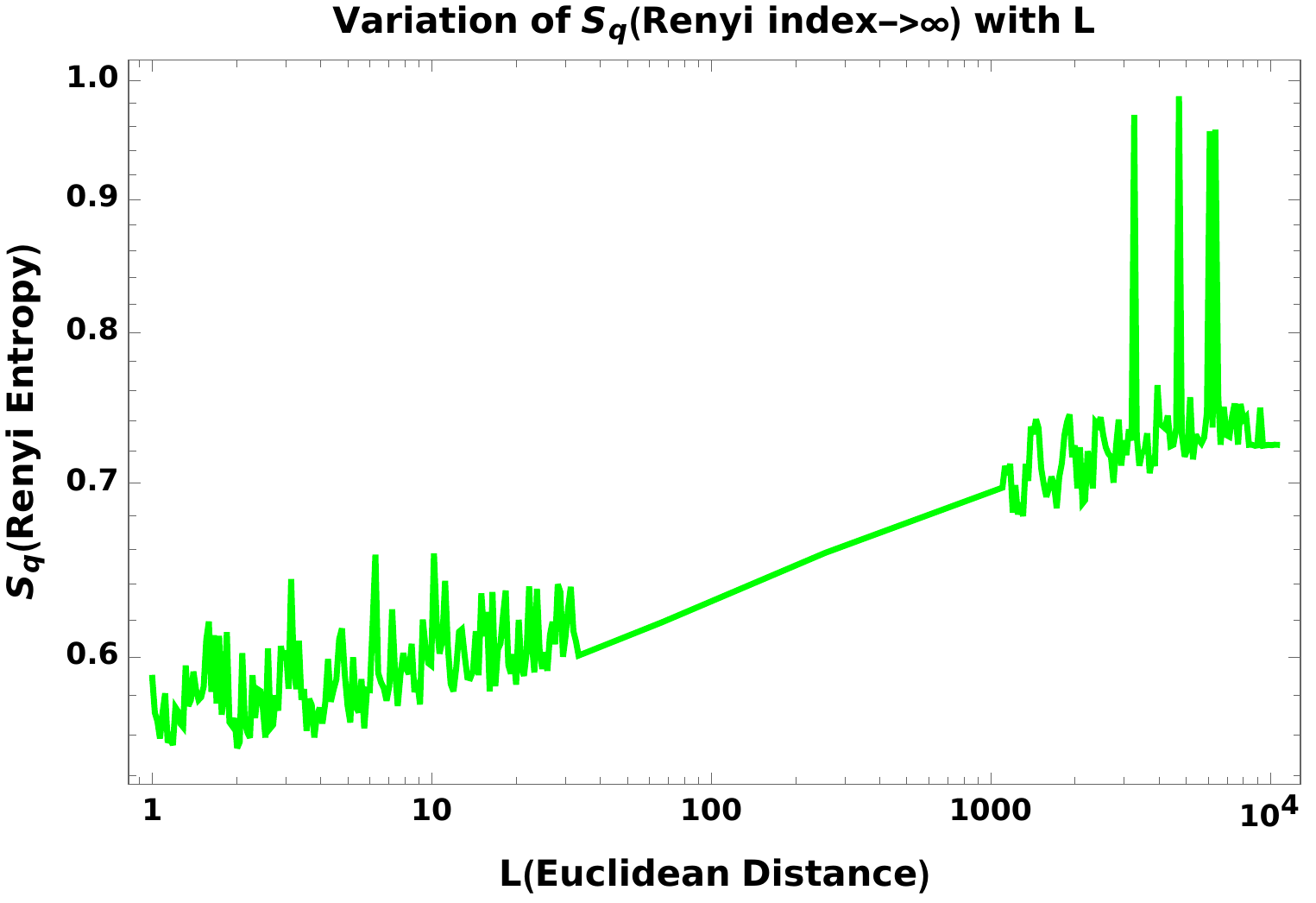}
	\label{minl}
	}
	\subfigure[Min Entanglement entropy vs k profile.]{
	\includegraphics[width=7.8cm,height=4cm] {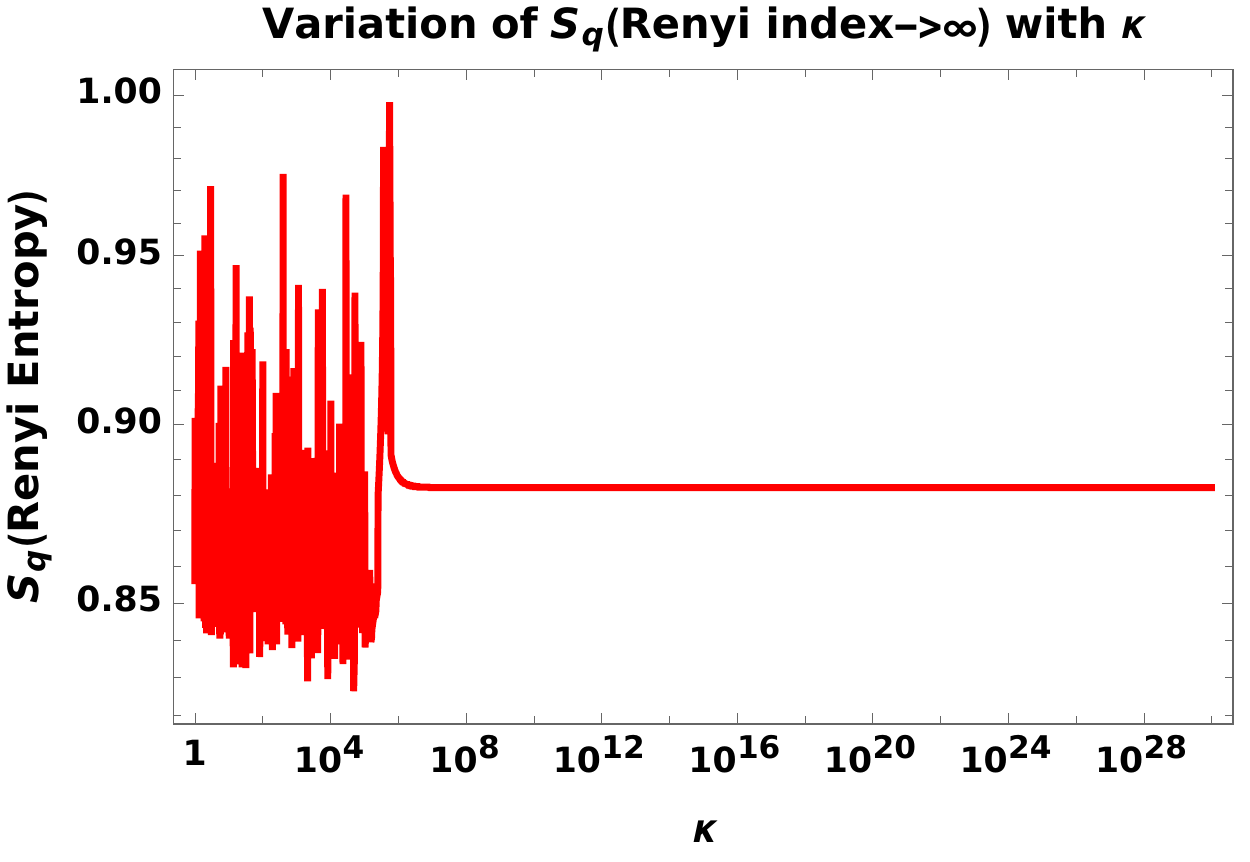}
	\label{mink}
	}
	\caption{Min entropy variation with various parameters are shown here.}
\end{figure}
\begin{figure}[htb]
\centering
{
	\includegraphics[width=12cm,height=5cm] {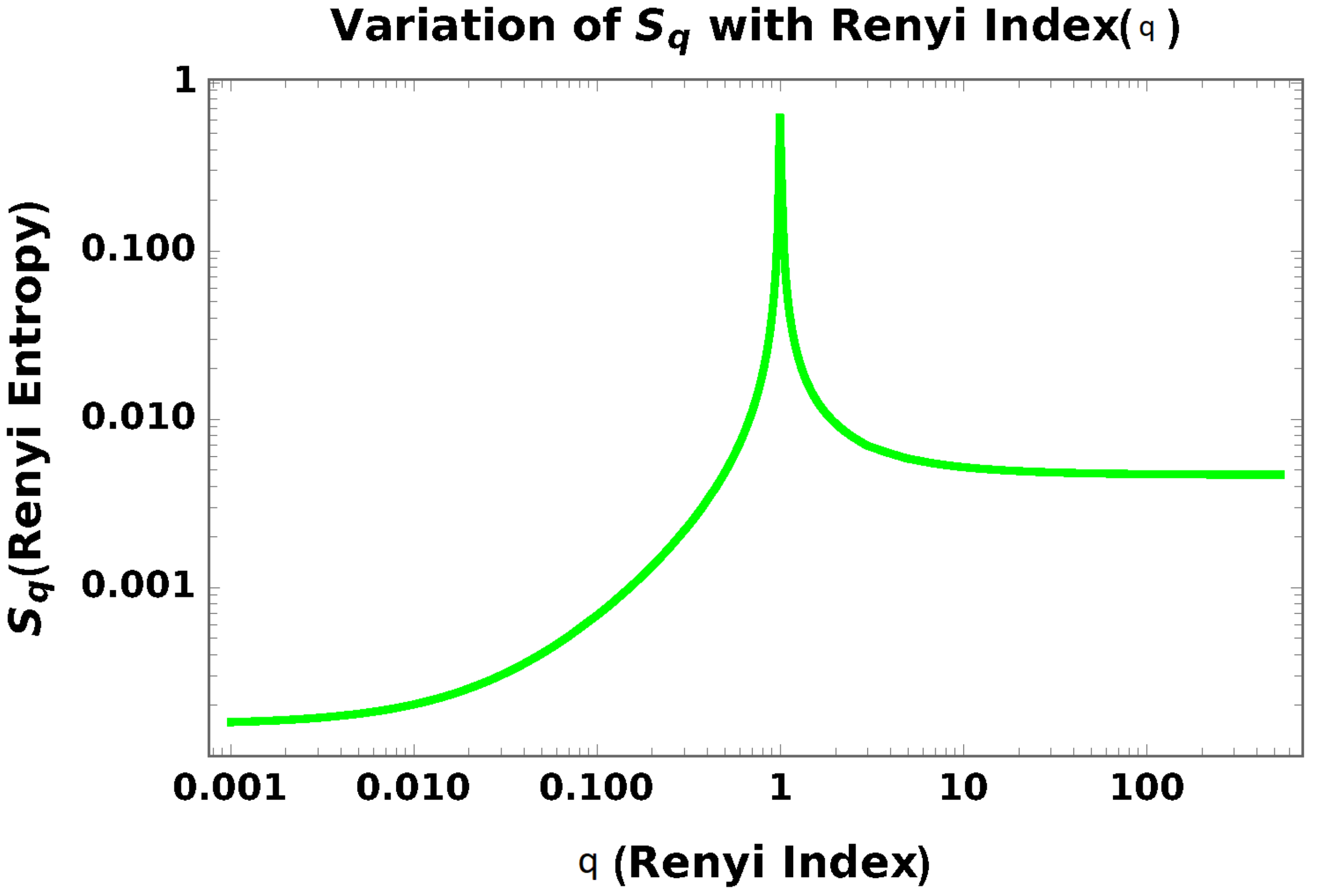}
	}
	\caption{Renyi Entropy variation with Renyi Index is shown here.}
\end{figure}

\subsection{\textcolor{blue}{ Logarithmic Negativity}}
\label{logneg}

Let us start with \textit{Peres-Horodeski criterion} which is the necessary condition for the joint density matrix of two quantum mechanical system A and B to be separable. This is sometimes called \textit{Positive Partial Transpose (PPT) criterion}. This is mainly used to decide the separability of mixed states, where \textit{Schmidt decomposition} does not apply. Using this criteria one can define\textit{Logarithmic Negativity} as:
\be
E_{N}(\rho_{\bf System} )=\ln||\rho^{T_A}||
\ee
where, $||\rho^T_A||$ is the trace norm and is defined as:
\bea
||\rho^T_A|| &=& {\rm Tr} \left( \sqrt{ (\rho^T_A)^\dagger (\rho^T_A)}\right)=\sum_i |\lambda_i|= \sum_{\lambda_i>0} \lambda_i+\sum_{\lambda_i<0}|\lambda_i|=2\sum_{\lambda_i<0}|\lambda_i|+1=2N+1.~~~~~~
\eea
Then the Logarithmic Negativity can be recast as:
\be E_{N}(\rho_{\bf System} )=\ln(2N+1).\ee
If we fix $N=0$ then it represents no entanglement in the quantum system. For this non entangled case one can write the system density matrix as:
\be
\rho_{\bf System} =\sum_i \lambda_i \rho_i^A \otimes \rho_i^B=\rho^{T_A},\ee
where we define the sub system density matrix as:
\be \rho_i^q=|i\rangle_{q}{}_{q}\langle i| ~~\forall ~~q=A,B~~~{\rm  with}~~ \lambda_i \geq 0
\ee
For $N\neq 0$ which represents the entangled case the system density matrix can be expressed as:

\be
\rho_{\bf System} = \sum_{i,j,k,l} {\cal C}_{ijkl}\left(|i\rangle_{AA}\langle j|\right) \otimes \left(|k\rangle_{AA}\langle l|\right)=\sum_{i,j,k,l} {\cal C}_{ijkl}(|j\rangle_{AA}\langle i|) \otimes (|k\rangle_{AA}\langle l|)
\ee

So from this computation  if we get negative eigen values for a quantum mechanical system then we can say that the corresponding quantum states are entangled.

For the case of two entangled atoms, using the solutions of the bloch vector components, the logarithmic negativity is given by
\bea
E_N(\rho)&=&\ln\left[\frac{17}{100}\left\{(2+8a^2_{03})+a_{33}(4+2a_{33})+a^2_{-+}+a^2_{+-}\right.\right.\nonumber\\&& \left.\left. .~~~~~~~~~ -\sqrt{(4+10a_{33}+(a_{-+}-a_{+-})^2)(16a^2_{03})+(a_{-+}+a_{+-})^2)}\right\}^{1/2}
\right. \nonumber\\ && \left.
+\frac{17}{100}\left\{(2+8^2a_{03})+a_{33}(4+2a_{33})+a^2_{-+}+a^2_{+-}\right.\right. \nonumber\\&& \left.\left..~~~~~~~~~ +\sqrt{(4+10a_{33}+(a_{-+}-a_{+-})^2)(16a^2_{03}+(a_{-+}+a_{+-})^2)}\right\}^{1/2}
\right. \nonumber \\ && \left. 
+\frac{17}{100}\left\{(2+a_{33}(-4+2a_{33})+a^2_{--}+a^2_{++}\right.\right. \nonumber \\ && \left.\left. .~~~~~~~~~ -\sqrt{(4+6a_{33}+(a_{--}-a_{++})^2)((a_{--}+a_{++})^2))}\right\}^{1/2}
\right.  \nonumber\\ && \left. 
+\frac{17}{100}\left\{(2+a_{33}(-4+2a_{33})+a^2_{--}+a^2_{++} \right. \right. \nonumber \\ &&\left.\left.  +\sqrt{(4+6a_{33}+(a_{--}-a_{++})^2)(4(\beta^2-a_{-+}a_{+-}+(a_{--}+a_{++})^2))}\right\}^{1/2}\right]
\eea

\begin{figure}[htb]
\centering
\subfigure[Log Negativity vs Time profile.]{
	\includegraphics[width=7.8cm,height=4cm] {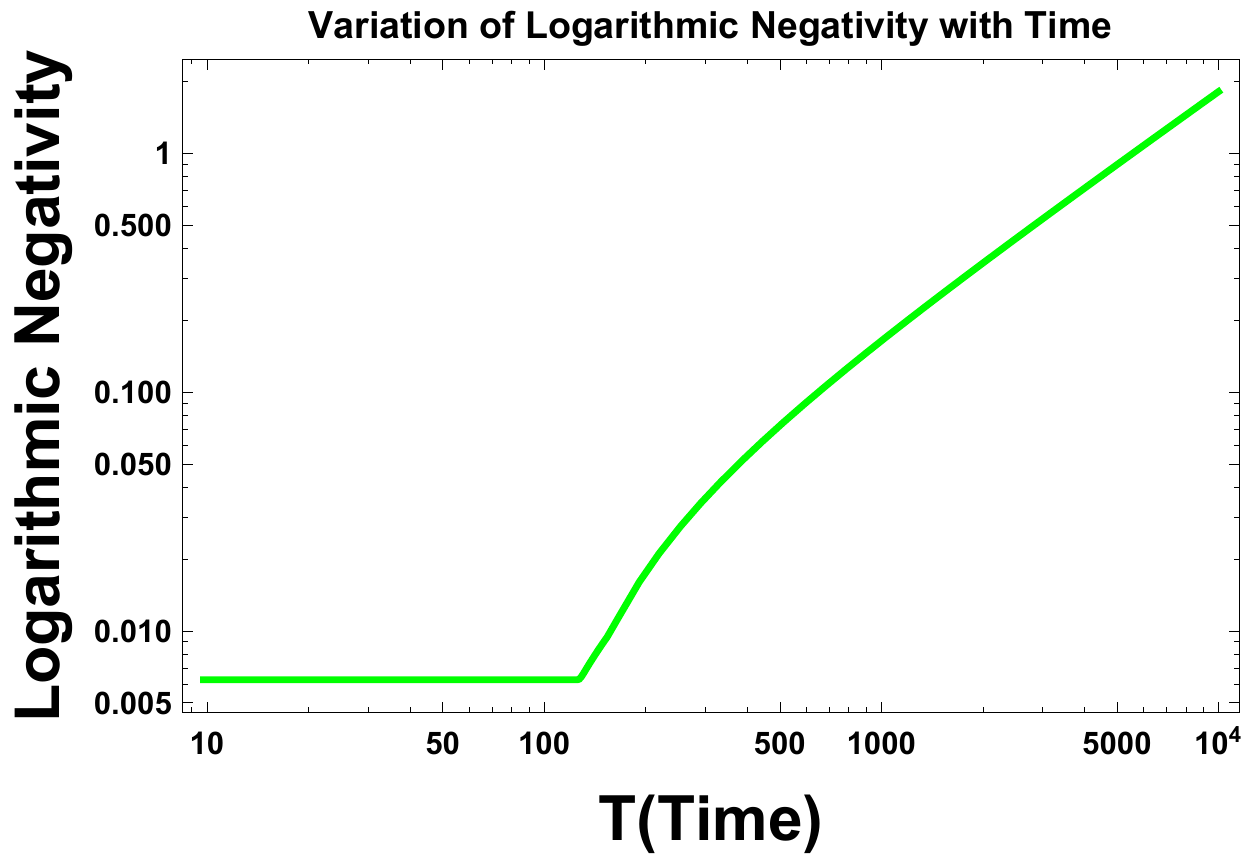}
	\label{lognt}
	}
\subfigure[Log Negativity vs $\rm |Frequency|$ profile..]{
	\includegraphics[width=7.8cm,height=4cm] {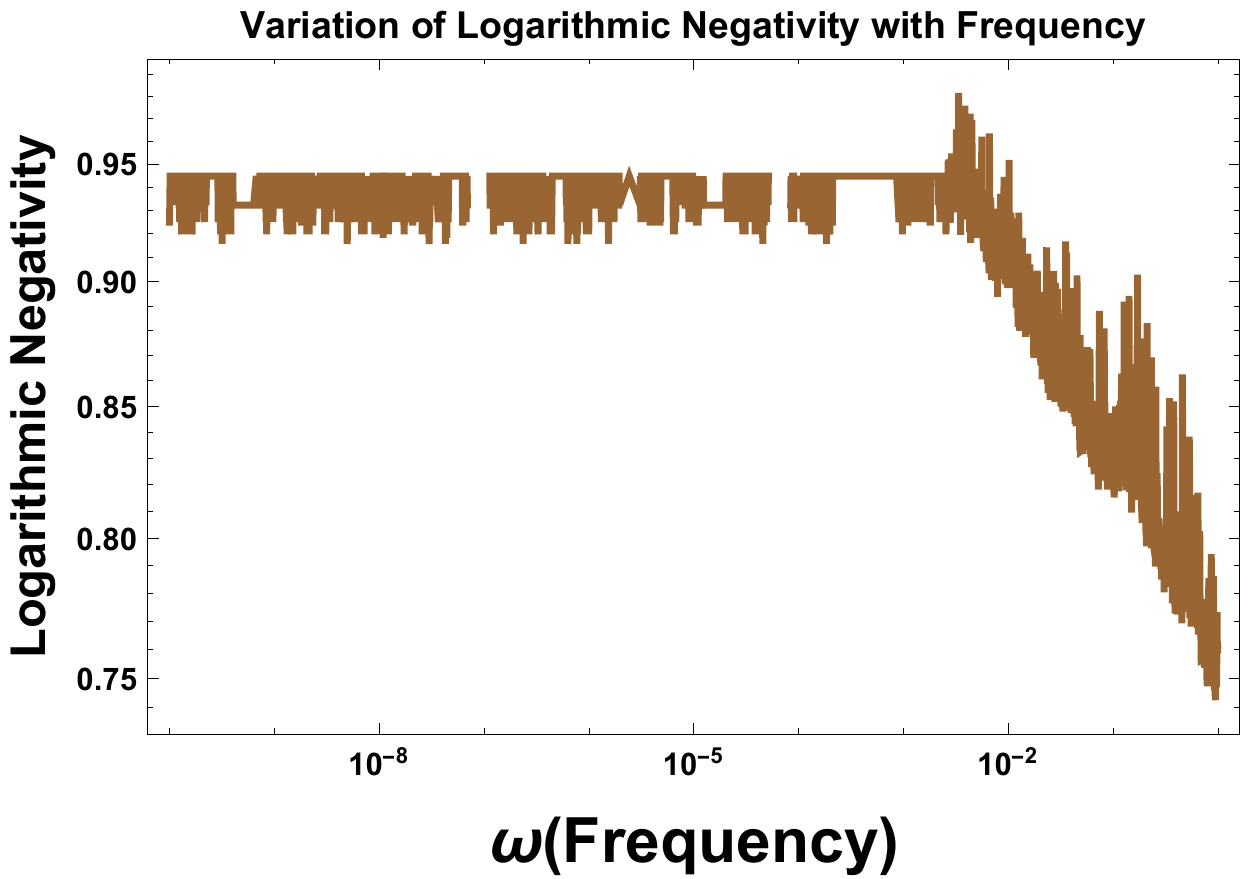}
	\label{lognw}
	}
	\subfigure[Log Negativity vs Euclidean Distance profile.]{
	\includegraphics[width=7.8cm,height=4cm] {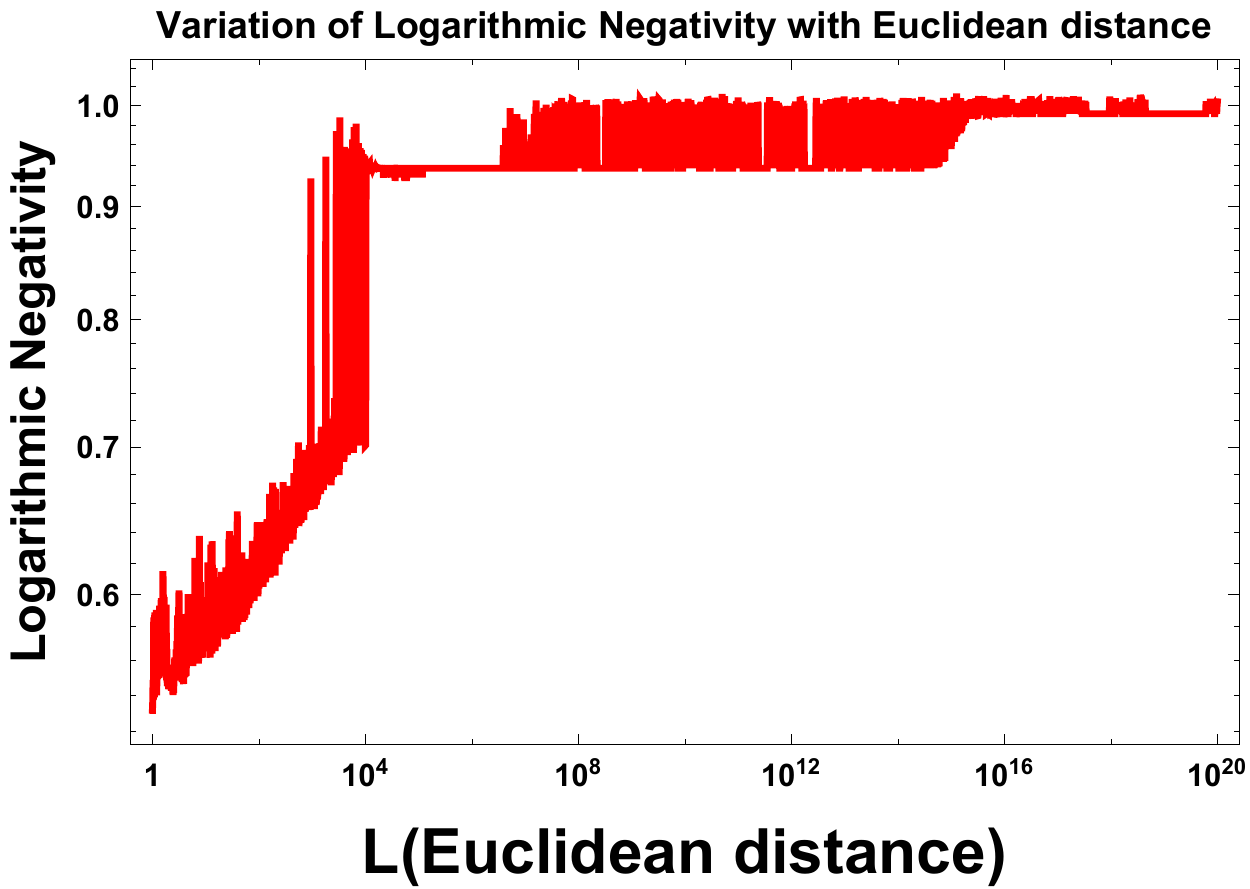}
	\label{lognl}
	}
	\subfigure[[Log Negativity vs k profile.]{
	\includegraphics[width=7.8cm,height=4cm] {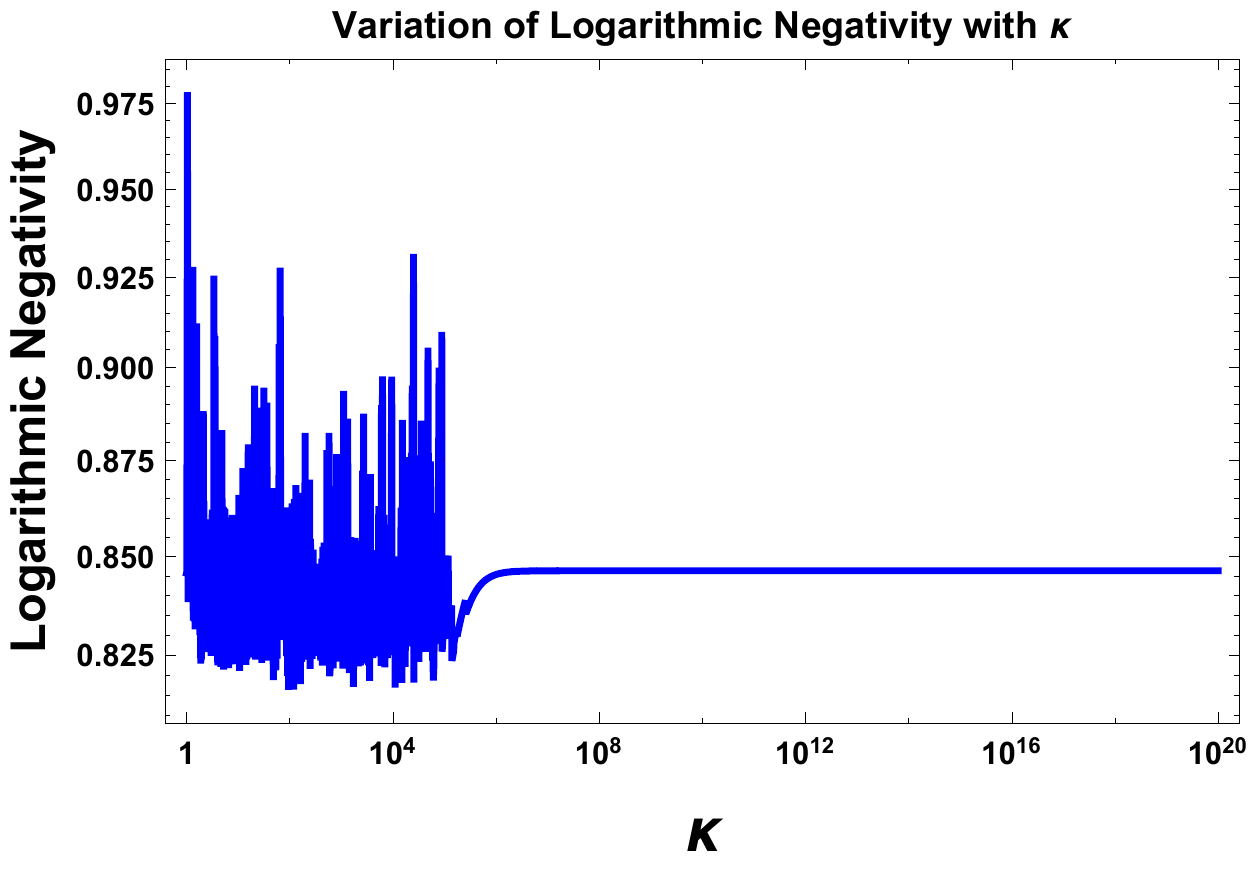}
	\label{lognk}
	}
	\caption{Log Negativity variation with various parameters are shown here.}
\end{figure}

\subsection{\textcolor{blue}{ Entanglement of formation and Concurrence }}
\label{concur}
Both of the measures studied in earlier sections are used to quantify the resources needed to create a given entangled state. Each of them are used as an entanglement measure for bipartite quantum state in quantum information theory. The \textit{entanglement of formation} for pure and mixed states takes the following form:
 
  \be\begin{array}{lll}\label{kg0}
\displaystyle E_{f}(\rho) =\left\{\begin{array}{ll}
\displaystyle   -{\rm Tr}(\rho_{A}\ln \rho_{A})=- {\rm Tr}(\rho_{B}\ln \rho_{B})~~~~ & \mbox{\small {\bf for ~pure~state}}  \\
\displaystyle  
{\bf inf}\left(\sum_j p_jE_f(\Phi_j)\right)~~~~ &
\mbox{\small {\bf for ~mixed~state}}  \\ 
 \end{array}
\right.
\end{array}\ee

For mixed states the infimum is taken over all possible decompositions of the density matrix $\rho$ into pure states.		      		   		      		   
The quantities used in the above equation are defined below:
\bea
\rho_A&=& Tr_B \rho,~~~
\rho_B= Tr_A \rho,
~~~
\rho_{\bf System}= |\Phi\rangle \langle \Phi|~~~~~~~ \mbox{\small {\bf for ~pure~state}}
\\
E_f(\Phi_j)&=&-{\rm Tr}(\Phi_jln \Phi_j),
~~~~
\rho_{\bf System}= \sum_j p_j |\Phi_j\rangle \langle \Phi_j|~~~~~~ \mbox{\small {\bf for ~mixed~state}}
\eea

 Relation between entanglement of formation and \textit{Concurrence} can be written as:
\be
E_f(\rho_{\bf System})=\mathcal{E}(C(\rho_{\bf System}))=h\left(\frac{1+\sqrt{1-C^2(\rho_{\bf System})}}{2}\right)
\ee

Where $h(x)$ is known as the {\bf Binary Entropy function} which is defined as:

\be
h(x)=-x\ln x-(1-x)\ln(1-x)
\ee
\begin{figure}[htb]
\centering
\subfigure[Binary entropy function h(x) vs x]{
	\includegraphics[width=7.8cm,height=4.5cm] {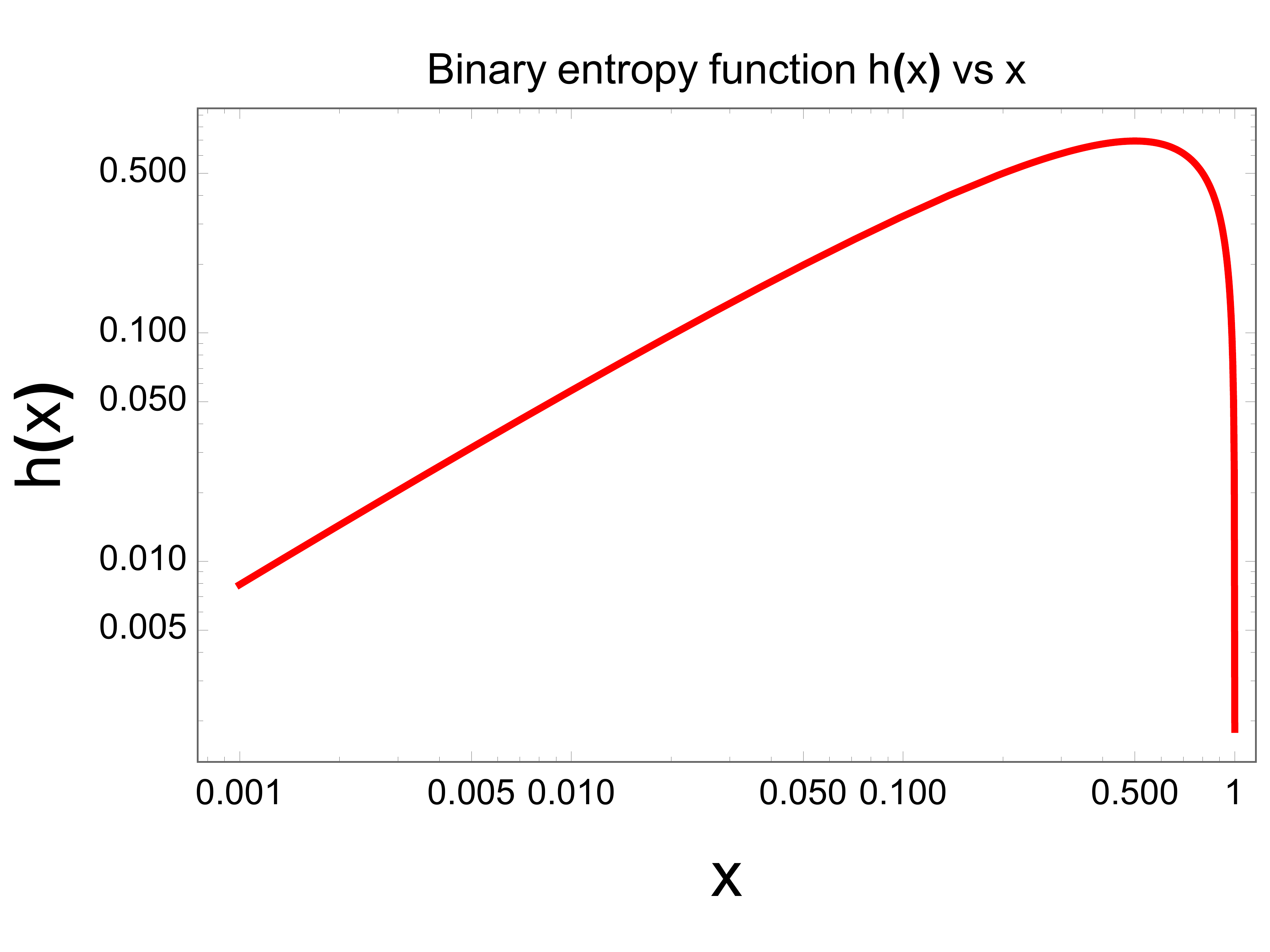}
	\label{11a}
	}
\subfigure[Binary entropy function h(x) vs x]{
	\includegraphics[width=7.8cm,height=4.5cm] {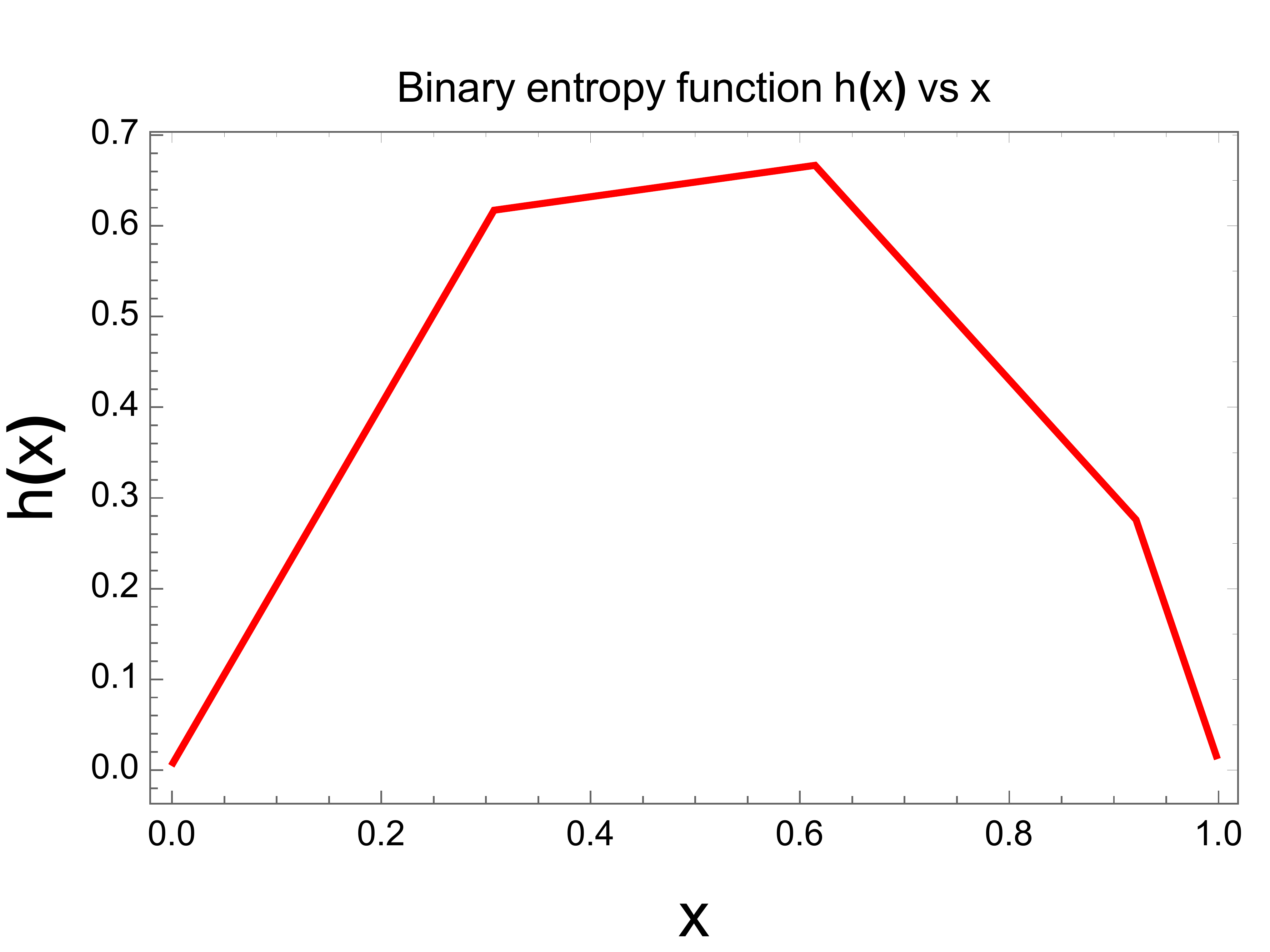}
	\label{11b}
	}

	\caption{Variation of the binary entropy function $h(x)$ with $x$ is shown here.}
\end{figure}
Here we study about concurrence in the context of two entangled atoms from the perspective of OQS. For X type states it is defined analytically by the following expression:
\be
\label{conn}
C(\rho_{\bf System})={\rm max}[0,\lambda_1-\lambda_2-\lambda_3-\lambda_4]
\ee
where the $\lambda_i$'s are the square roots of the eigenvalues of the matrix:
 \be\begin{array}{lll}\label{kg0}
		      		   		      		    \displaystyle \mathcal{R}\equiv\left\{\begin{array}{ll}
		      		   		      		                       \displaystyle  \sqrt{\sqrt{\rho_{\bf System}}~\tilde{\rho}_{\bf System}~\sqrt{\rho_{\bf System}}}~~~~ &
		      		   		      		    \mbox{\small {\bf for ~Hermitian}}  \\ 
		      		   		      		   	\displaystyle  
		      		   		      		   	\sqrt{\rho_{\bf System} ~\tilde{\rho}_{\bf System}}~~~~ &
		      		   		      		    \mbox{\small {\bf for ~Non~Hermitian}} \\ 
		      		   		      		             \end{array}
		      		   		      		   \right.
		      		   		      		   \end{array}\ee	
where $\tilde{\rho}_{\bf System}$ is the {\it spin flip (Werner type) quantum states} \cite{Werner:1989zz} given by~\footnote{\underline{\textcolor{red}{\bf Werner State:}}~A Werner state is a $d \times d$ dimensional bipartite quantum state density matrix that is invariant under all unitary operators of the form ($U\otimes U$). That is, it is a bipartite quantum state that satisfies the following condition:
\be \rho_{AB}=(U\otimes U)\rho_{AB}(U^\dagger \otimes U^\dagger,)\ee  
 for all unitary operators $U$ acting on $d$-dimensional Hilbert space.}: 
\bea
\label{rht}
\tilde{\rho}_{\bf System}&=&(\sigma_2 \otimes \sigma_2) \rho^*(\sigma_2 \otimes \sigma_2)=[((\sigma_--\sigma_+)\otimes(\sigma_--\sigma_+)\rho^*(\sigma_--\sigma_+)\otimes(\sigma_--\sigma_+))]~~~~~~~~
\eea
and $\rho^*_{\bf System}$ in the above expression indicates complex conjugate of $\rho_{\bf System}$.

The eigenvalues $\lambda$'s follow the following sequence:
\be
\lambda_1>\lambda_2>\lambda_3>\lambda_4,  {\rm where}~~~
\lambda_1-\lambda_2-\lambda_3-\lambda_4>0
\ee
\begin{figure}[htb]
\centering
\subfigure[Entanglement of formation vs Concurrence]{
	\includegraphics[width=7.8cm,height=5cm] {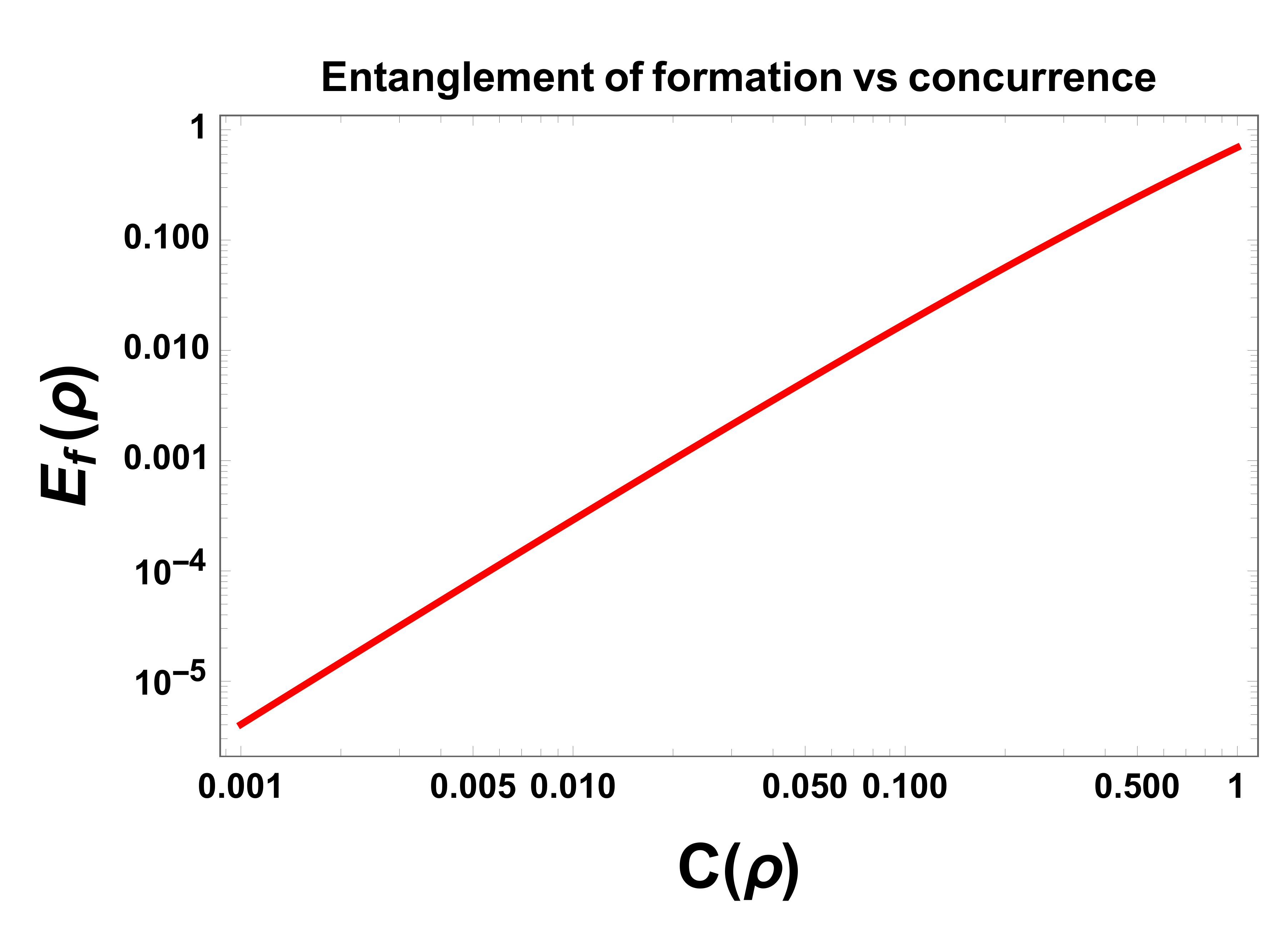}
	\label{12a}
	}
\subfigure[Entanglement of formation vs Concurrence]{
	\includegraphics[width=7.8cm,height=5cm] {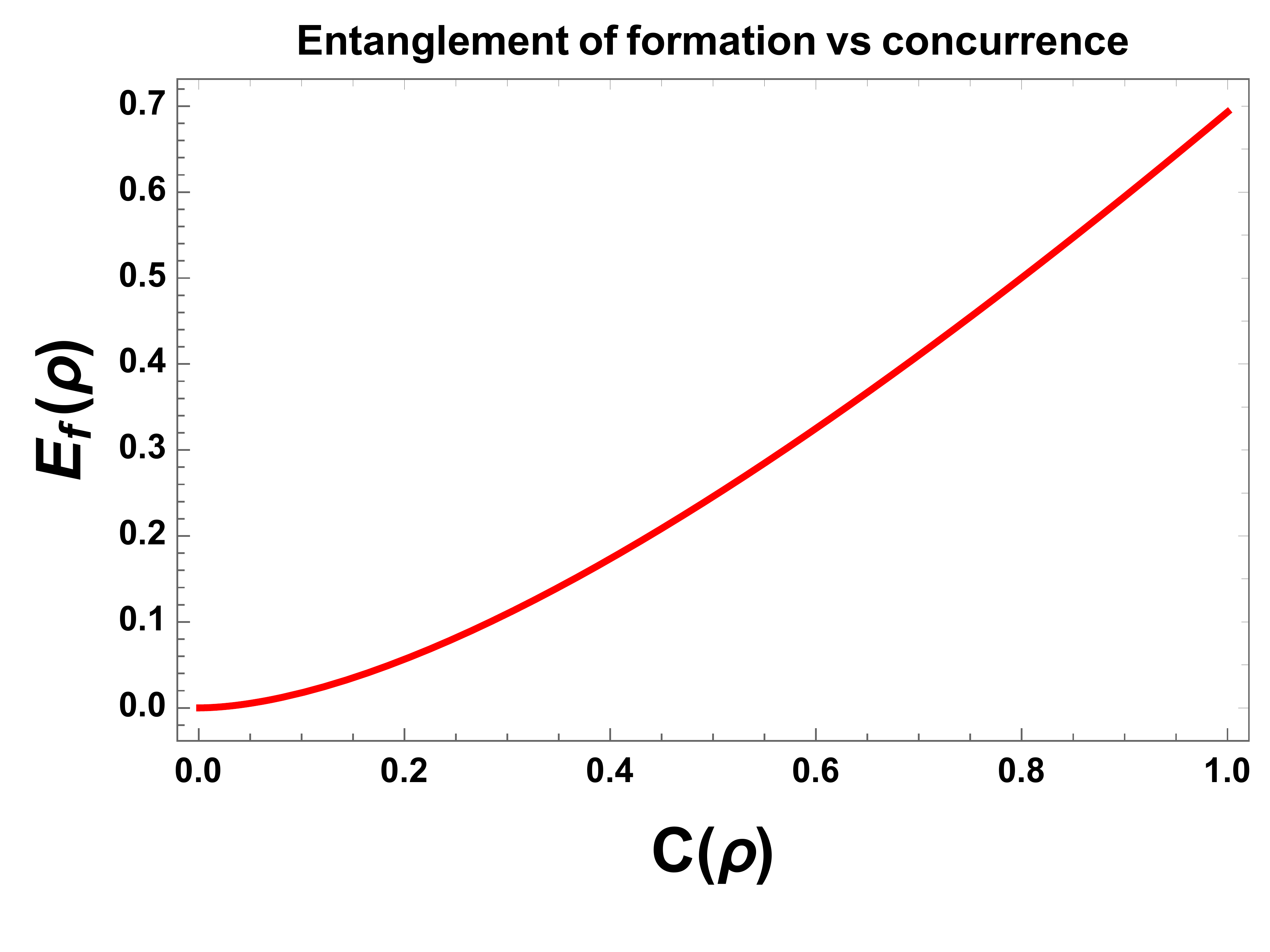}
	\label{12b}
	}

	\caption{Relation between the entanglement of formation and concurrence is shown here.}
\end{figure}

For the case of two entangled atoms relevant to our system the solutions of the Bloch vector components ${ a_{30}}$ and ${ a_{03}}$ obtained are identical. Thus the eigenvalues $\lambda_1$, $\lambda_2$, $\lambda_3$ and $\lambda_4$ are given by the following expressions:
\bea
\lambda_1&=&\frac{1}{4}\sqrt{{\cal S}-2{\cal F}}
,~~~
\lambda_2=\frac{1}{4}\sqrt{{\cal S}+2{\cal F}},~~~
\lambda_3=\frac{1}{4}\sqrt{{\cal X}-2{\cal U}},
~~~
\lambda_4=\frac{1}{4}\sqrt{{\cal X}+2{\cal U}}.~~~
\eea
where we define ${\cal S}$, ${\cal X}$, ${\cal F}$ and ${\cal U}$ as:

\be
\bal
{\cal S}&= 1-2a_{33}+a^2_{33}+a_{-+}a_{+-} \\
{\cal X}&=1-4a^2_{03}+2a_{33}+a^2_{33}+a_{--}a_{++} \\
{\cal F}&= \sqrt{a_{-+}a_{+-}-2a_{33}a_{-+}a_{+-}+a^2_{33}a_{-+}a_{+-}} \\
{\cal U}&= \sqrt{a_{--}a_{++}-4a^2_{03}a_{--}a_{++} +2a_{33}a_{--}a_{++}+a^2{33}a_{--}a_{++}} 
\eal
\ee

 Therefore, using equation \ref{conn} the concurrence for two entangled atoms relevant to our system can be calculated as:

\be
C(\rho_{\bf System})={\rm max}\left[0,\frac{1}{4}\left(\sqrt{{\cal S}-2{\cal F}}-\sqrt{{\cal S}+2{\cal F}}-\sqrt{{\cal X}-2{\cal U}}-\sqrt{{\cal X}+2{\cal U}}\right)\right]
\ee

\begin{figure}[htb]
\centering
\subfigure[Concurrence vs Time profile.]{
	\includegraphics[width=7.8cm,height=4cm] {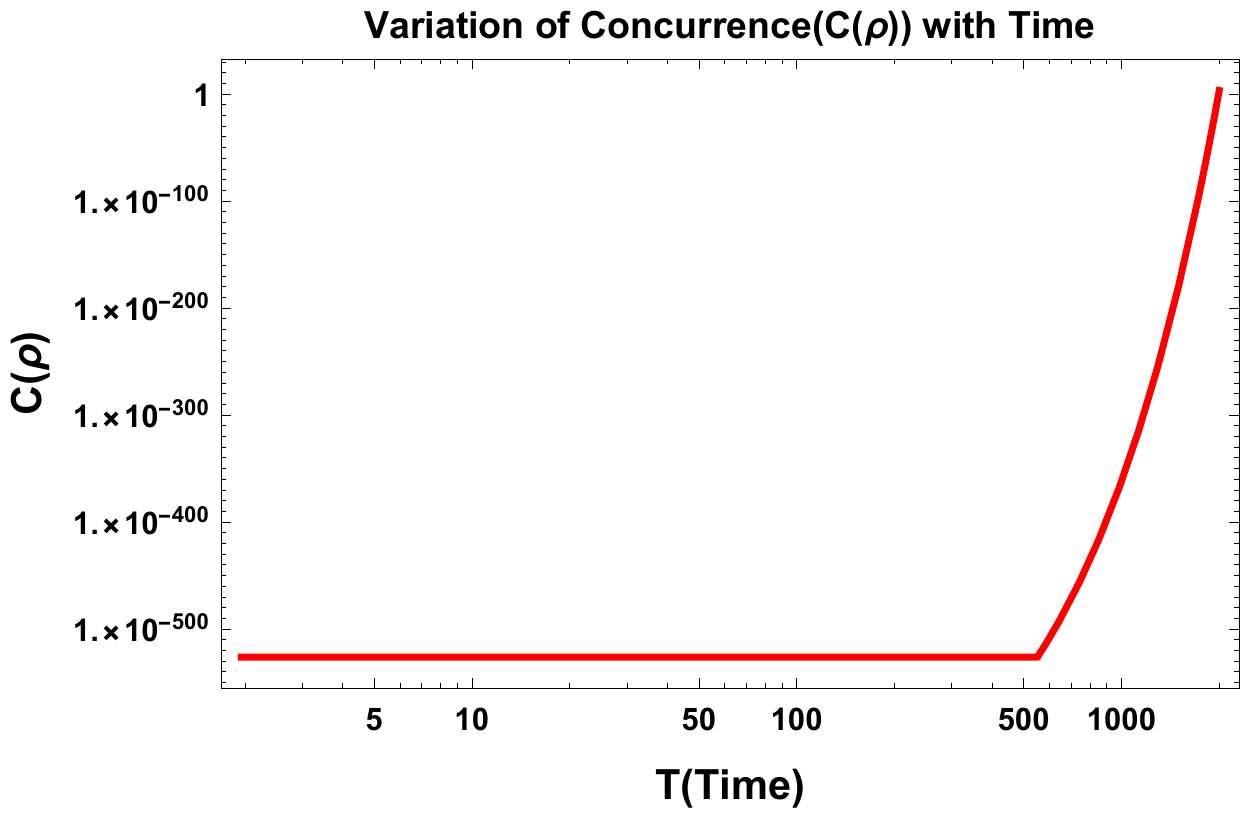}
	\label{concurt}
	}
\subfigure[Concurrence vs $|Frequency|$ profile.]{
	\includegraphics[width=7.8cm,height=4cm] {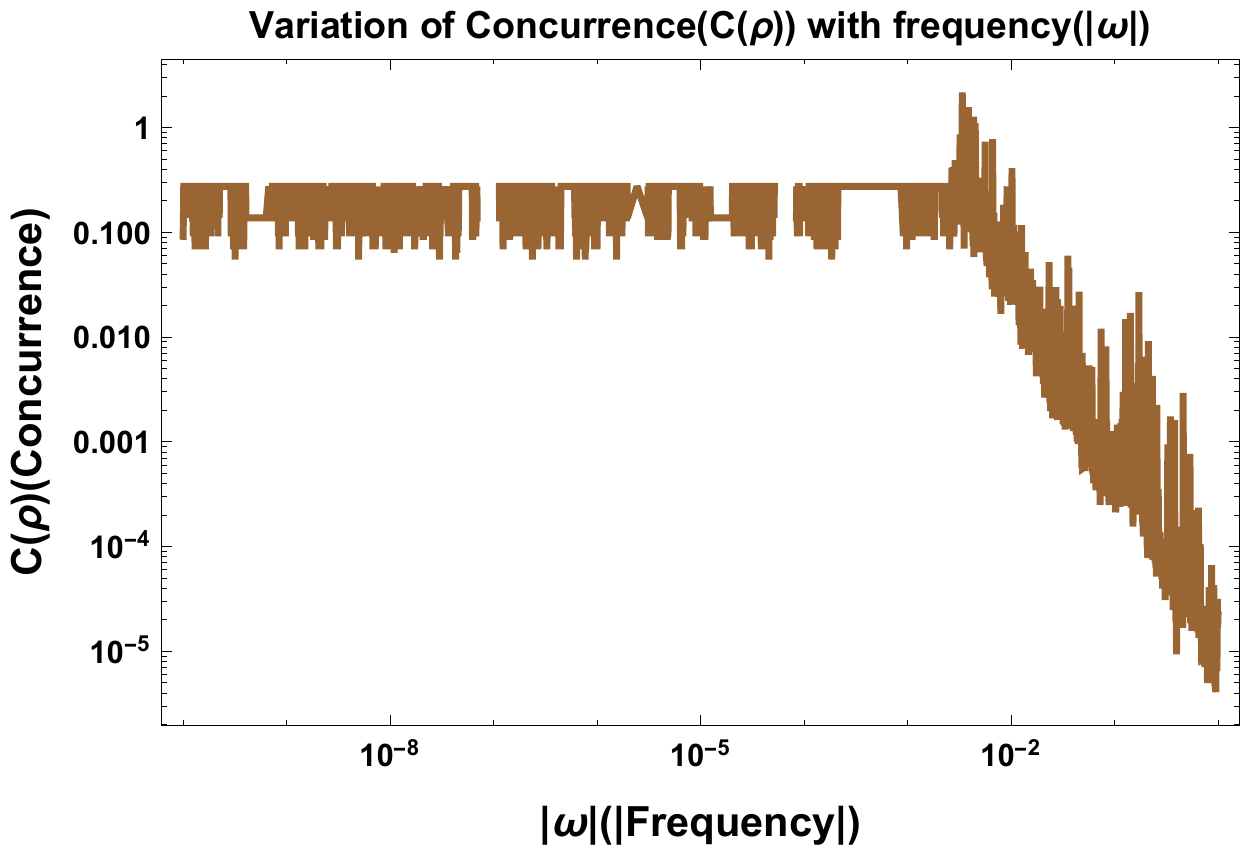}
	\label{concurw}
	}
	\subfigure[Concurrence vs Euclidean distance profile.]{
	\includegraphics[width=7.8cm,height=4cm] {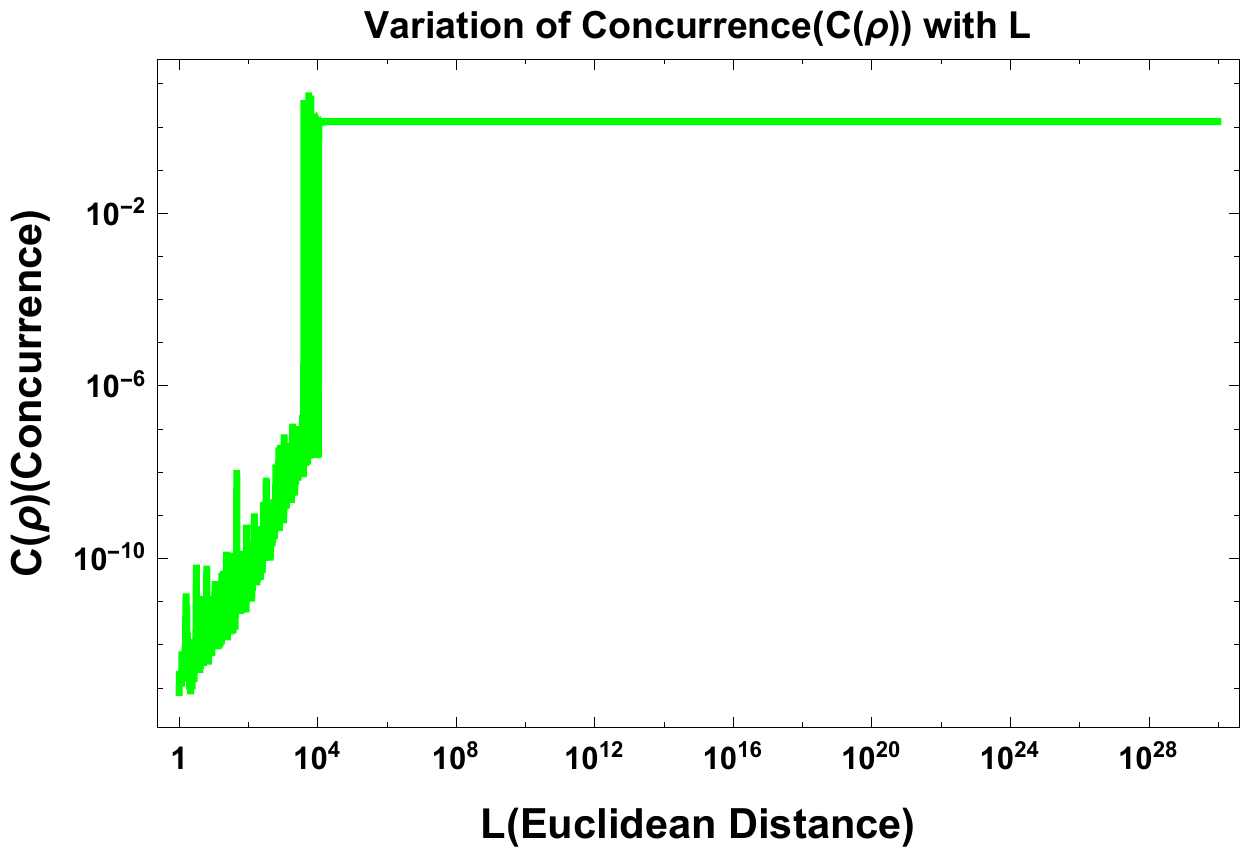}
	\label{concurL}
	}
	\subfigure[Concurrence vs k profile.]{
	\includegraphics[width=7.8cm,height=4cm] {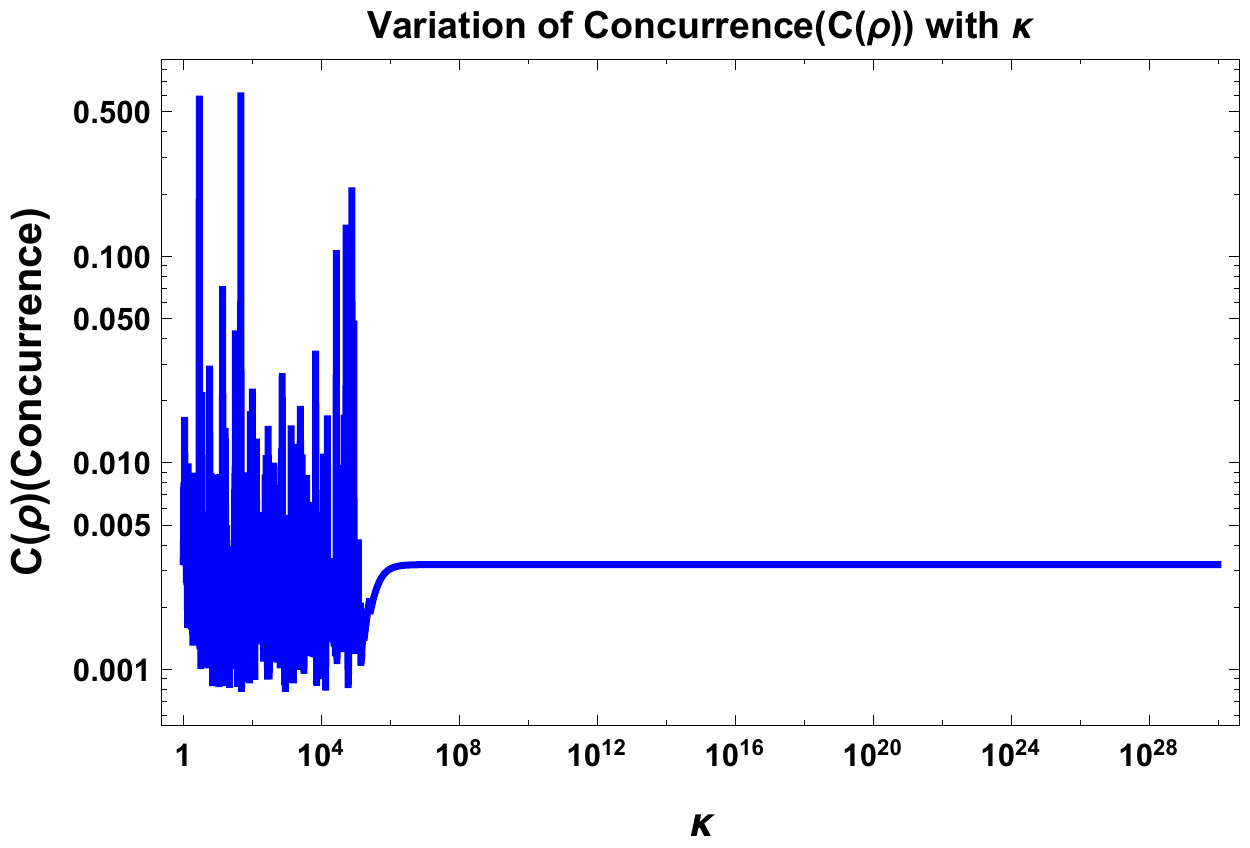}
	\label{concurk}
	}
	\caption{Variation of Concurrence with various parameters is shown here. }
\end{figure}

The explicit expression for the entanglement of formation for our system the is given by the following expression:
\bea
E_{f}(\rho)&=&\frac{1}{2}\left(-1-\sqrt{1-(C(\rho))^2}\right)\log \left[\frac{1}{2}\left(1+\sqrt{1-C(\rho)}\right)\right]\nonumber\\
&&~~~ -\left(1+\frac{1}{2}\left(-1-\sqrt{1-C(\rho)}\right)\right) \log \left[1+\frac{1}{2}\left(-1-\sqrt{1-C(\rho)}\right)\right]
\eea
The following section shows various plots of entanglement of formation calculated from concurrence of our model.

\begin{figure}[htb]
\centering
\subfigure[Entanglement of formation vs Time profile.]{
	\includegraphics[width=7.8cm,height=4cm] {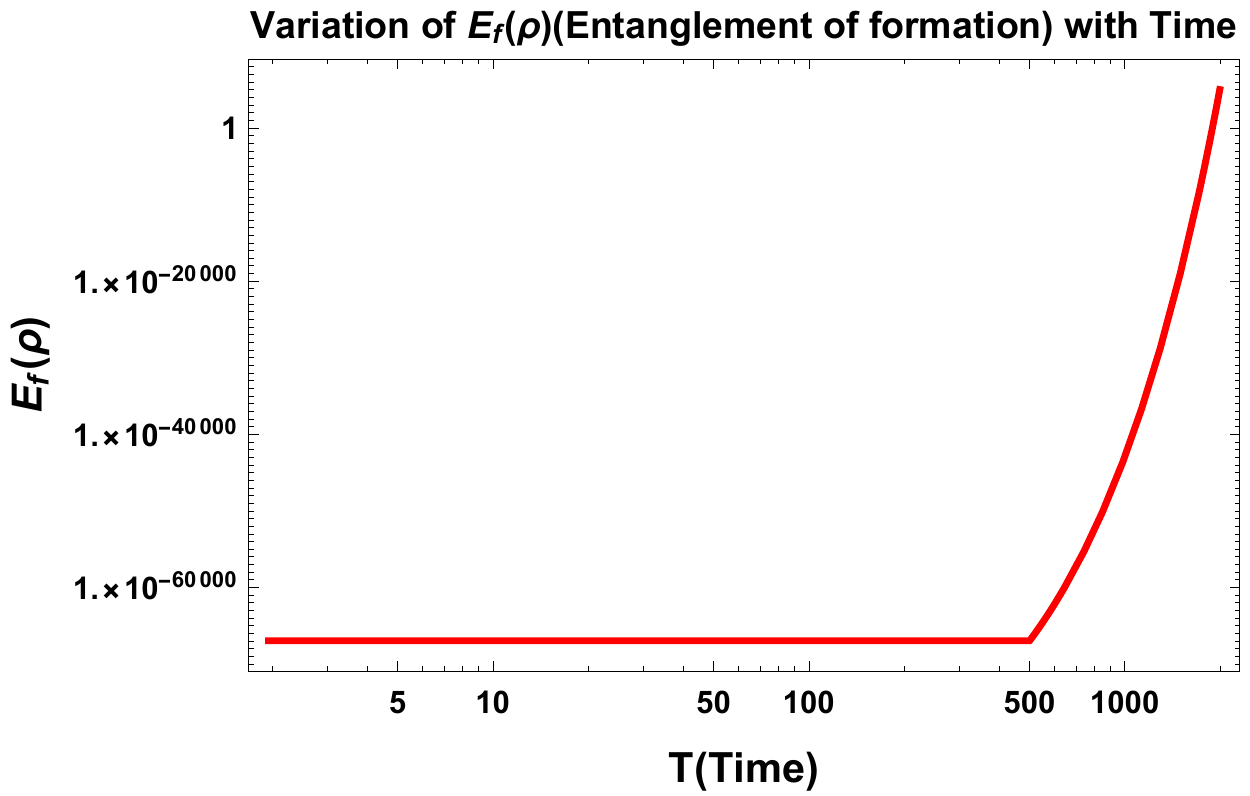}
	\label{eformt}
	}
\subfigure[Entanglement of formation vs $|Frequency|$ profile.]{
	\includegraphics[width=7.8cm,height=4cm] {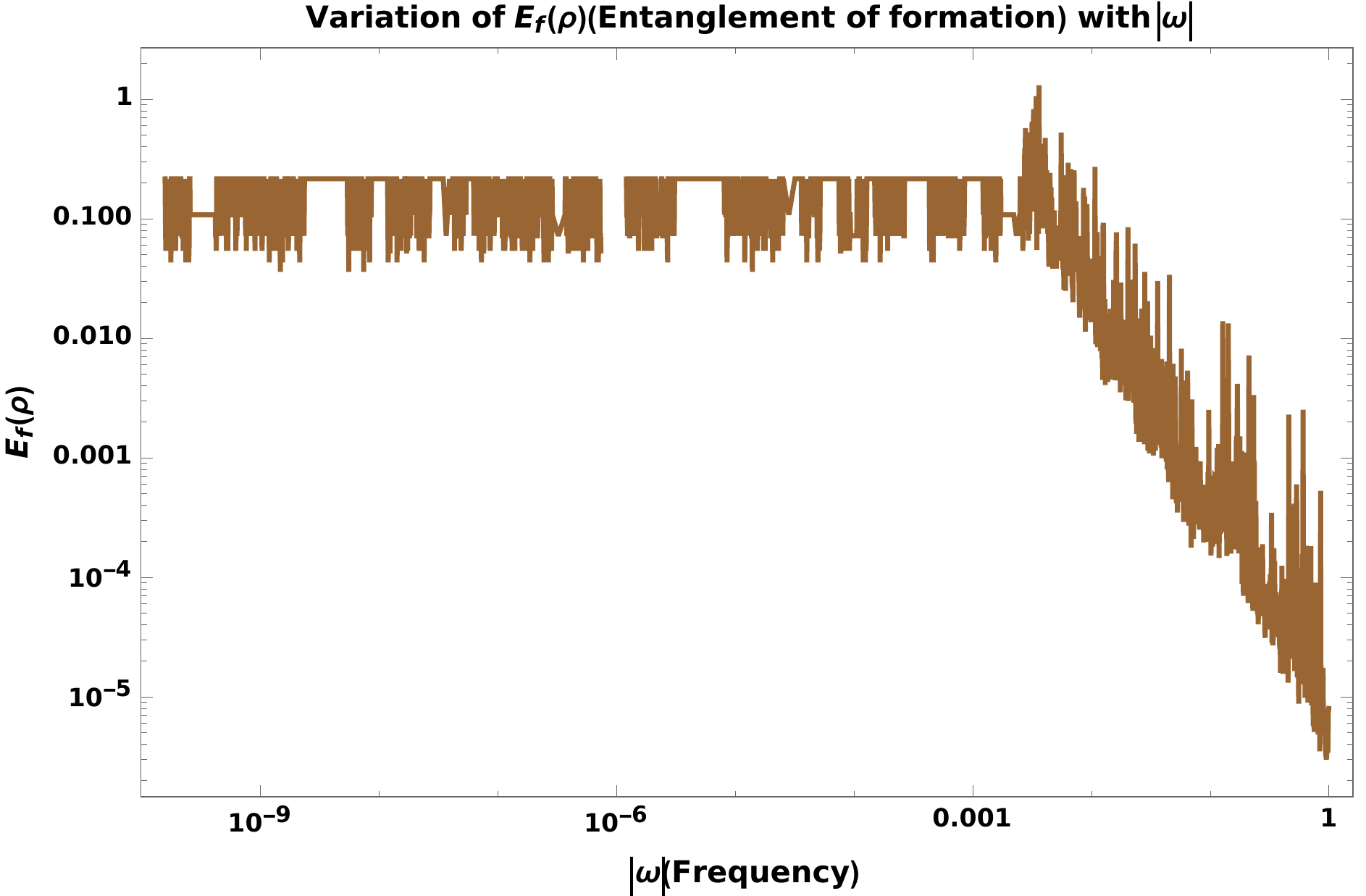}
	\label{eformw}
	}
	\subfigure[Entanglement of formation vs vs Euclidean distance profile.]{
	\includegraphics[width=7.8cm,height=4cm] {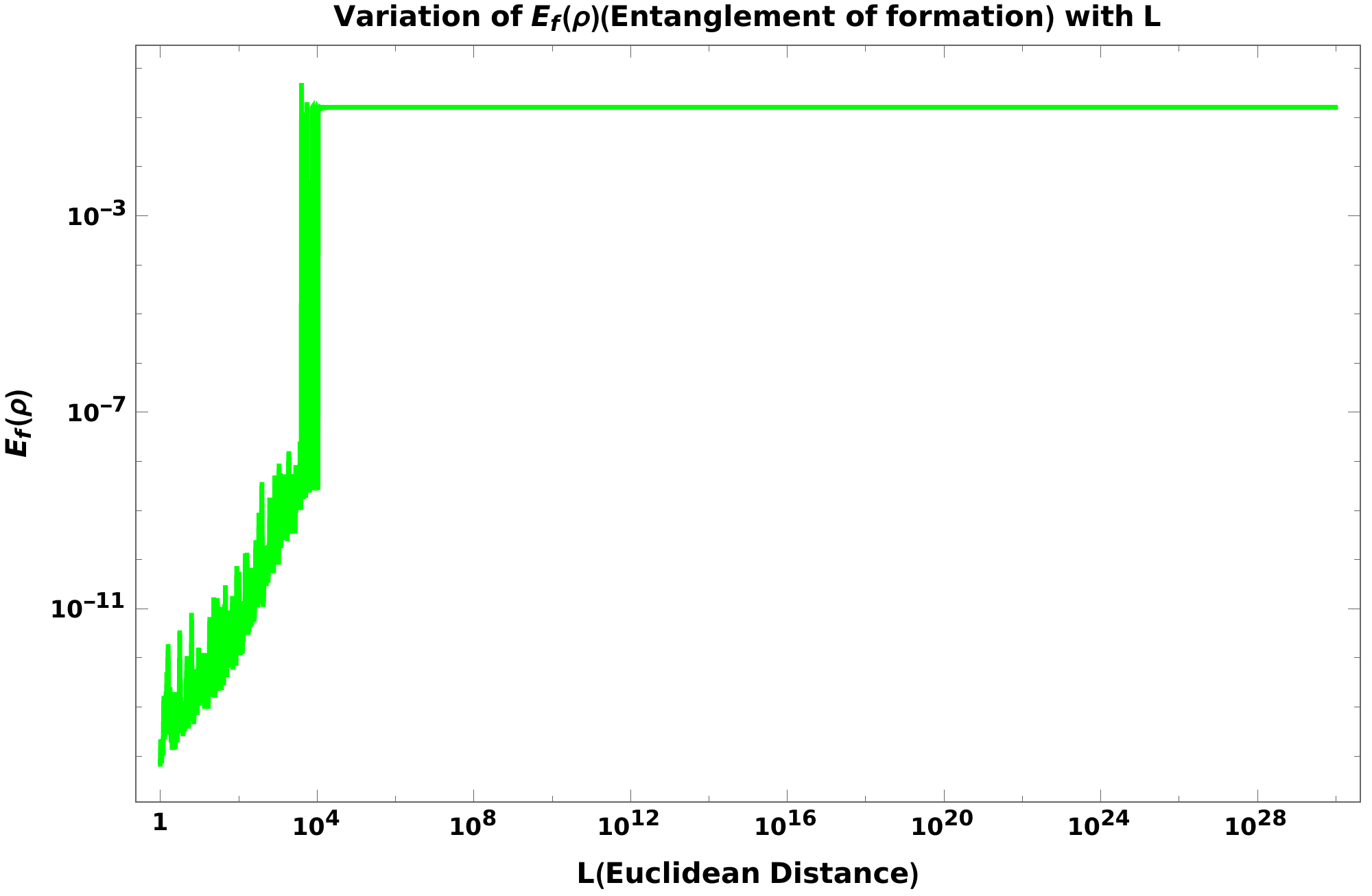}
	\label{eformL}
	}
	\subfigure[Entanglement of formation vs $k$ profile.]{
	\includegraphics[width=7.8cm,height=4cm] {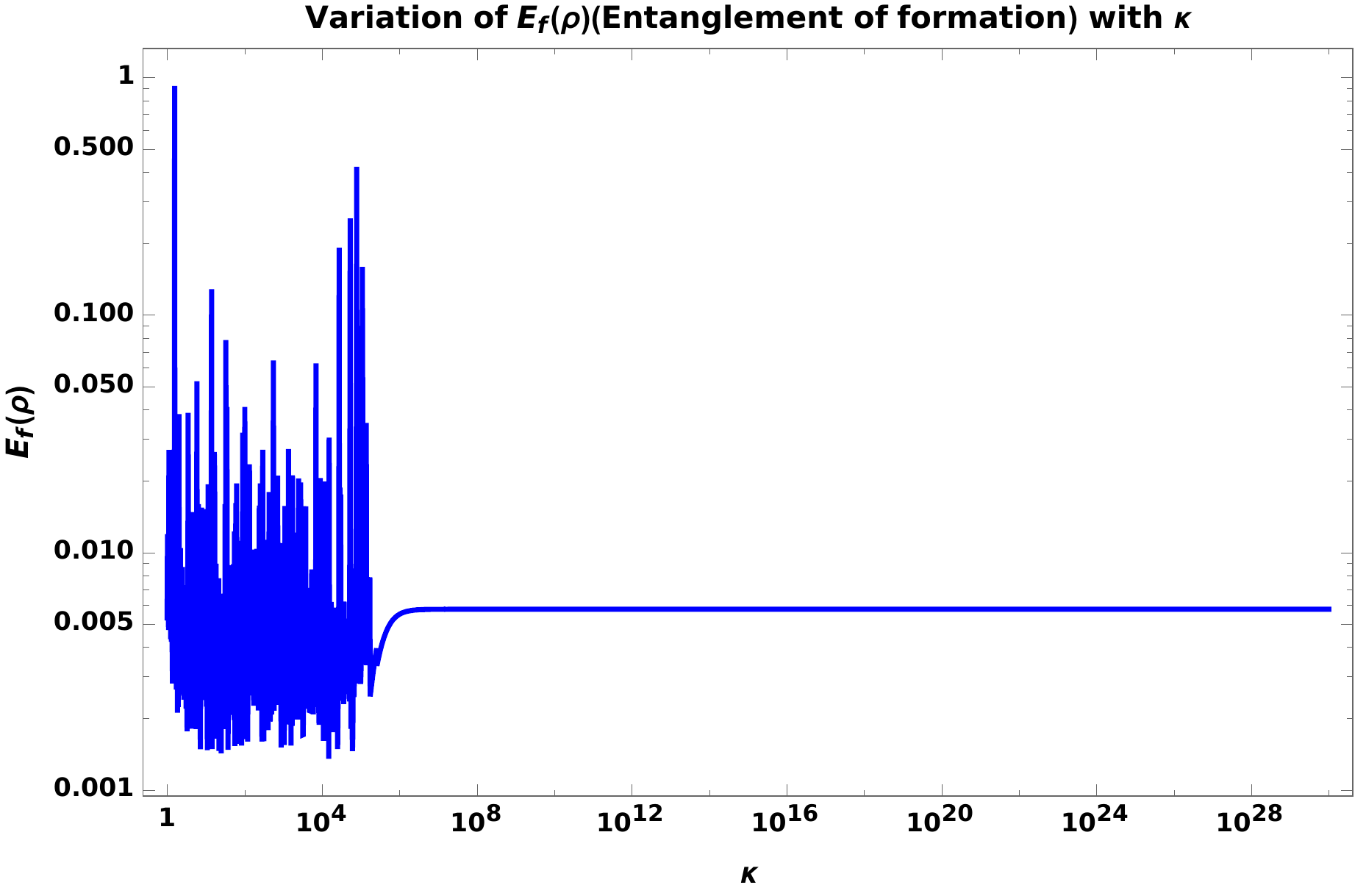}
	\label{eformk}
	}
	\caption{Variation of Entanglement of formation with various parameters is shown here. }
\end{figure}

\subsection{\textcolor{blue}{Quantum Discord }}
\label{Quandis} 

It is a measure of non classical correlation between two subsystems of a quantum system. It includes correlations that are due to quantum physical effects, but do not necessarily involve the concept of quantum entanglement. Sometimes it is also identified as measure of quantumness of correlation functions. If the two quantum states are separable then it does not imply that the quantum correlations exist between them. It is defined as
\be
\mathcal{D}_A(\rho_{\bf System})=\mathcal{I}(\rho_{\bf System})-{\bf max}_{\Pi_j^A}\mathcal{T}_{\Pi_j^A(\rho_{\bf System})}
\ee
where mutual information $\mathcal{I}(\rho_{\bf System})$ is defined as:
\be \mathcal{I}(\rho_{\bf System})=S(\rho_A)+S(\rho_B)-S(\rho_{\bf System}) \ee
where $S(\rho_A)$, $S(\rho_B)$ and $S(\rho_{\bf System})$ represent the Von Neumann entropy of system $A$ ({\bf Atom~1}), $B ({\bf Atom~2})$ and the combined system respectively. On the other hand, the part of the correlation that can be attributed to the classical correlation, which is represented by, $_{\Pi_j^A}\mathcal{T}_{\Pi_j^A}(\rho_{\bf System})$, is defined as:
\bea _{\Pi_j^A}\mathcal{T}_{\Pi_j^A}(\rho_{\bf System})&=& S(\rho_B)-\Pi_j^A S(\rho_B|\Pi_j^A)
\eea
Now we know that for any two qubit state the density matrix is given by the following expression:
\be
\rho=\frac{1}{4}\left(I^a \otimes I^b+\sum_{i=1}^3(a_i\sigma_i \otimes I^b+I^a \otimes b_i\sigma_i)+\sum_{i,j=1}^3 T_{ij}\sigma_i \otimes \sigma_j \right).
\ee
Then the geometric measure of quantum discord is evaluated as:
\be
\label{discal}
{\cal D}(\rho)=\frac{1}{4}(||a||^2+||T||^2-\lambda_{max})
\ee
where $a$ is the column vector, which is defined as, 
$a=(a_1+a_2+a_3)^t$. Here the superscript $t$ stands for the transpose of the vectors or matrices, the trace norm square is defined as:$||a||^2=\sum_i a_i^2$ and $T=(t_{ij})$ is a matrix which one can compute for a specific quantum system and $\lambda_{max}$ is the largest eigenvalue of the matrix $(aa^t+TT^t)$. The matrix $T$ and the column vector relevant to our system are given by:
\begin{eqnarray}
 T=\begin{pmatrix}
a_{++} & & a_{+-} & & 0\\
a_{-+} & & a_{--} & & 0 \\
0 & & 0 & & a_{33}  \\
\end{pmatrix}~~~~~~~~~~
 a=\begin{pmatrix}
 0\\
 0 \\
 a_{30}  \\
\end{pmatrix}
\end{eqnarray}
Using the above definition, the square of the norm of the matrix $T$ (The norm of any matrix M is given by:
$||M||$=$\sqrt{{\rm Tr}(M^\dagger M)}$) and the column vector $a$ relevant to the system studied is given by: 
\bea
||T||^2=a^2_{33}+a^2_{--}+a^2_{-+}+a^2_{+-}+a^2_{++}~~~~~~~~~
||a||^2= a^2_{30}.
\eea
The matrix $(\rm{aa^t+TT^t})$ then becomes

\begin{eqnarray}
(\rm{aa^t+TT^t}) =\begin{pmatrix}
a^2_{+-}+a^2_{++} & &a_{--}a_{+-}+a_{-+}+a_{++} & & 0\\
a_{--}a_{+-}+a_{-+}+a_{++} & &a^2_{--}+a^2_{-+} & & 0 \\
0 & & 0 & & a^2_{33}+a^2_{30}  \\
\end{pmatrix}
\end{eqnarray}
with eigenvalues
\bea
\lambda_1=a^2_{33}+a^2_{03}~~~~
\lambda_2=\frac{1}{2}({\cal P}-{\cal Q})~~~~
\lambda_3=\frac{1}{2}({\cal P}+{\cal Q})
\eea
where ${\cal P}$ and ${\cal Q}$ are
\bea {\cal P}&=&a^2_{--}+a^2_{-+}+a^2_{+-}+a^2_{++} \\
{\cal Q}&=& \sqrt{(-a^2_{--}-a^2_{-+}-a^2_{+-}-a^2_{++})^2-4(a^2_{-+}a^2_{+-}-2a_{--}a_{-+}a_{+-}a_{++}+a^2_{--}a^2_{++})} ~~~~~~~\eea

It has already been mentioned in the earlier sections that the solution of the Bloch vector components $a_{30}$ and $a_{03}$ are identical. 

Thus the quantum discord for the two entangled atoms relevant to our system calculated using equation \ref{discal} is therefore given by the following expression:
\be
{\cal D}(\rho)=\frac{1}{4}\left[a^2_{03}+a^2_{33}+\mathcal{Q}\right]-{\rm max}\left[a_{03}^2+a^2_{33},\frac{1}{2}(\mathcal{Q}-\sqrt{\mathcal{Q}^2-4\mathcal{W}}),\frac{1}{2}(\mathcal{Q}+\sqrt{\mathcal{Q}^2-4\mathcal{W}})\right]
\ee
where the symbols $\mathcal{Q}$ and $\mathcal{W}$ are given as follows:

\bea
\mathcal{Q}=a^2_{--}+a^2_{-+}+a^2_{+-}+a^2_{++}~~~~
\mathcal{W}=a^2_{-+}a^2_{+-}-2a_{--}a_{-+}a_{+-}a_{++}+a^2_{--}a^2_{++}
\eea

For a given set of parameters the maximum eigenvalue is calculated and the following set of plots is obtained by varying various parameters appearing in our model.

\begin{figure}[htb]
\centering
\subfigure[Quantum discord vs Time profile.]{
	\includegraphics[width=7.8cm,height=4cm] {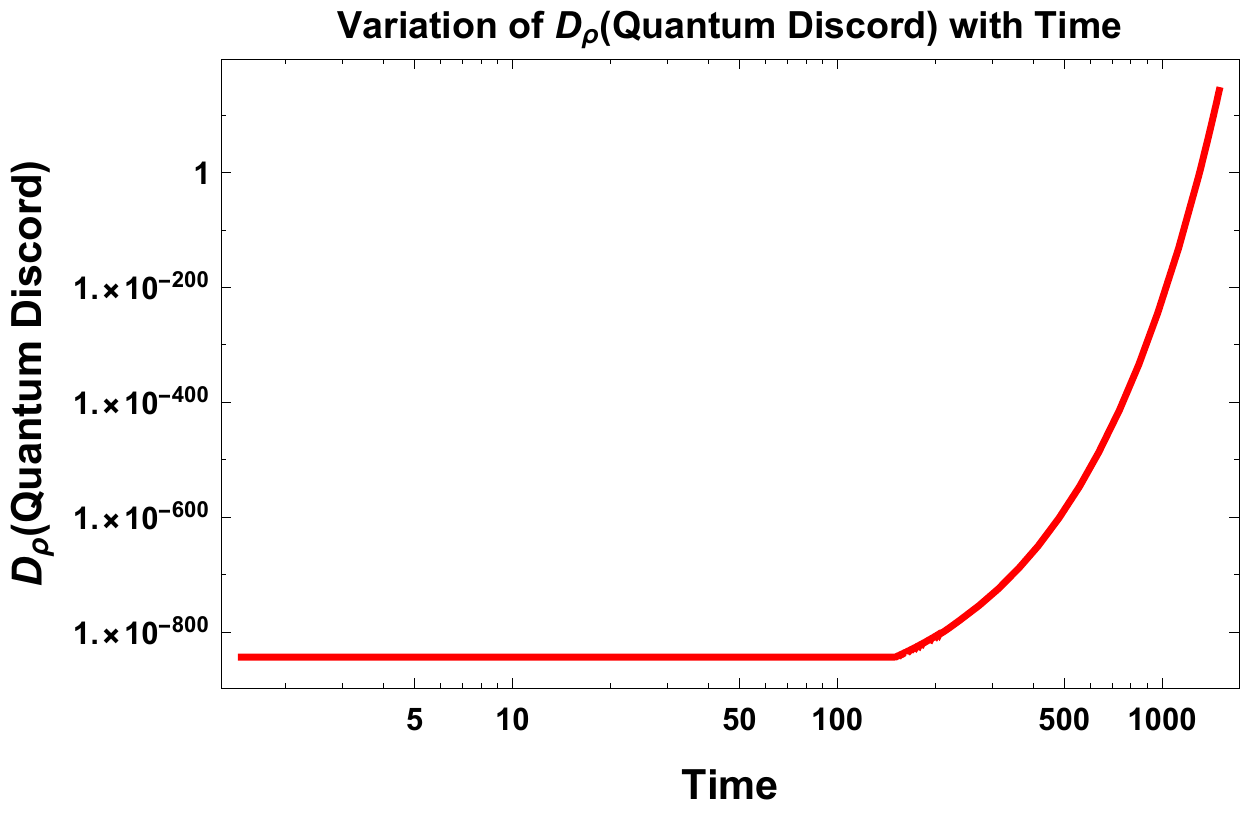}
	\label{15a}
	}
\subfigure[Quantum Discord vs $|Frequency|$ profile.]{
	\includegraphics[width=7.8cm,height=4cm] {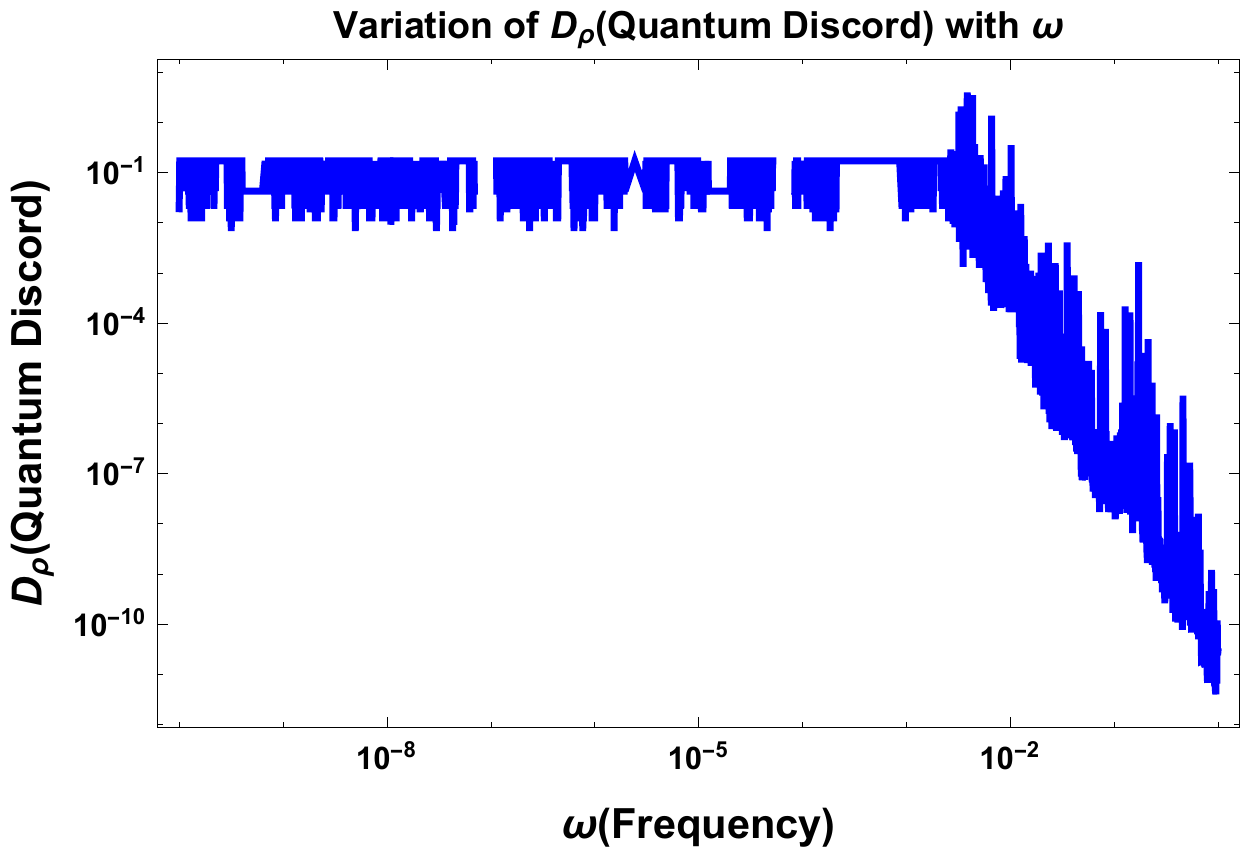}
	\label{15b}
	}
	\subfigure[Quantum discord vs Euclidean distance profile.]{
	\includegraphics[width=7.8cm,height=4cm] {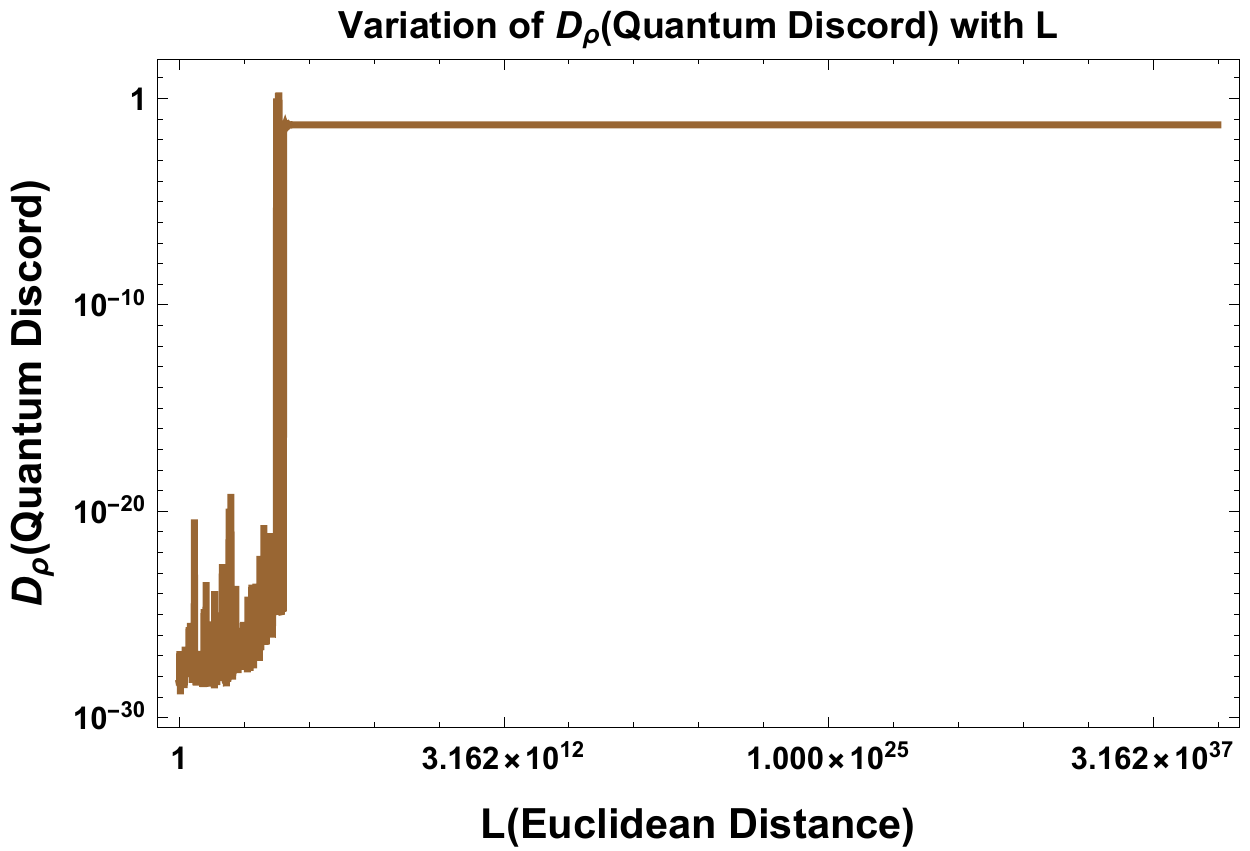}
	\label{15c}
	}
	\subfigure[Quantum discord vs k profile.]{
	\includegraphics[width=7.8cm,height=4cm] {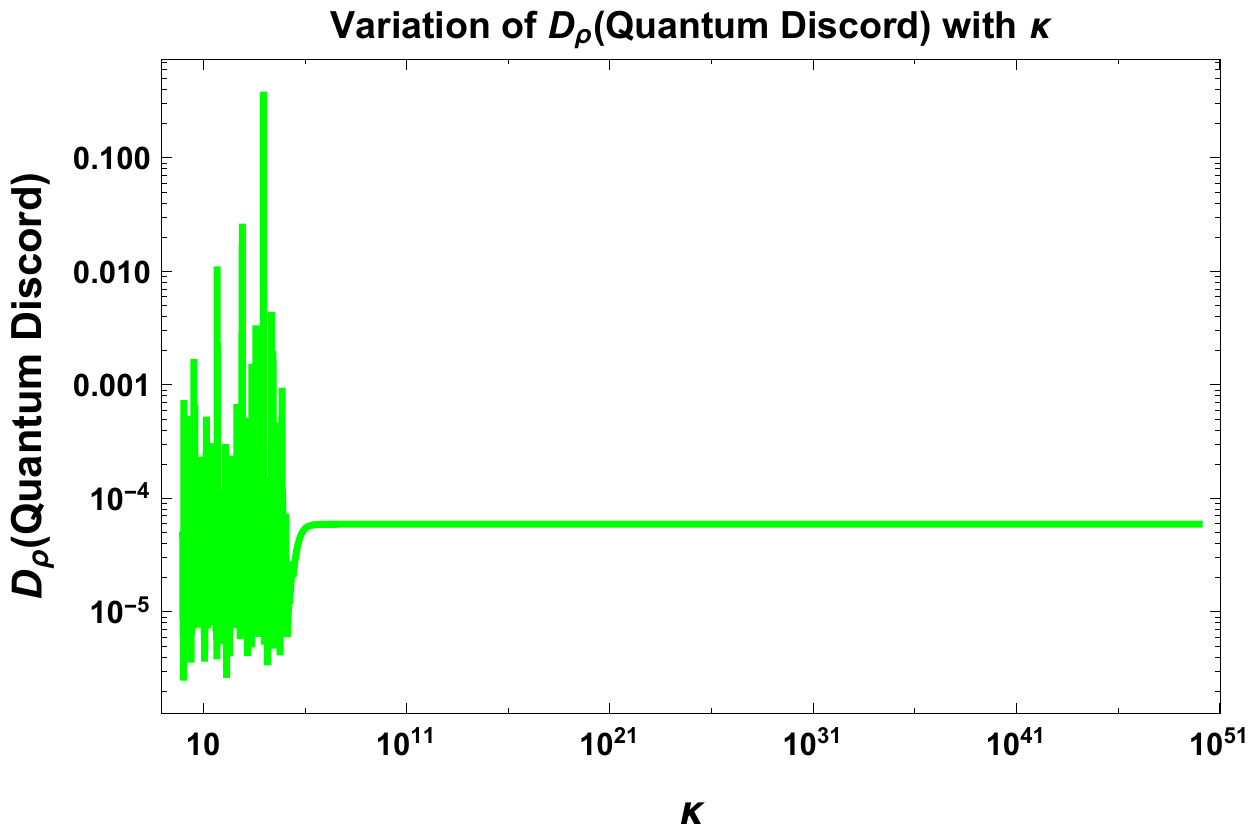}
	\label{15d}
	}
	\caption{Variation of the geometric measure of Quantum discord with various parameters is shown here. }
\end{figure}

In Fig. \ref{15a}, we have explicitly shown the behaviour of quantum discord of our two atomic OQS set up in static patch of de Sitter space with respect to rescaled time $T$. We have normalized the values with the result obtained from $T=10000$ to properly interpret the obtained result from our model. We consider both small and large time scale limiting situations in this context. It is clearly observed that initially the value of quantum discord is almost zero implying that our two atomic OQS set up do not show any signature of quantum correlations. As time goes on, the sub-system gets more and more quantum mechanically correlated and at a very late time scale, it almost saturates to unity, which implies the maximum measure one can obtain from our model to get quantum correlation. Now as we have passed the test for Von Neumann entropy for our model i.e. that this measure is non-zero then one can surely say that non-zero value of quantum discord and Von Neumann entropy together imply the existence of quantum entanglement in the context of our present model of discussion.

In Fig. \ref{15b}, we have shown the behaviour of quantum discord with respect to frequency $\omega$ by keeping all other parameters of the model fixed. We have normalized the values with the result obtained from $\omega=5\times 10^{-2}$. For better understanding the underlying physics we consider both small and large frequency scale limiting situations. It is clearly observed that initially for a finite frequency scale, the value of quantum discord is almost constant showing maximum measure of quantum correlations obtained from quantum discord. From this figure it is easily observed that, as frequency is further increased the obtained measure of quantum discord decreases for our system implying reduction in quantum correlations and when it gets close to $\omega_0$ the value of quantum discord is almost approximately zero indicating no quantum correlation in our system.

In Fig. \ref{15c}, we have shown the behaviour of quantum discord with respect to the Euclidean distance $L$. Again like previous plots, we have normalized the values with the result obtained from $L=10^4$ to explain the underlying physics from this system. We consider both small and large length scale limiting situations. It is observed that initially there is some fluctuation with increase in quantum discord for small length scale. But as the Euclidean distance between the two atoms increase, the subsystem becomes quantum correlated to a maximum saturated value. With further increase in Euclidean distance between the two atoms, there is no change in value implying that the quantum correlation between the atoms in our OQS set up reaches its maximum value.

Last but not the least, in Fig. \ref{15d}, we have shown the behaviour of quantum discord with respect to the parameter $k$, which is basically proportional to the curvature scalar or the Ricci scalar of the static patch of de Sitter space. Like previous plots here also we have normalized the obtained value of quantum discord with the result obtained from k=10$^4$. Similarly like previously mentioned all the plots we consider both small and large values of the k scale limiting situations to interpret the obtained  result from our OQS set up . It is clearly observed that initially there are fluctuations in quantum spectrum discord with increase in the parameter k. With further increase in the value of k, the value remains constant implying no change in quantum correlations which reaches its accessible stable very small value. This further implies that the effect of quantum correlation is extremely small for the large value of the parameter k.

\section{\textcolor{blue}{\bf \large Non Locality from Bell CHSH inequality in de Sitter space}}
\label{nonlocal}

The authors have studied Bell violation in Quantum Mechanics and Primordial cosmology in their earlier papers \cite{Choudhury:2017bou, Choudhury:2017qyl, Choudhury:2016cso, Choudhury:2018fpj, Choudhury:2016pfr}. Here, in analogy to those, we generalise to the Bell violation in dS spacetime.  


%

To establish the concept of non-locality in de Sitter space let us start with a quantum mechanical Bell CHSH operator, which can be defined in the following form:
\be
{\cal B}_{\bf CHSH} = \left({\bf a}\cdot \sigma \right) \otimes \left\{({\bf b} + {\bf b}').\sigma\right\} + \left({\bf a'}\cdot \sigma\right) \otimes \left\{({\bf b}-{\bf b}') \cdot \sigma \right\}
\ee
where ${\bf a},{\bf b},{\bf a'}$ and ${\bf b'}$ are real unit vectors which play significant role to establish non-locality in the present context.

Now the Bell CHSH inequality states that~\footnote{For local hidden variable (LHV) description of correlation, CHSH inequality holds. Violation of CHSH inequality indicates existence of non-locality.}, 
$|\langle B_{\bf CHSH}\rangle| \leq 2$. To establish the non-locality we necessarily have to violate CHSH inequality in de Sitter spacetime, which is of course not a trivial task to do. The main problem in de Sitter space to generate the effect of long range quantum correlation at late time scale from a non-local Bell's inequality violating set up. However, in ref.~\cite{Choudhury:2017bou,Choudhury:2017qyl,Choudhury:2016cso,Choudhury:2018fpj,Choudhury:2016pfr} we and other authors have explicitly shown that in case of axion one can construct such a Bell's inequality violating set up, where the axion effective potential is generated from string theory. In the present set up instead of choosing axion as a Bell's inequality violating candidate to establish non-locality in quantum mechanics we establish this in a more general and model independent way. We use the density matrix formalism from quantum statistical mechanics where the general density matrix for a quantum system can be parametrized as:
\be
\rho=\frac{1}{4}\left[I\otimes I+{\bf a \cdot \sigma}\otimes I+I\otimes {\bf b.\sigma}+\sum_{j,k=1}^3 c_{jk}\sigma_j \otimes \sigma_k\right]
\ee
where ${\bf a}$, ${\bf b}$ and $c_{jk}$ all are in general time dependent quantities which can be explicitly obtained by solving the GSKL master equation in presence of the {\it effective Hamiltonian} and the quantum dissipator {\it Lindbladian} operator . Here, the vectors ${\bf a}$ and ${\bf b}$ are given by the following general representation in terms of the elements of the density matrix:
\bea
{\bf a}=(0,0,\rho_{11}+\rho_{22}-\rho_{33}-\rho_{44})~~~~
b=(0,0,\rho_{11}+\rho_{33}-\rho_{22}-\rho_{44})
\eea
and the $c_{jk}$ matrix takes the form

\begin{eqnarray}
 c_{jk}=\frac{1}{4}\begin{pmatrix}
2\rho_{23}-2(\rho_{14}) & & 2(\rho_{14})_I & & 0\\
2(\rho_{14})_I & & 2\rho_{23}-2(\rho_{14}) & & 0 \\
0 & & 0 & &  \rho_{11}+\rho_{44}-\rho_{22}-\rho_{33} \\
\end{pmatrix}
\end{eqnarray}

In the context of OQS described by the two entangled atoms the vectors {\bf a} and {\bf b} are given by the simplified form, 
${\bf a}=(0,0,a_{30})$, 
${\bf b}=(0,0,a_{03}),$
and the coefficient matrix $c_{jk}$ can be written as: 
\begin{eqnarray}
 c_{jk}=\frac{1}{4}\begin{pmatrix}
2a_{+-}-a_{++}-a_{--} & & i(a_{--}-a_{++}) & & 0\\
i(a_{--}-a_{++}) & & 2a_{+-}+a_{++}+a_{--} & & 0 \\
0 & & 0 & &  a_{33} \\
\end{pmatrix}
\end{eqnarray}

Now, Bell CHSH inequality is violated in OQS if and only if when the sum of the two largest eigenvalues of the following matrix satisfy the following constraint condition, $c~ c^\dagger>1$, 
where the eigenvalues are:
\bea
\lambda_1= 4\left(\rho_{11}+\rho_{44}-\frac{1}{2}\right)^2=a^2_{33}
~~~~~
\lambda_{2,3}= 4(|\rho_{14}|\pm \rho_{23})^2=\frac{1}{4}(|a_{--}|\pm a_{+-})^2
\eea
For the initial separable state $\rho_0=|00\rangle \langle 00|$, we cannot expect these eigenvalues to exceed unity after evolution because only non-zero component of the initial state is $\rho_{44}=1$. However, the Bell CHSH inequality provides only a necessary condition for the LHV(local hidden variable) description and does not guarantee existence of an LHV. For this reason we need to pass each detector through a local filter, which transforms the matrix c and $\rho$ in the following form.
\bea
c'=(f_{A}\otimes f_{B})~c~(f_{A}\otimes f_{B})~~~~
\rho'= (f_{A}\otimes f_{B})~\rho~(f_{A}\otimes f_{B})
\eea
where the two local filters $f_A$ and $f_B$ are described by the following square matrix:
\begin{eqnarray}
 f_{A}=f_{B}=\begin{pmatrix}
1 & & & & & & 0\\
0 & & & & & & \eta\\
\end{pmatrix}
\end{eqnarray} 

The matrices $\rho'$ and $c'$ can be represented as 

\begin{eqnarray}
 \rho'&&=\frac{1}{\rho_{11}+\eta^2(\rho_{22}+\rho_{33})+\eta^4 \rho_{44}}\begin{pmatrix}
\rho_{11} & & 0 & & 0 & & \eta^2 \rho_{14}\\
 0 & & \eta^2 \rho_{22} & & \eta^2 \rho_{22} & & 0 \\
0 & & \eta^2 \rho_{23} & & \eta^2 \rho_{33} & &  0 \\
\eta^2 \rho^*_{14} & & 0 & & 0 & & \eta^4 \rho_{44}
\end{pmatrix},
\\
 c'&&=\frac{2\eta^2}{\rho_{11}+\eta^2(\rho_{22}+\rho_{33})+\eta^4 \rho_{44}}\begin{pmatrix}
\rho_{23}-(\rho_{14})_R & & (\rho_{14})_R & & 0 \\
 (\rho_{14})_R & & \rho_{23}+(\rho_{14})_R & & 0 \\
0 & & 0 & & \frac{\rho_{11}-\eta^2(\rho_{22}+\rho_{33})+\eta^4 \rho_{44}}{2\eta^2}
\end{pmatrix}.~~~~~~~~~~
\end{eqnarray}

The eigenvalues of the new matrix $c'(c')^{\dagger}$ are appended below:

\be
\bal
\lambda_1' &=&\frac{\rho_{11}-\eta^2(\rho_{22}+\rho_{33})+\eta^4 \rho_{44}}{\rho_{11}+\eta^2(\rho_{22}+\rho_{33})+\eta^4 \rho_{44}}=\frac{(1-2\eta^2)+(1-\eta^4)(a_{03}+a_{30})+(1+\eta^2)^2}{(1+2\eta^2)+(1-\eta^4)(a_{03}+a_{30})+(1-\eta^2)^2}
\\
\lambda'_{2,3}&=&\frac{2\eta^2(\rho_{23} \pm |\rho_{14}|)}{\rho_{11}+\eta^2(\rho_{22}+\rho_{33})+\eta^4 \rho_{44}}=\frac{2\eta^2(a_{+-}+|a_{--}|)}{(1+2\eta^2)+(1-\eta^4)(a_{03}+a_{30})+(1-\eta^2)^2}
\eal
\ee

Now, to implement the violation of the Bell-CHSH inequality in de Sitter we follow the following steps:

\begin{enumerate}
	
	\item  \underline{\textcolor{red}{\bf Step 1:}}\\ After passing through the local filter in the new basis we have to satisfy the following necessary constraint condition:
	$$ c'(c')^{\dagger}>1 $$
	\item  \underline{\textcolor{red}{\bf Step 2:}}\\ The above condition directly implies the following inequality:
	$$ 0>\eta^4-\frac{(\rho_{23} + |\rho_{14}|)^2} {\rho_{44} (\rho_{22}+\rho_{33})}\eta^2 + \frac{\rho_{11}}{\rho_{44}} $$
	
	where each of the quantities is appearing previously in the density matrix after passing through local filter.
	
	\item  \underline{\textcolor{red}{\bf Step 3:}}\\ The above inequality can further be simplified to
	$$ (\rho_{23}+|\rho_{14}|)^4> 4\rho_{11}\rho_{44}(\rho_{22}+\rho_{33})^2 $$
	
	\item  \underline{\textcolor{red}{\bf Step 4:}}\\  For real parameter $\eta$, if we substitute the entries of the local filter transformed density matrix in terms of the time dependent Bloch coefficients in the $(+,-,3)$ transformed basis, then the above inequality becomes:
	$$ (a_{+-}+|a_{--}|)^4 >(1-a_{33})^2[(1+a_{33})^2-(a_{03}+a_{30})^2] $$
	
	\item \underline{\textcolor{red}{\bf Step 5:}}\\
	Now, our job is to explicitly verify this inequality for our open quantum system described by two entangled atoms. To serve this purpose we plot the following functions:
	\bea {\cal J}_1(t)&=&(a_{+-}(t)+|a_{--}(t)|)^4 \\
	{\cal J}_2(t)&=&(1-a_{33}(t))^2[(1+a_{33}(t))^2-(a_{03}(t)+a_{30}(t))^2]\eea
	separately. In the plot we represent ${\cal J}_1(t)$ and ${\cal J}_2(t)$ with green and red color. \\

\end{enumerate}

\begin{figure}[htb]
\centering
\subfigure[]{
	\includegraphics[width=7.8cm,height=4cm] {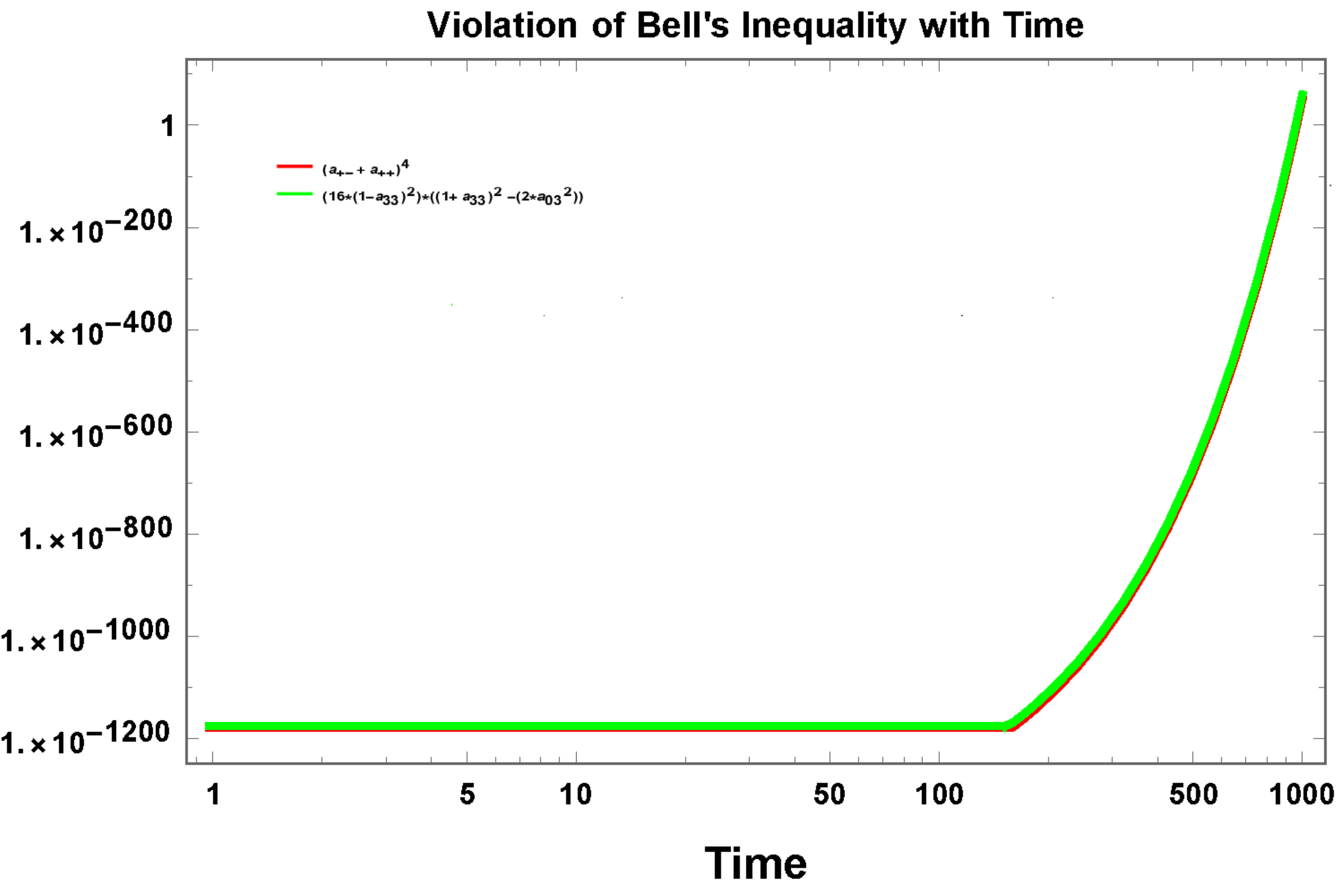}
	\label{17a}
	}
\subfigure[]{
	\includegraphics[width=7.8cm,height=4cm] {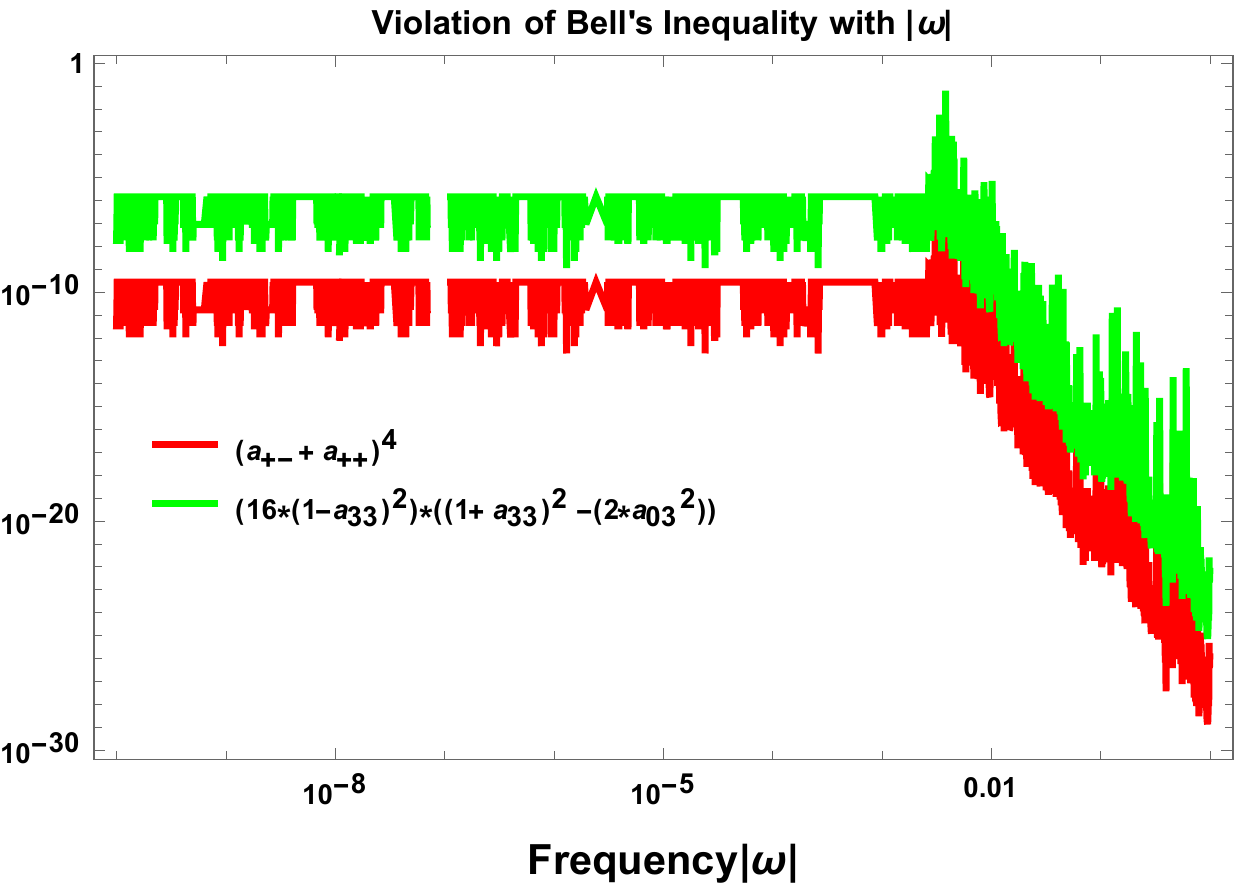}
	\label{17b}
	}
\subfigure[]{
	\includegraphics[width=7.8cm,height=4cm] {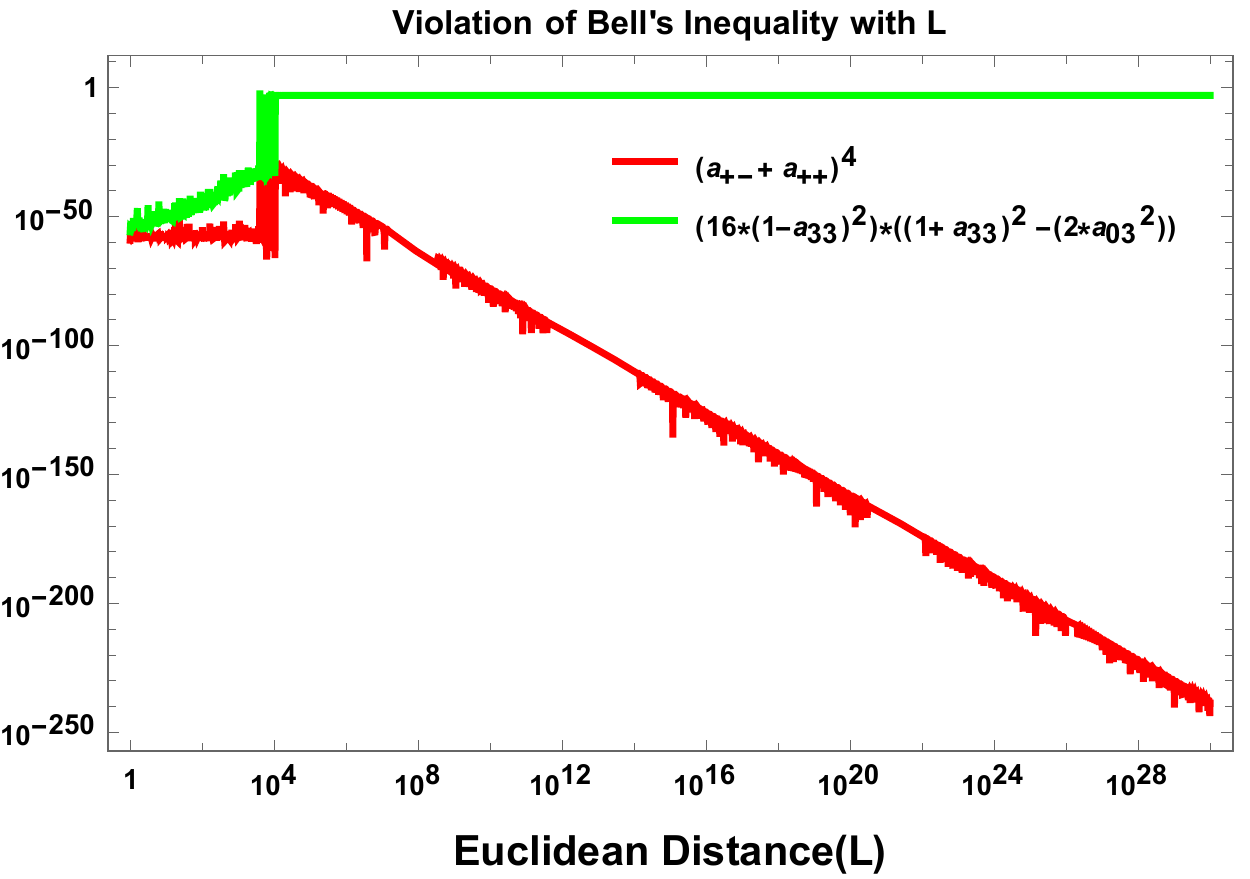}
	\label{17c}
	}
	\subfigure[]{
	\includegraphics[width=7.8cm,height=4cm] {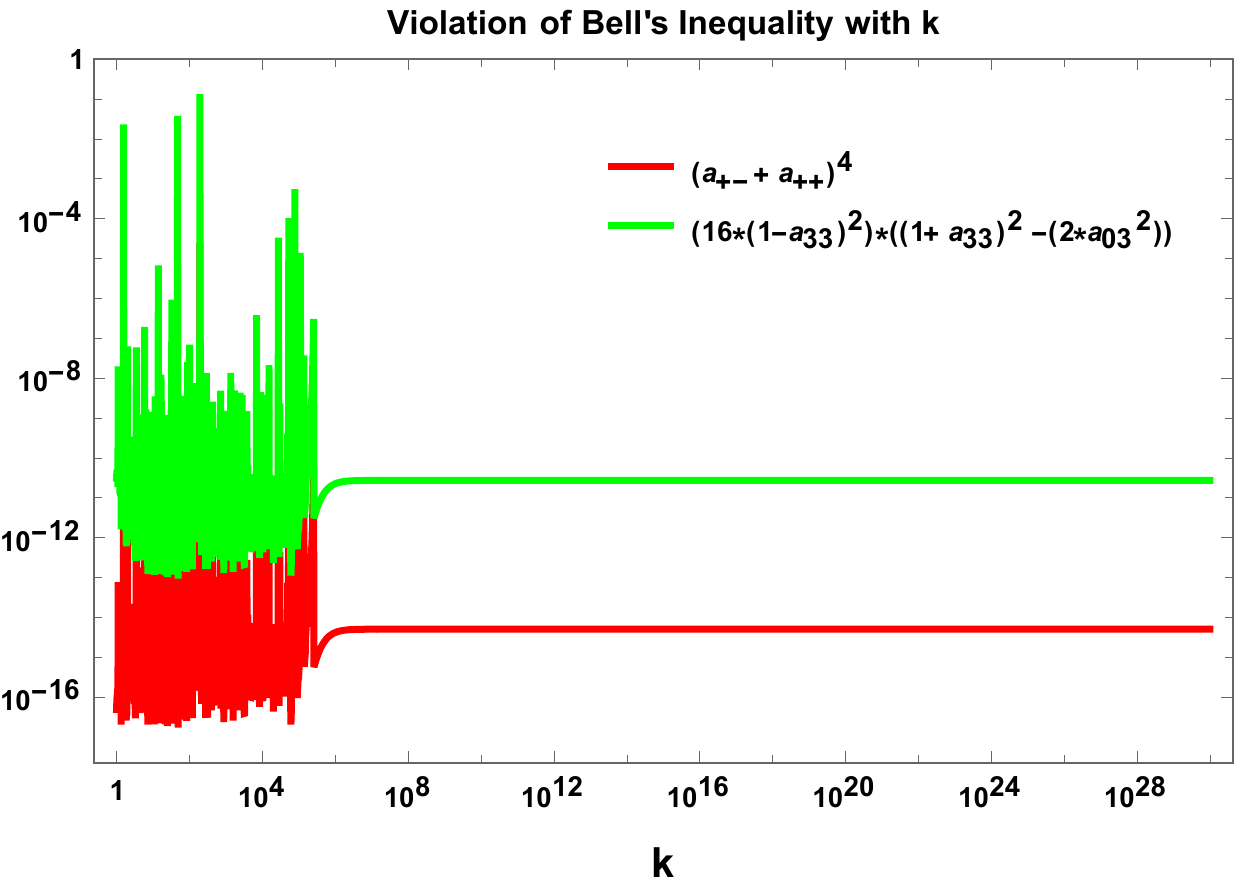}
	\label{17d}
	}
	\caption{Verification of the Violation of the Bell-CHSH inequality with various parameters is shown here }
\end{figure}

The plots shows the behaviour of the two functions appearing on either sides of the Bell's inequality. The above plots clearly shows that the function represented by green colour is always greater than that of red colour i.e the condition $ {\cal J}_1(t)>{\cal J}_2(t) $ is always satisfied irrespective of the parameter chosen to test the inequality. This establishes Bell-CHSH inequality violation and non locality in De-Sitter Space with the present two atomic OQS setup.

\section{\textcolor{blue}{\bf \large General discussion}}
\label{disc}

As seen above, all the entanglement measures give almost similar behaviour with respect to different parameters. In this section we will analyze those plots together.  

In figures \ref{vont},\ref{renq1t},\ref{colt},\ref{mint},\ref{lognt}, \ref{concurt},\ref{eformt} we have plotted different entanglement measures  of our two atomic OQS set up in the static patch of de-Sitter space with respect to Time($T$) keeping all other parameters fixed. Here {\it T} is defined as the time difference, $T=\tau-\tau^{'}$, which is very useful for further analysis. Here $\tau$ is the usual time scale and $\tau^{'}$ is the time scale where the equilibrium boundary condition is imposed. These plots have been normalized with $T=10^4$. It is clearly observed that initially the value of normalized measure is almost zero implying that our two atomic OQS set up do not show any signature of quantum entanglement. This is consistent with our assumption that initially there is no correlation and the quantum states being represented by pure states only. As time goes on, the sub-system gets more and more entangled due to more correlation and at a late time scale, it saturates to unity, the maximum value of the measure. The late time scale behaviour is also consistent with the prediction from the present system i.e. as time goes on the system is represented by mixed quantum states due to getting more quantum correlation.

In figures \ref{vonw}, \ref{renq1w}, \ref{colw}, \ref{minw} \ref{lognw}, \ref{concurw}, \ref{eformw} we have plotted different entanglement measures with respect to frequency($\omega$). These plots have been normalized with $\omega=5\times10^{-2}$. It is can be seen that initially for a finite frequency scale, the value of the measures are almost constant showing entanglement due to the appearance of mixed quantum states. As frequency is further increased the value of the measures decrease implying reduction in entanglement and when it gets close to $\omega_0$ the value of entropy is almost zero indicating that the joint density matrix of our subsystem becomes separable.

In figures \ref{vonl}, \ref{renq1l}, \ref{coll}, \ref{minl}, \ref{lognl}, \ref{concurL}, \ref{eformL}  we have plotted different entanglement measures with respect to Euclidean distance($L$). These plots have been normalized with ${\it L}=5000$. It is clearly observed that initially there are some fluctuations with increase in entropy for small length scale. The maximum value of the fluctuation implies the maximum value of the entangled entropy one can obtain once we want to analyse the behaviour with respect to the Euclidean distance $L$. This is because at small $L$ value the quantum states are dominated by pure states and there is no corresponding quantum correlation. But as the distance between the two atoms increases, the subsystem gets entangled to a maximum value close to 1. With further increase in distance, there is no change in entropy implying that the correlation between the atoms in our OQS set up is maximum and the corresponding quantum state is dominated by mostly mixed states. It can be observed that the measures increase with increase in the distance between the two atoms, indicating rise in entanglement with rise in {\it L}. As the distance between the two atoms increases, the subsystem gets entangled to a maximum value.

In figures \ref{vonk}, \ref{renq1k}, \ref{colk}, \ref{mink}, \ref{lognk}, \ref{concurk}, \ref{eformk} we have plotted different entanglement measures with respect to the parameter $k$. These plots have been normalized with $k=5\times 10^4$. We know that the curvature of static patch of the de Sitter space can be expressed in terms of the parameter k as:
\be R=\frac{12}{\alpha^2}=\frac{12}{k^2+r^2}\approx \frac{12}{k^2}~~{\rm for}~~ k=\sqrt{\alpha^2-r^2}\approx \alpha >>r\ee which implies we actually have considered both flat and static de Sitter space by varying the parameter k.  It is clearly observed that initially there are fluctuations in entanglement measures with increase in the value of k. However, the maximum value the fluctuation implies the maximum value of the entangled entropy one can obtain once we want to analyse the behaviour with respect to the parameter k. This is because at small k or non-zero effect of curvature value the quantum states are dominated by mixed states and the corresponding quantum correlation is non zero. With further increase in the value of the parameter k, the entanglement measures remain constant and very very small implying almost getting no quantum correlation and described by the pure quantum states for the large values of k, which corresponds to the flat space time situation.

\begin{table}[h!]
\centering
\footnotesize
\begin{tabular}{|||c||c||c|||}\hline\hline\hline
\textcolor{red}{\bf Parameters} & \textcolor{red}{\bf Best entanglement measure} & \textcolor{red}{\bf Remarks}\\ \hline \hline
\textcolor{blue}{\bf Rescaled time ($T$)} & Collision Entropy (${\cal H}_{2}$)& The long range \\
 &  &  correlation at\\
& & late time scale can   \\ 
& & be best understood.  \\ \hline \hline
\textcolor{blue}{\bf Frequency ($|\omega|$)} & Min entropy  (${\cal H}_{\infty}$)& In the chosen frequency \\
&  &  range it has the\\
& &  maximum amplitude \\
& & showing maximum  \\ 
& & entanglement. \\\hline \hline
\textcolor{blue}{\bf Euclidean distance ($L$)} & Except Log negativity ($E_{N}$) & In the large length scale\\
& others are appropriate & it shows fluctuations \\ 
& & unlike other \\ 
& & entanglement measures.\\ \hline \hline
\textcolor{blue}{\bf Inverse curvature (k) }& Collision entropy (${\cal H}_{2}$),   &  In the chosen $k$ \\
& Min entropy (${\cal H}_{\infty}$) &  range it has the \\
& and Logarithmic negativity  ($E_{N}$) &  maximum normalized \\
& & amplitude showing \\
& & maximum \\ 
& & entanglement.\\   \hline \hline \hline
\end{tabular}
\label{tab1}
\caption{A comparative study of various entanglement measures for our two atomic OQS setup}
\end{table}

\begin{table}[ht]
\centering
\footnotesize
\begin{tabular}{|||c||c||c|||}\hline\hline\hline
\textcolor{red}{\bf Parameters} & \textcolor{red}{\bf One atomic system} & \textcolor{red}{\bf Two atomic system}\\ \hline \hline
\textcolor{blue}{\bf Reduced subsystem} & Not entangled & Entangled \\
\textcolor{blue}{\bf density matrix}  &  & \\
 \hline \hline
\textcolor{blue}{\bf Equilibrium tempe-} &$\displaystyle T= {T_{\bf GH}}=\frac{1}{2\pi\alpha}$& $\displaystyle T=\sqrt{T^2_{\bf GH}+T^2_{\bf Unruh}}$\\
\textcolor{blue}{\bf rature of bath} &  & $\displaystyle=\frac{1}{2\pi \sqrt{\alpha^2-r^2}}$ \\
\hline \hline
\textcolor{blue}{\bf Inverse curvature $k$} & k=$\alpha$ & $k=\sqrt{\alpha^2-r^2}\neq \alpha$\\
\hline \hline
\textcolor{blue}{\bf Nature of} & One body   & Many body \\
\textcolor{blue}{\bf  Wightman function}& Wightman function & Wightman function \\ 
& $G(\tau-\tau^{'})=\langle \Phi(\tau)\Phi(\tau^{'})\rangle $ & $G_{\alpha\beta}(\tau-\tau^{'})=\langle \Phi(\tau,x_{\alpha})\Phi(\tau^{'},x_{\beta})\rangle $\\
 \hline \hline
\textcolor{blue}{\bf Quantum}& Ground ($|G\rangle =|g\rangle$), & Ground ($|G\rangle =|g_1\rangle \otimes |g_2 \rangle$), \\
\textcolor{blue}{\bf states} &  &  Excited ($|E\rangle= |e_1\rangle \otimes |e_2 \rangle$),  \\
\textcolor{blue}{\bf $|\Psi \rangle$} & Excited ($|E\rangle=|e\rangle $).  &  Symmetric  ($\displaystyle |S\rangle=\frac{1}{\sqrt{2}}(|e_1\rangle \otimes |g_2\rangle $,\\
&  &  $~~~~~~~~~~~~~~~~~~~~~~~~~~~~~~\displaystyle+|g_1\rangle \otimes |e_2\rangle) $),\\
&  &  Anti-symmetric ($\displaystyle|A\rangle=\frac{1}{\sqrt{2}}(|e_1\rangle \otimes |g_2\rangle $. \\
&  &  $~~~~~~~~~~~~~~~~~~~~~~~~~~~~~~~~~~~\displaystyle-|g_1\rangle \otimes |e_2\rangle) $). \\
\hline \hline
\textcolor{blue}{\bf Position of atom} & The bath correlation  function & The bath correlation \\
& does not depend  on the  & function depends on the \\
& position of the atom.  &  position  of the atoms. \\
\hline \hline
\textcolor{blue}{\bf Difference in} & Can be observed with & Observed in larger proportions \\
\textcolor{blue}{\bf entanglement measures}&  very less magnitude, & due to entanglement between \\
& due to interaction with bath. &  the two atoms besides \\
& & interaction with the bath.\\
\hline \hline
\textcolor{blue}{\bf Lamb shift} & At {\rm $\omega$} $\gg$ {\rm$\omega_c$ (Bethe cut-off)} &  At $\omega \gg \omega_c$ (Bethe cut-off)\\
\textcolor{blue}{\bf $\delta E_{\bf LS}=\langle \Psi|H_{\bf LS}|\Psi \rangle$}& no Lamb shift observed. & a finite Lamb shift is observed \\
  \hline \hline \hline
\end{tabular}
\label{table2}
\caption{Comparative study of one atomic and two atomic OQS setup.}
\end{table}

In table 1 a comparative study of different entanglement measure is done for our two atomic OQS setup. Various entanglement measures calculated in this context are compared taking into account various parameters like {\it rescaled time}, {\it Frequency}, {\it Euclidean distance}, {\it Inverse curvature}. The best entanglement measure is found after studying the entanglement with respect to the respective parameter in the entire chosen range and then the conclusion is given. In table 2 a comparative study of the one atomic and two atomic systems are done and the main highlighting differences have been noted. The reduced subsystem density matrix shows entanglement between the two atoms constituting the system whereas the one atomic subsystem in not entangled. It is also noted that for the case of one atomic case, the equilibrium temperature of the bath is exactly equal to the Gibbons Hawking temperature whereas for the two atomic case the equilibrium temperature is written in terms of Gibbons Hawking and Unruh temperature. Due to this difference in temperature, the curvature of the background spacetime appears to be different for the two atomic cases. The Wightman function has only one component ($G_{11}$) for the one atomic case as the correlation function is independent of the position of the atom, but for the two atomic case the Wightman function has four components given by $G_{11}$, $G_{12}$, $G_{21}$ and $G_{22}$ corresponding to the dependence of the bath correlation functions on the position of the atoms. Entanglement measure observed (if any) in case of one atomic system is mainly due to its interactions with the bath whereas for the two atomic case significant amount of entanglement measures are observed due to their mutual entanglement besides interaction with the bath.

\section{\textcolor{blue}{\bf \large Conclusion}}

To summarize, in this work, we have addressed the following issues to study the quantum entanglement phenomenon from two entangled atomic OQS set up:

\begin{itemize}
    \item To begin with, we have started our discussion with two entangled atomic OQS set up. In this framework, the two entangled atoms mimic the role of Unruh-De-Witt detectors, conformally coupled to a thermal bath which is modelled by a massless scalar field in this specific problem. Apart from that, within this OQS set up, a non adiabatic Resonant Casimir Polder Interaction(RCPI) takes place between the Unruh-De-Witt detectors and the thermal bath. Most importantly this interaction is effected by the background De-Sitter space time in which the probe massless scalar field is fluctuating.
    
    \item Since we are only interested in the dynamics of the reduced two entangled atomic subsystem, we partially trace over the probe massless scalar field or thermal bath degrees of freedom. Consequently, without solving the total(system+bath+interaction) quantum liouville equation for the total density matrix, we actually solve the Gorini-Kossakowski-Sudarshan-Lindblad equation (Master equation) to explicitly know about the time evolution of the reduced subsystem density matrix. However, solving GSKL master equation  with proper initial condition is an extremely complicated task, as it involves two non trivial components in the equation of motion. These are the effective Hamiltonian and Quantum dissipator or Lindbladian operator. Due to the complicated structure of both of them it is obvious that the analytical solution of the GSKL master equation is not possible for all type of OQS setup. For our two atomic entangled OQS setup, we represent the density matrix corresponding to each of the atom through bloch sphere representation. However, the reduced subsystem density matrix, which can be constructed by taking the tensor product of two atomic density matrices, cannot be parametrized by a bloch sphere. Instead of that we found the reduced subsystem density matrix is actually parametrized by three time dependent coefficients $a_{0i}(\tau), a_{i0}(\tau)$ and $a_{ij}(\tau)$ $\forall i,j=1,2,3$. this implies that solving GSKL master equation for the present OQS set up is actually determining the time dependent behaviour of the above mentioned coefficients. We found that this leads to huge number of coupled differential equations of all these time dependent coefficients, which are not analytically solvable for  given appropriate initial condition. To solve this problem next we transformed the basis from $\left\{1,2,3 \right\}$ to $\left\{+,-,3\right\}$. In this new basis, we get simplified form of the linear differential equation which are less in number compared to the previous case. Also using the large time equilibrium behaviour of the reduced density matrix, which plays the role of initial condition in our problem, we have found the explicit analytical solution of $ a_{ij} \ \forall \ \  i,j= +,-,3 $.
    
    \item Using the analytical density matrix of the reduced subsystem, we further computed various measures of quantum entanglement $viz.$ Von Neumann entropy, R${e'}$nyi entropy, Logarithmic negativity, Quantum discord, Entanglement of formation and Concurrence, which are commonly used in the context of quantum information theory these days. From the time dependent behaviour of all the measure of quantum entanglement we have found out almost the similar behaviour which states that for initial time $t=0$, the measure of quantum entanglement is 0 and as time goes on the subsystem gets more entangled and after a certain time it increases very slowly $i.e.$ it almost saturates. Apart from these as we have obtained the similar feature both in the case of Von Neumann entropy and in the case of Quantum discord, this imply existence of long range quantum correlation at the late time scale, satisfying the necessary and sufficient condition for quantum entanglement. Similarly we have obtained the time dependent feature of logarithmic negativity, entanglement of formation and concurrence which strongly imply that at initial time $t=0$ our two atomic OQS setup do not show any signature of quantum entanglement as in all the cases the quantum measure is zero. As time goes on we have found out all of these measures significantly increase and at very late time scale it almost saturates. This is obviously a significant finding of our two atomic entangled OQS setup from which one can extract the existence of long range quantum correlation in late time scale, which is a very common topic of study in the context of quantum information theory.
   
    \item Last but not the least we have studied Bell-CHSH inequality \footnote{Bell CHSH inequality in quantum mechanics is the most generalized version of the Bell's inequality.} violation from our present setup in de-Sitter space. Though these kind of violation of Bell-CHSH inequality in de-Sitter space is not very trivial. Most importantly, without introducing any axion in a more model independent way we have established the violation of Bell-CHSH inequality in de-sitter space, which is the necessary ingredient to study the non local effect in correlations in quantum mechanics. 
\end{itemize}

The future prospects of this work is as follows.

\begin{itemize}

    \item In this work, we have restricted our subsystem which is made up of two entangled atoms. One can further generalize the same problem with arbitrary  number of atoms within the framework of OQS.
  
    \item In this work, we did our computation in the background static De-Sitter metric. However this framework can be implemented in any curved spacetime metric. One can even carry forward the calculations in other patches of De-Sitter space like the global and inflationary (planer) patch. It is expected to get significant modifications in the cosmological correlation functions. Within the framework of OQS and particularly for inflationary patch \cite{Choudhury:2013iaa, Choudhury:2012ib, Choudhury:2011sq, Choudhury:2012yh, Choudhury:2018rjl, Choudhury:2018bcf} of the De-Sitter space, one can exactly compare these results with the known cosmological correlation functions computed within the framework of closed quantum system. Such comparative analysis between the obtained cosmological correlation functions obtained from OQS and CQS will help us to know about the correct quantum mechanical picture of early universe cosmology. Additionally, by comparing this result with the data obtained from the observational probe for the early universe cosmology one may further rule out one of the possible quantum pictures mentioned here.
 
    \item For our work, we have computed the signature of quantum entanglement from various quantum information theoretic measures. However, we have not done the calculation for all possible measures. For completeness and also to conclude about the existence of long range quantum correlations at the late time scale one can further compute squashed entangled entropy \cite{	arXiv:quant-ph/0308088}, entanglement of distillation \cite{arXiv:quant-ph/0202144}, relative entropy \cite{Nakagawa:2018kvo} etc. Additionally, one can also compute fisher information from the present OQS set up to know about the difference between the results obtained in the classical and quantum limiting situations.
    
    \item This kind of study can be also be used in parameter estimation theory and quantum metrology to estimate various parameters relevant to the theory but cannot be treated as any quantum mechanical observable.\cite{Choudhury:2020dpf}
    
    \item Very recently the phenomena of quantum teleportation has been observed  for the first time with qutrit states \cite{news,teleport}, which is obviously an outstanding finding in the context of quantum information theory. The authors have succeeded in teleporting three-dimensional quantum states for the first time. High-dimensional teleportation could play an important role in future quantum computers. In this direction, our future plan is to study such possibilities from the present OQS set up in de Sitter space.

\end{itemize}

\section*{\textcolor{blue}{\bf \large Acknowledgements}}

SC would like to thank Quantum Gravity and Unified Theory and Theoretical Cosmology
Group, Max Planck Institute for Gravitational Physics, Albert Einstein Institute (AEI) for providing the Post-Doctoral Research Fellowship. SC take this opportunity to thank sincerely to
Jean-Luc Lehners for their constant support and inspiration. SC thank the organisers of XXXI Workshop Beyond the Standard Model, Physikzentrum Bad Honnef, Workshop in String Theory and Cosmology, NISER, Bhubansewar for providing the local
hospitality during the work. SC also acknowledge the discussion regarding this work with Shiraz Minwalla, Sandip Trivedi, Joseph Samuel, Madhavan Varadarajan, Aninda Sinha, Justin Raju David, Arnab Kundu, Shibaji Roy, Soumitra Sengupta, which help us significantly to improve the discussions as well as the presentation of the work. Most significantly, SC is thankful to Shiraz Minwalla for giving the opportunity to present this work in very prestigious seminar series, ``Quantum Space-time Seminar" at DTP, TIFR. SP acknowledges the J. C. Bose National Fellowship for support of his research. Last but not the least, we would like to acknowledge our debt to the people belonging to the various part of the world for their generous and steady support for research in natural sciences. Satyaki and Abinash are grateful to NISER, Bhubaneswar for providing the excellent atmospehere for this research. Finally, we thank the reviewer and the editor for their valuable comments and suggestions to help us improve this work considerably.

\appendix

\section{\textcolor{blue}{\bf  \large Wightman function for probe massless scalar field in static de Sitter Space}}\label{sk}
In this section we  compute the two atomic Wightman correlation function for a massless probe scalar field in static de Sitter spacer characterised by the following infinitesimal line element:
\be
ds^{2}=\left(1-\frac{r^{2}}{\alpha^{2}}\right)dt^{2}-\left(1-\frac{r^{2}}{\alpha^{2}}\right)^{-1}dr^{2}-r^{2}(d\theta^{2}+\sin^{2}\theta d\phi^{2})~~~{\rm where}~~\alpha=\sqrt{\frac{3}{\Lambda}}>0.
\ee
 To compute the expression for the each of the entries of the two body Wightman function of the probe scalar field present in the external thermal bath we use the four dimensional static de Sitter geometry of our space-time. In this set of coordinate system in four dimension, the Klein-Gordon field equation for the massless conformally coupled external probe scalar field for the non-adiabatic environment can be expressed as:
\bea \left[\frac{1}{\cosh^3\left(\frac{t}{\alpha}\right)}\frac{\partial}{\partial t}\left(\cosh^3\left(\frac{t}{\alpha}\right)\frac{\partial}{\partial t}\right)-\frac{1}{\alpha^2\cosh^2\left(\frac{t}{\alpha}\right)}{\bf L}^2\right]\Phi(t,\chi,\theta,\phi)=0~~~,\eea
where ${\bf L}^2$ is the {\it Laplacian differential operator} in the three dimensions characterised by the coordinate $(\chi,\theta,\phi)$ , which is explicitly defined as:
\be{\bf L}^2=\frac{1}{\sin^2\chi}\left[\frac{\partial}{\partial\chi}\left(\sin^2\chi\frac{\partial}{\partial\chi}\right)+\frac{1}{\sin\theta}\frac{\partial}{\partial\theta}\left(\sin\theta\frac{\partial}{\partial\theta}\right)+\frac{1}{\sin^2\theta}\frac{\partial^2}{\partial\phi^2}\right]~,\ee
where we introduce a new coordinate $\chi$ which is related to the radial coordinate $r$ as $r=\sin\chi$.
The corresponding two body Wightman function between two space-time points for massless probe scalar field can be expressed as:
\bea G(x,x^{'})&=&\begin{pmatrix} ~
G^{11}(x,x')~~~ &~~~ G^{12}(x,x')~ \\
~G^{21}(x,x')~~~ &~~~ G^{22}(x,x') ~
\end{pmatrix}
=\begin{pmatrix} ~
\langle \Phi({\bf x_{1}},\tau)\Phi({\bf x_{1}},\tau')\rangle~~~ &~~~ \langle \Phi({\bf x_{1}},\tau)\Phi({\bf x_{2}},\tau')\rangle~ \\
~\langle \Phi({\bf x_{2}},\tau)\Phi({\bf x_{1}},\tau')\rangle~~~ &~~~ \langle \Phi({\bf x_{2}},\tau)\Phi({\bf x_{2}},\tau')\rangle ~
\end{pmatrix}~\nonumber\\
&=&\frac{1}{Z_{\bf Bath}}\begin{pmatrix} ~
{\rm Tr}\left[\rho_{\bf Bath}(\tau-\tau^{'}) \Phi({\bf x_{1}},\tau)\Phi({\bf x_{1}},\tau')\right]~ &~{\rm Tr}\left[\rho_{\bf Bath}(\tau-\tau^{'})  \Phi({\bf x_{1}},\tau)\Phi({\bf x_{2}},\tau')\right]~ \\
~{\rm Tr}\left[ \rho_{\bf Bath}(\tau-\tau^{'}) \Phi({\bf x_{2}},\tau)\Phi({\bf x_{1}},\tau')\right]~ &~ {\rm Tr}\left[ \rho_{\bf Bath}(\tau-\tau^{'}) \Phi({\bf x_{2}},\tau)\Phi({\bf x_{2}},\tau')\right] ~
\end{pmatrix},~~~~~~~\eea
where the Partition function for the bath is given by:
\be Z_{\bf Bath}={\rm Tr}\left[\rho_{\bf Bath}(\tau-\tau^{'})\right].\ee
Also, the components of the two atomic Wightman function can be expressed as~\footnote{A similar computation of the two atomic Wightman function can be done in the flat space limit case. The expressions of the Wightman function in the flat space limit is given by 
\begin{eqnarray*}
		G^{11} (x,x') &=& G^{22} (x,x') = - \frac{1}{4 \pi^2} \sum_{n=-\infty}^{\infty} \frac{1}{(\Delta \tau - in/T - i\epsilon)^2} \\
		G^{12} (x,x') &=& G^{21} (x,x') = - \frac{1}{4 \pi^2} \sum_{n=-\infty}^{\infty} \frac{1}{(\Delta \tau - in/T - i\epsilon)^2 - L^2}
\end{eqnarray*} 
A computation of the non zero spectroscopic integral kernel using the above written wightman function will give a result proportional to $\cos(\omega_0L)$ which is exactly what we obtain when we take L$\ll$k limit in the spectroscopic integral kernel computed from Wightman functions of the curved de Sitter spacetime. \cite{Tian:2016uwp, Bhattacherjee:2019eml, Banerjee:2020ljo}}:
\bea
G^{11}(x,x')  = G^{22}(x,x')
            =\langle \Phi({\bf x_{1}},\tau)\Phi({\bf x_{1}},\tau')\rangle
           &=&\langle \Phi({\bf x_{2}},\tau)\Phi({\bf x_{2}},\tau')\rangle=- \frac{1}{16\pi^{2}k^{2}}\frac{1}{\sinh^2\left(\frac{\Delta \tau}{2k}-i\epsilon\right)}, ~~~~~~~~~   
\\
G^{12}(x,x')   = G^{21}(x,x')
             \nonumber = \langle \Phi({\bf x_{1}},\tau)\Phi({\bf x_{2}},\tau')\rangle
             &=& \langle \Phi({\bf x_{2}},\tau)\Phi({\bf x_{1}},\tau')\rangle\nonumber\\
              & =&-\frac{1}{16\pi^{2}k^{2}}\frac{1}{\left\{\sinh[2](\frac{\Delta \tau}{2k}-i\epsilon)-\frac{r^{2}}{k^{2}}\sin[2](\frac{\Delta \theta}{2})\right\}}.~~~~~~~~~
\eea
Here we have introduced few parameters, which are defined as:
\bea k=\sqrt{g_{00}}\alpha =  \sqrt{\alpha^{2}-r^{2}},~~~ \Delta \tau=\tau-\tau'=\sqrt{g_{00}}(t-t')=k\left(\frac{t-t'}{\alpha}\right).~~~~~~~~~~\eea

\section{\textcolor{blue}{\bf  \large  Gorini~Kossakowski~Sudarshan~Lindblad (GSKL) ($C^{\alpha \beta}_{ij})$ matrix }}
\label{gskl}
In this appendix, we explicitly write down the entries of the Gorini~Kossakowski~Sudarshan~Lindblad (GSKL) ($C^{\alpha \beta}_{ij})$ matrix which is appearing in the expression for the quantum dissipator or {\it Lindbladian} operator as given by the following expression:
\be
\mathcal{L[\rho_{\bf System}(\tau)]}=\frac{1}{2}\sum_{i,j=\pm}^3 \sum_{\alpha \beta=1}^2 C_{ij}^{\alpha \beta}\left[2 \sigma_j^{\beta}\rho_{\bf System}(\tau) \sigma_i^\alpha-\left\{\sigma_i^\alpha \sigma_j^\beta,\rho_{\bf System}(\tau)\right\}\right], 
\ee
where we have written the expression in a transformed basis span by $(+,-,3)$ for two entangled OQS set up.
 The components of $C^{\alpha \beta}_{ij}$ matrix are very crucial to solve the time evolution equation of the reduced subsystem density matrix when we partially trace out the bath degrees of freedom from our two atomic entangled OQS set up. In general, Gorini~Kossakowski~Sudarshan~Lindblad (GSKL) ($C^{\alpha \beta}_{ij})$ can be expressed as:
 \be C^{\alpha\beta}_{ij}=\tilde{A}^{\alpha\beta}\delta_{ij}-i\tilde{B}^{\alpha\beta}\epsilon_{ijk}\delta_{3k}-\tilde{A}^{\alpha\beta}\delta_{3i}\delta_{3j}~~~~~~~\forall ~~i,j=+,-,3~~{\rm and}~~\forall \alpha,\beta=1(\textcolor{red}{\bf Atom~1}),2(\textcolor{red}{\bf Atom~2}).\ee
 In terms of explicit components the entries of the $C^{\alpha \beta}_{ij}$ matrix can be written as:
\bea
C_{++}^{\alpha \beta}=\tilde{A}^{\alpha \beta},
~
C_{+-}^{\alpha \beta}=-i \tilde{B}^{\alpha \beta},~
C_{+3}^{\alpha \beta}= C_{-3}^{\alpha \beta}=C_{3+}^{\alpha \beta}=C_{3-}^{\alpha \beta}=C_{33}^{\alpha \beta}=0,~
C_{-+}^{\alpha \beta}=i \tilde{B}^{\alpha \beta},~
C_{--}^{\alpha \beta}= \tilde{A}^{\alpha \beta}~~~~~
\eea
In the next two subsections, we will explicitly compute each of the components of $\tilde{A}^{\alpha\beta}$ and $\tilde{B}^{\alpha\beta}$ for the two atomic OQS in $(+,-,3)$ transformed basis.

Now we are interested in the following limit ~\footnote{To simplify the solutions we have considered this assumption. However, the similar kind of feature one can get in the sub horizon time scale in de Sitter space itself. Particularly in sub horizon scale one can neglect the contribution for the mode momentum which is appearing in the frequency of the fluctuating modes and for de Sitter inflationary patch of the metric one can explicitly show that $\omega$ is imaginary and controlled by conformal time scale. In our case the parameter $k=\sqrt{\frac{3}{\Lambda}-r^2}>0$ as for de Sitter space the cosmological constant $\Lambda>0$. This implies that in our case we get imaginary frequency if we respect this specific constraint condition and it is physically justifiable.}:

\be  
{\rm coth}(\pi k \omega_0)=0~\Longrightarrow~k\omega_0=i\left(n+ \frac{1}{2}\right)~~~~~{\rm where}~n\in \mathbb{Z}
\ee

Finally, substituting the explicit forms of $\tilde{A}^{\alpha\beta}$ and $\tilde{B}^{\alpha\beta}$ which we have derived in the previous sub section, we get the following expressions for the entries of the GSKL matrix:
\bea
C_{+3}^{\alpha \beta}&=& C_{-3}^{\alpha \beta}=C_{3+}^{\alpha \beta}=C_{3-}^{\alpha \beta}=C_{33}^{\alpha \beta}=0
\\
C_{-+}^{11}&=&C_{-+}^{22}=-C_{+-}^{11}=-C_{+-}^{22}=i \tilde{B}^{11}=i\tilde{B}^{22}=\frac{\mu^2\omega_0 i}{8\pi},
\\
C_{-+}^{12}&=&C_{-+}^{21}=-C_{+-}^{12}=-C_{+-}^{21}=i \tilde{B}^{12}=i \tilde{B}^{21}=\frac{\mu^2\omega_0 i}{8\pi}~f\left(\omega_0,\frac{L}{2}\right)
\\
C_{++}^{11}&=&C_{++}^{11}=C_{--}^{11}=C_{--}^{22}= \tilde{A}^{11}= \tilde{A}^{22}=\frac{\mu^2\omega_0}{8\pi}{\rm coth}\left(\pi k\omega_0\right),\\
C_{++}^{12}&=&C_{++}^{21}=C_{--}^{12}=C_{--}^{21}= \tilde{A}^{12}=\tilde{A}^{21}=\frac{\mu^2\omega_0}{8\pi}{\rm coth}\left(\pi k\omega_0\right)~f\left(\omega_0,\frac{L}{2}\right).
\eea
Further, using the assumption, $k\omega_0=i\left(n+ \frac{1}{2}\right)$ where $n\in \mathbb{Z}$, we get the following simplified expressions for the entries of the GSKL matrix:
\bea
C_{+3}^{\alpha \beta}&=& C_{-3}^{\alpha \beta}=C_{3+}^{\alpha \beta}=C_{3-}^{\alpha \beta}=C_{33}^{\alpha \beta}=0
\\
C_{-+}^{11}&=&C_{-+}^{22}=-C_{+-}^{11}=-C_{+-}^{22}=i \tilde{B}^{11}=i\tilde{B}^{22}=-\frac{\mu^2}{8\pi k}\left(n+\frac{1}{2}\right),
\\
C_{-+}^{12}&=&C_{-+}^{21}=-C_{+-}^{12}=-C_{+-}^{21}=i \tilde{B}^{12}=i \tilde{B}^{21}=-\frac{\mu^2}{8\pi k}\left(n+\frac{1}{2}\right)~f\left(\omega_0,\frac{L}{2}\right)
\\
C_{++}^{11}&=&C_{++}^{11}=C_{--}^{11}=C_{--}^{22}= \tilde{A}^{11}= \tilde{A}^{22}=0,\\
C_{++}^{12}&=&C_{++}^{21}=C_{--}^{12}=C_{--}^{21}= \tilde{A}^{12}=\tilde{A}^{21}=0,
\eea
where, the parameter \be k=\sqrt{\frac{3}{\Lambda}-r^2}=\sqrt{\left(\frac{12}{R_{\bf DS}}\right)^2-r^2}>0,\ee for de Sitter space. Here we use the fact that, the curvature of static de Sitter space is given by, $R_{\bf DS}=\sqrt{48\Lambda}>0$  as $\Lambda>0$ for De~Sitter.

\section{\textcolor{blue}{\bf  \large   Effective Hamiltonian ($H_{ij}^{\alpha \beta}$) matrix }}
\label{effham}
In this appendix, we explicitly write down the entries of the Effective Hamiltonian ($H_{ij}^{\alpha \beta}$) matrix which is appearing in the expression for the {\it Lamb Shift} part of the Hamiltonian as given by the following expression:
\be H_{\bf Lamb~Shift}=-\frac{i}{2}\sum^{2}_{\alpha,\beta=1}\sum^{3}_{i,j=1}H_{ij}^{\alpha \beta} (n^{\alpha}_{i}.\sigma^{\alpha}_{i})(n^{\beta}_{j}.\sigma^{\beta}_{j}).\ee 
where this Hamiltonian is constructed by partially tracing over bath degrees of freedom, which is in our problem a probe massless scalar field. Technically, the Effective Hamiltonian ($H_{ij}^{\alpha \beta}$) matrix represent here the strength of the spin chain interaction mentioned above. The general expression of $H_{ij}^{\alpha \beta}$ matrix is given by the following expression:
\be
H^{ij}_{\alpha \beta}=\mathcal{A}^{\alpha \beta} \delta_{ij}-i \mathcal{B}^{\alpha \beta} \epsilon_{ijk}\delta_{3k}-\mathcal{A}^{\alpha \beta} \delta_{3i} \delta_{3j}~~~~~~~\forall ~~i,j=+,-,3~~{\rm and}~~\forall \alpha,\beta=1(\textcolor{red}{\bf Atom~1}),2(\textcolor{red}{\bf Atom~2}).\ee
In terms of explicit components the entries of the ${H}_{ij}^{\alpha \beta}$ matrix can be written as:
\bea
H_{++}^{\alpha \beta}&=&H_{--}^{\alpha \beta}=A^{\alpha \beta}~~~\forall~\alpha,\beta=1,2,
\\
H_{+-}^{\alpha \beta}&=& -i B^{\alpha \beta}~~~\forall~\alpha,\beta=1,2,
\\
H_{-+}^{\alpha \beta}&=& i B^{\alpha \beta}~~~\forall~\alpha,\beta=1,2,
\\
H_{+3}^{\alpha \beta}&=&H_{-3}^{\alpha \beta}=0~~~\forall~\alpha,\beta=1,2,
\\
 H_{3j}^{\alpha \beta}&=&0~~~\forall ~j=+,-,3~~{\rm and}~~\forall~\alpha,\beta=1,2.
\eea
Finally, substituting the explicit forms of ${A}^{\alpha\beta}$ and ${B}^{\alpha\beta}$ which we have derived in the previous sub section, we get the following expressions for the entries of the effective Hamiltonian matrix:
\bea
H_{ij}^{11}&=&H_{ij}^{22}=\frac{\mu^2P}{4\pi^2 i} \left[(\delta_{ij}-\delta_{3i}\delta_{3j})\Theta_1-i \epsilon_{ijk}\delta_{3k}\Theta_2 \right],\\
H_{ij}^{12}&=&H_{ij}^{21}=\frac{\mu^2P}{4\pi^2 i}  \left[(\delta_{ij}-\delta_{3i}\delta_{3j})\Theta_3-i \epsilon_{ijk}\delta_{3k}\Theta_4 \right].
\eea
Therefore, all the various entries of ${H}_{ij}^{\alpha \beta}$ can be written as:
\bea
H_{++}^{11}=H_{++}^{22}=H_{--}^{11}=H_{--}^{22}&=&A^{11}=A^{22}=\frac{\mu^2 P}{4\pi^2 i}\Theta_1~~~~~~~~~~
\\
H_{-+}^{11}=H_{-+}^{22}=-H_{+-}^{11}=-H_{+-}^{22}&=&i B^{11}=i B^{22}=\frac{\mu^2 P}{4\pi^2 }\Theta_2
\\
H_{++}^{12}=H_{++}^{21}=H_{--}^{12}=H_{--}^{21}&=&A^{12}=A^{21}=\frac{\mu^2 P}{4\pi^2 i}\Theta_3~~~~~~~~~
\\
H_{-+}^{12}=H_{-+}^{21}=-H_{+-}^{12}=-H_{+-}^{21}&=&i B^{12}=i B^{21}=\frac{\mu^2 P}{4\pi^2 }\Theta_4~~~~~~~~~~
\eea
The function $f\left(\omega,\frac{L}{2}\right)$ has already been defined in the previous section. Here we introduce four integral functions $\Theta_i\forall i=1,2,3,4$, which is defined as:

\bea \underline{\textcolor{red}{\bf Integral ~I:}}~~~
\Theta_{1}:&&=\int^{\infty}_{-\infty}d\omega~\frac{\omega^2}{\left(1-e^{-2\pi k\omega}\right)\left(\omega+\omega_0\right)\left(\omega-\omega_0\right)} \\
  \underline{\textcolor{red}{\bf Integral ~II:}}~~~
  \Theta_{2}:&&=\int^{\infty}_{-\infty}d\omega~\frac{\omega\omega_0}{\left(1-e^{-2\pi k\omega}\right)\left(\omega+\omega_0\right)\left(\omega-\omega_0\right)}
\\ 
\underline{\textcolor{red}{\bf Integral ~III:}}~~~
\Theta_{3}:&&=\int^{\infty}_{-\infty}d\omega~\frac{\omega^2 }{\left(1-e^{-2\pi k\omega}\right)\left(\omega+\omega_0\right)\left(\omega-\omega_0\right)}f\left(\omega,\frac{L}{2}\right)~~~~~~
\\ 
\underline{\textcolor{red}{\bf Integral ~IV:}}~~~
\Theta_{4}:&&=\int^{\infty}_{-\infty}d\omega~\frac{\omega\omega_0}{\left(1-e^{-2\pi k\omega}\right)\left(\omega+\omega_0\right)\left(\omega-\omega_0\right)}f\left(\omega,\frac{L}{2}\right) .~~~~~~\eea
In the next section we explicitly compute the contributions from all of these integrals.

\section{\textcolor{blue}{\bf \large Calculation of useful integrals}}
In this section, we explicitly compute the analytical expression for the useful integrals $\Theta_i\forall i=1,2,3,4$, which are very useful to compute the expressions for the effective Hamiltonian matrix elements. 
\subsection{\textcolor{blue}{ Integral~I}}
In this subsection we explicitly compute the finite contribution from the following integral:
\bea {\textcolor{red}{\Theta_{1}:=\int^{\infty}_{-\infty}d\omega~\frac{\omega^2}{\left(1-e^{-2\pi k\omega}\right)\left(\omega+\omega_0\right)\left(\omega-\omega_0\right)}}}~.\eea
In the limiting approximation $2\pi k\omega>>1$, one can further expand the integrand by taking large $k\omega$ approximation as:
\be {\cal V}(\omega_0,\omega,k):=~\frac{\omega^2}{\left(1-e^{-2\pi k\omega}\right)\left(\omega+\omega_0\right)\left(\omega-\omega_0\right)}\xrightarrow[2\pi k\omega>>1]{}\frac{\omega^2}{\left(\omega+\omega_0\right)\left(\omega-\omega_0\right)}:={\cal V}(\omega_0,\omega).\ee
This implies that, after taking large $k\omega$ approximation the integrand of $\Theta_1$ becomes independent of the parameter $k$. Now, further using this approximation the integral $\Theta_1$ can be further simplified as:
\bea \Theta_{2}&&\approx \int^{\infty}_{-\infty}d\omega~{\cal V}(\omega_0,\omega)=\underbrace{\int^{0}_{-\infty}d\omega~{\cal V}(\omega_0,\omega)}_{\textcolor{red}{\equiv~ {\cal D}_1(\omega_0)}}~+~\underbrace{\int^{\infty}_{0}d\omega~{\cal V}(\omega_0,\omega)}_{\textcolor{red}{\equiv~ {\cal D}_2(\omega_0)}},\eea
where we have written the integrals into two parts,represented by 
$\textcolor{red}{{\cal D}_1(\omega_0)}$ and $\textcolor{red}{{\cal D}_2(\omega_0)}$. Now, here we see that in the $2\pi k>>1$ limit we get:
\bea {\textcolor{red}{{\cal D}_1(\omega_0)}=\int^{0}_{-\infty}d\omega~{\cal V}(\omega_0,\omega)=-\int^{\infty}_{0}d\omega~{\cal V}(\omega_0,\omega)=-\textcolor{red}{{\cal D}_2(\omega_0)}}~.\eea
Now, here $\textcolor{red}{{\cal D}_1(\omega_0)}$ and $\textcolor{red}{{\cal D}_2(\omega_0)}$ gives divergent contributions in the frequency range, $-\infty<\omega<0$ and $0<\omega<\infty$. To get the finite regularised contributions from these integrals we introduce a cut-off regulator $\omega_c$, by following {\it Bethe regularisation} procedure. After introducing this cut-off we get the following result:
\bea {\textcolor{red}{{\cal D}_1(\omega_0,\omega_c)}=\int^{0}_{-\omega_c}d\omega~{\cal V}(\omega_0,\omega)=\int^{\omega_c}_{0}d\omega~{\cal V}(\omega_0,\omega)=\textcolor{red}{{\cal D}_2(\omega_0,\omega_c)}=\frac{1}{2}\left[\omega_c-\omega_0\tanh^{-1}\left(\frac{\omega_c}{\omega_0}\right)\right]}.~~~~~~~\eea
Consequently, we get the following regularised expression for the integral $\Theta_1$, as given by:
\bea {\textcolor{red}{~~~\Theta_{1}=\textcolor{red}{{\cal D}_1(\omega_0,\omega_c)}+\textcolor{red}{{\cal D}_2(\omega_0,\omega_c)}=\left[\omega_c-\omega_0\tanh^{-1}\left(\frac{\omega_c}{\omega_0}\right)\right]}}~.\eea
Now, if we further use the approximation that the cut-off is small compared to $\omega_0$  i.e. $\omega_c<<\omega_0$, then we get~\footnote{In the limit, $\omega_c<<\omega_0$ we can approximate the Taylor series expansion of the following function as:
	\be \tanh^{-1}\left(\frac{\omega_c}{\omega_0}\right)=\left(\frac{\omega_c}{\omega_0}\right)+\frac{1}{3}\left(\frac{\omega_c}{\omega_0}\right)^3+\cdots\approx \left(\frac{\omega_c}{\omega_0}\right)<<1.\ee}:
\bea {\textcolor{red}{\boxed{\boxed{{\bf Integral ~I:}~~~\Theta_{2}=\textcolor{red}{{\cal D}_1(\omega_0,\omega_c)}+\textcolor{red}{{\cal D}_2(\omega_0,\omega_c)}=\left[\omega_c-\omega_0\left(\frac{\omega_c}{\omega_0}\right)\right]\approx 0}}}}~.\eea

\subsection{\textcolor{blue}{Integral~II}}
In this subsection we explicitly compute the finite contribution from the following integral:
\bea \textcolor{red}{\Theta_{2}:=\int^{\infty}_{-\infty}d\omega~\frac{\omega\omega_0}{\left(1-e^{-2\pi k\omega}\right)\left(\omega+\omega_0\right)\left(\omega-\omega_0\right)}}~.\eea
It is important to note that in the limit, $2\pi k\omega>>1$, one can further expand the integrand given by the following approximation as:
\be {\cal M}(\omega_0,\omega,k):=~\frac{\omega_0~\omega}{\left(1-e^{-2\pi k\omega}\right)\left(\omega+\omega_0\right)\left(\omega-\omega_0\right)}\xrightarrow[2\pi k\omega>>1]{}\frac{\omega_0~\omega}{\left(\omega+\omega_0\right)\left(\omega-\omega_0\right)}:={\cal M}(\omega_0,\omega).\ee
This implies that, after taking limit $2\pi k\omega>>1$ the integrand of $\Theta_2$ becomes independent of the parameter $k$. Now, further using this approximation the integral $\Theta_1$ can be expressed as:
\bea \Theta_{1}&&\approx \int^{\infty}_{-\infty}d\omega~{\cal M}(\omega_0,\omega)=\underbrace{\int^{0}_{-\infty}d\omega~{\cal M}(\omega_0,\omega)}_{\textcolor{red}{\equiv~ {\cal N}_1(\omega_0)}}~+~\underbrace{\int^{\infty}_{0}d\omega~{\cal M}(\omega_0,\omega)}_{\textcolor{red}{\equiv~ {\cal N}_2(\omega_0)}},\eea
where we have written the integrals into two parts, indicated by 
$\textcolor{red}{{\cal N}_1(\omega_0)}$ and $\textcolor{red}{{\cal N}_2(\omega_0)}$. Now, in the limit $2\pi k\omega>>1$, we get:
\bea \textcolor{red}{{\cal N}_1(\omega_0)}=\int^{0}_{-\infty}d\omega~{\cal M}(\omega_0,\omega)=-\int^{\infty}_{0}d\omega~{\cal M}(\omega_0,\omega)=-\textcolor{red}{{\cal N}_2(\omega_0)}~.\eea
Now, here $\textcolor{red}{{\cal N}_1(\omega_0)}$ and $\textcolor{red}{{\cal N}_2(\omega_0)}$ gives divergent contributions in the frequency range, $-\infty<\omega<0$ and $0<\omega<\infty$. To get the finite contributions from these integrals we introduce a cut-off regulator $\omega_c$, by following {\it Bethe regularisation} procedure. After introducing this cut-off we get the following finite contribution:
\bea{\textcolor{red}{{\cal N}_1(\omega_0,\omega_c)}=\int^{0}_{-\omega_c}d\omega~{\cal M}(\omega_0,\omega)=-\int^{\omega_c}_{0}d\omega~{\cal M}(\omega_0,\omega)=-\textcolor{red}{{\cal N}_2(\omega_0,\omega_c)}=-\frac{\omega_0}{2}\ln\left[1-\left(\frac{\omega_c}{\omega_0}\right)^2\right]}.~~~~~~~\eea
Consequently, we get the following expression for the integral $\Theta_2$, as given by:
\bea {\textcolor{red}{\boxed{\boxed{{\bf Integral ~II:}~\Theta_{1}=\textcolor{red}{{\cal U}_1(\omega_0,\omega_c)}+\textcolor{red}{{\cal U}_2(\omega_0,\omega_c)}=\frac{\omega_0}{2}\ln\left[1-\left(\frac{\omega_c}{\omega_0}\right)^2\right]-\frac{\omega_0}{2}\ln\left[1-\left(\frac{\omega_c}{\omega_0}\right)^2\right]=0}}}}.~~~~~~~~\eea

\subsection{\textcolor{blue}{ Integral~III}}
In this subsection we explicitly compute the finite contribution from the following integral:
\bea {\textcolor{red}{\Theta_{3}:=\int^{\infty}_{-\infty}d\omega~\frac{\omega^2}{\left(1-e^{-2\pi k\omega}\right)\left(\omega+\omega_0\right)\left(\omega-\omega_0\right)}f\left(\omega,\frac{L}{2}\right)}}~,\eea
where, we define the function $f\left(\omega,\frac{L}{2}\right)$ given by the following expression:
\bea { \textcolor{red}{f\left(\omega,\frac{L}{2}\right)=\frac{1}{L\omega\sqrt{1+\left(\frac{L}{2k}\right)^2}}\sin(2k\omega \sinh^{-1}\left(\frac{L}{2k}\right))}}~.\eea
In the limit $2\pi k>>1$, one can further expand the integrand as:
\be {\cal Z}(\omega_0,\omega,k):=~\frac{\omega^2~f\left(\omega,\frac{L}{2}\right)}{\left(1-e^{-2\pi k\omega}\right)\left(\omega+\omega_0\right)\left(\omega-\omega_0\right)}\xrightarrow[2\pi k\omega\>>1]{}\frac{\omega^2~f\left(\omega,\frac{L}{2}\right)}{\left(\omega+\omega_0\right)\left(\omega-\omega_0\right)}:={{\cal Z}(\omega_0,\omega,k)}.\ee
This implies that, in the limit $2\pi k>>1$ the integrand of $\Theta_3$ is not independent of the parameter $k$. Now, further using this approximation the integral $\Theta_3$ can be further simplified as:
\bea \Theta_{3}&&\approx \int^{\infty}_{-\infty}d\omega~{{\cal Z}(\omega_0,\omega,k)}=\underbrace{\int^{0}_{-\infty}d\omega~\widetilde{{\cal Z}(\omega_0,\omega,k)}}_{\textcolor{red}{\equiv~ {\cal R}^{l}_1(\omega_0,k)}}~+~\underbrace{\int^{\infty}_{0}d\omega~\widetilde{{\cal Z}(\omega_0,\omega,k)}}_{\textcolor{red}{\equiv~ {\cal R}^{l}_2(\omega_0,k)}},\eea
where we have written the integrals into two parts, indicated by 
$\textcolor{red}{{\cal R}^{l}_1(\omega_0,k)}$ and $\textcolor{red}{{\cal R}^{l}_2(\omega_0,k)}$. Now, here in the limit $2\pi k>>1$ we get:
\bea \textcolor{red}{{\cal R}^{l}_1(\omega_0,k)}=\int^{0}_{-\infty}d\omega~{{\cal Z}(\omega_0,\omega,k)}&=&\int^{\infty}_{0}d\omega~{{\cal Z}(\omega_0,\omega,k)}=\textcolor{red}{{\cal R}^{l}_2(\omega_0,k)}\nonumber\\
&=&\frac{\pi}{2L\sqrt{1+\left(\frac{L}{2k}\right)^2}}\cos(2k\omega_0 \sinh^{-1}\left(\frac{L}{2k}\right))~.~~~~~\eea
Consequently, we get the following expression for the integral $\Theta_3$, given by:
\bea {\textcolor{red}{\boxed{\boxed{{\bf Integral ~III:}~~~\Theta_{3}=\textcolor{red}{{\cal R}^{l}_1(\omega_0,k)}+\textcolor{red}{{\cal R}^{l}_2(\omega_0,k)}=\frac{\pi}{L\sqrt{1+\left(\frac{L}{2k}\right)^2}}\cos(2k\omega_0 \sinh^{-1}\left(\frac{L}{2k}\right))}}}}~.~~~~~~~~\eea
Further, substituting $k\omega_0=i\left(n+\frac{1}{2}\right)~\forall~n\in \mathbb{Z}$, we get the following simplified expression for the integral  $\Theta_3$, given by:
\bea {\textcolor{red}{\boxed{\boxed{{\bf Integral ~III:}~~~\Theta_{3}=\textcolor{red}{{\cal R}_1(\omega_0,k)}+\textcolor{red}{{\cal R}_2(\omega_0,k)}=\frac{\pi}{L\sqrt{1+\left(\frac{L}{2k}\right)^2}}\cosh(\left(2n+1\right) \sinh^{-1}\left(\frac{L}{2k}\right))}}}}~.~~~~~~~~\eea
\subsection{\textcolor{blue}{ Integral~IV}}
In this subsection we explicitly compute the finite contribution from the following integral:
\bea {\textcolor{red}{\Theta_{4}:=\int^{\infty}_{-\infty}d\omega~\frac{\omega\omega_0}{\left(1-e^{-2\pi k\omega}\right)\left(\omega+\omega_0\right)\left(\omega-\omega_0\right)}f\left(\omega,\frac{L}{2}\right)}}~,\eea
where, we define the function $f\left(\omega,\frac{L}{2}\right)$ given by the following expression:
\bea { \textcolor{red}{f\left(\omega,\frac{L}{2}\right)=\frac{1}{L\omega\sqrt{1+\left(\frac{L}{2k}\right)^2}}\sin(2k\omega \sinh^{-1}\left(\frac{L}{2k}\right))}}~.\eea
In the limit $2\pi k>>1$, one can further expand the integrand as:
\be {\cal S}(\omega_0,\omega,k):=~\frac{\omega\omega_0~f\left(\omega,\frac{L}{2}\right)}{\left(1-e^{-2\pi k\omega}\right)\left(\omega+\omega_0\right)\left(\omega-\omega_0\right)}\xrightarrow[2\pi k\omega\>>1]{}\frac{\omega\omega_0~f\left(\omega,\frac{L}{2}\right)}{\left(\omega+\omega_0\right)\left(\omega-\omega_0\right)}:={{\cal S}(\omega_0,\omega,k)}.\ee
This implies that, in the limit $2\pi k>>1$ the integrand of $\Theta_3$ is not independent of the parameter $k$. Now, further using this approximation the integral $\Theta_4$ can be further simplified as:
\bea \Theta_{4}&&\approx \int^{\infty}_{-\infty}d\omega~{{\cal S}(\omega_0,\omega,k)}=\underbrace{\int^{0}_{-\infty}d\omega~{{\cal S}(\omega_0,\omega,k)}}_{\textcolor{red}{\equiv~ {\cal L}_1(\omega_0,k)}}~+~\underbrace{\int^{\infty}_{0}d\omega~{{\cal S}(\omega_0,\omega,k)}}_{\textcolor{red}{\equiv~ {\cal L}_2(\omega_0,k)}},\eea
where we have written the integrals into two parts, indicated by 
$\textcolor{red}{{\cal L}_1(\omega_0,k)}$ and $\textcolor{red}{{\cal L}_2(\omega_0,k)}$. Now, here in the limit $2\pi k>>1$ we get:
\bea \textcolor{red}{{\cal L}_1(\omega_0,k)}=\int^{0}_{-\infty}d\omega~{{\cal S}(\omega_0,\omega,k)}&=&-\int^{\infty}_{0}d\omega~{{\cal S}(\omega_0,\omega,k)}=-\textcolor{red}{{\cal L}_2(\omega_0,k)}\nonumber\\
&=&-\frac{i\sqrt{\pi}}{2L\sqrt{1+\left(\frac{L}{2k}\right)^2}}\underbrace{G_{1,3}^{2,1}\left(-\omega^2_0 k^2 \sinh ^{-1}\left(\frac{L}{2 k}\right)^2|
	\begin{array}{c}
		\frac{1}{2} \\
		\frac{1}{2},\frac{1}{2},0 \\
	\end{array}
	\right)}_{\textcolor{red}{\bf Meijer~G~function}}~.~~~~~~~~~~~\eea
Consequently, we get the following expression for the integral $\Theta_4$, as given by:
\bea {\textcolor{red}{\boxed{\boxed{{\bf Integral ~IV:}~~~\Theta_{4}=\textcolor{red}{{\cal L}_1(\omega_0,k)}+\textcolor{red}{{\cal L}_2(\omega_0,k)}=0}}}}~.~~~~~~~~\eea

\end{document}